\documentclass[lettersize,journal]{IEEEtran}
\usepackage[T1]{fontenc}

\usepackage{amsmath,amssymb,amsfonts}
\usepackage{graphicx}
\usepackage{array}
\usepackage{booktabs}
\usepackage{multirow}
\usepackage{braket}
\usepackage[breakable,skins]{tcolorbox}

\usepackage{cite}

\usepackage[caption=false,font=footnotesize]{subfig}

\usepackage{algorithm}
\usepackage{algpseudocode}
\algrenewcommand\algorithmicrequire{\textbf{Input:}}
\algrenewcommand\algorithmicensure{\textbf{Output:}}
\algdef{SE}[INDENT]{Indent}{EndIndent}{}{}
\algtext*{Indent}
\algtext*{EndIndent}

\usepackage{stfloats} 
\usepackage[hyphens]{url}
\usepackage{verbatim}
\usepackage{xcolor}
\usepackage{amsthm}
\usepackage{makecell}
\newtheorem{theorem}{Theorem}
\newtheorem{proposition}{Proposition}
\newtheorem{corollary}{Corollary}
\newtheorem{definition}{Definition}
\DeclareMathOperator{\Tr}{Tr}

\newcommand{\ketbra}[2]{\lvert #1 \rangle\!\langle #2 \rvert}
\newcommand{\HS}{\mathcal{H}}
\newcommand{\Cone}{\mathrm{Cone}}
\newcommand{\zeron}{\ket{0^{n}}}

\usepackage{float}
\usepackage{tikz}
\usetikzlibrary{shapes.geometric,arrows,positioning,fit,backgrounds}
\usepackage{tcolorbox}
\usepackage{todonotes}
\usepackage{enumitem}
\usepackage{balance}

\tikzset{
    startstop/.style={
        rectangle, 
        rounded corners,
        minimum width=3cm, 
        minimum height=0.5cm,
        text centered, 
        draw=black
    },
    process/.style={
        rectangle, 
        minimum width=3cm, 
        minimum height=0.5cm, 
        text centered, 
        draw=black
    },
    decision/.style={
        diamond, 
        minimum width=3cm, 
        minimum height=0.5cm, 
        text centered, 
        draw=black
    },
    arrow/.style={
        ->,
        thick,
        >=stealth
    }
}

\begin{document}
\pagecolor{white}
\color{black}
\title{QMon: Monitoring the Execution of Quantum Circuits with Mid-Circuit Measurement and Reset}

\author{%
\IEEEauthorblockN{%
Ning Ma\IEEEauthorrefmark{1},
Jianjun Zhao\IEEEauthorrefmark{2},
Foutse Khomh\IEEEauthorrefmark{1},
Shaukat Ali\IEEEauthorrefmark{3},
Heng Li\IEEEauthorrefmark{1}}%

\IEEEauthorblockA{\IEEEauthorrefmark{1}Department of Computer and Software Engineering, Polytechnique Montreal, Montreal, Canada\\
\{ning.ma, foutse.khomh, heng.li\}@polymtl.ca}

\IEEEauthorblockA{\IEEEauthorrefmark{2}Faculty of Information Science and Electrical Engineering, Kyushu University, Fukuoka, Japan\\
zhao@ait.kyushu-u.ac.jp}

\IEEEauthorblockA{\IEEEauthorrefmark{3}Department of Engineering Complex Software Systems, Simula Research Laboratory, Oslo, Norway\\
shaukat@simula.no}
}

\date{}


\maketitle

\begin{abstract}
Monitoring quantum circuits is difficult because measurements can disturb
the computation and arbitrary unknown states cannot be copied for separate
inspection. These constraints complicate debugging and pre-deployment
validation.

This paper presents \textbf{\textsc{QMon}}, a development-time instrumentation method that leverages mid-circuit measurements, reset operations, and causal-cone replay to monitor selected intermediate values of quantum circuits while preserving the full state of the original circuit qubits under
the stated ideal conditions. \textsc{QMon} enables the instrumentation of monitoring operators at selected locations within the circuit, allowing comparisons between expected and observed one-qubit outcome probabilities at those locations.

Under an ideal noise-free model, we prove that \textsc{QMon} preserves the full circuit state when the monitored qubit is separable, replay is exact, and the measurement record does not control later operations. Across 310 benchmark circuits with a 24-qubit limit per run, \textsc{QMon} monitors 44.54\% of gates and 89.75\% of circuit qubits at least once, on average. On 2,860 simulated buggy circuits (mutated circuits that alter final outputs), it achieves a mean detection rate of 71.1\%. 
Under calibration-informed simulated noise, exploratory comparisons show that \textsc{QMon} detects more faults that could affect its candidate monitoring locations than statistical assertions at all five noise scales and dynamic-ancilla assertions at all four nonzero scales.

\end{abstract}




 
\section{Introduction}
\label{sec:introduction}

Over the last decade, quantum computing has attracted growing attention from
academia, governments, and industry. This interest has supported the
development of quantum software frameworks and languages, such as
Qiskit~\cite{javadi2024quantum} and Q\#~\cite{svore2018q}, as well as work on
practical quantum applications~\cite{lau2022nisq}. Recent experimental
progress has further increased expectations about the impact of quantum
computing~\cite{kim2023utility,zhong2020quantum}.

As quantum computing moves from laboratory prototypes to cloud-accessible
platforms such as IBM Quantum~\cite{steffen2011quantum,ibm2016quantumcloud} and Amazon
Braket~\cite{gonzalez2021cloud}, quantum software development is becoming an
active engineering discipline~\cite{javadi2024quantum,
paltenghi2024surveytestinganalysisquantum,
ramalho2024testingdebuggingquantumprograms}. This growth increases the need
for effective methods to test, debug, and observe quantum programs during
execution. Existing methods can have difficulty identifying
quantum-specific bugs and locating the circuit operations that caused
them~\cite{10.1145/3527330,ramalho2024testingdebuggingquantumprograms}.
Effective quality assurance for quantum software is therefore increasingly
important~\cite{metwalli2022tool,10.1145/3643667.3648226}.

Observing a quantum circuit during execution is more difficult than observing
a classical program. Classical logging and runtime tracing can inspect or
copy program data without changing the computation. In contrast, the
no-cloning theorem prevents an arbitrary unknown quantum state from being
copied, and measurement can change the state being observed
~\cite{nielsen2010quantum}. Intermediate observation therefore requires
conditions under which measurement and subsequent reconstruction do not
change the intended computation.

Quantum-program bugs also make diagnosis difficult. Examples include
incorrect gate sequences, misplaced measurements, faulty control logic, and
incorrect quantum phases
~\cite{ramalho2024testingdebuggingquantumprograms}. Moreover, quantum-program
outputs are probabilistic. A fault may change how often different outcomes
occur without making every execution fail~\cite{10.1145/3527330}. Final
measurements may therefore show that the output has changed while providing
little information about where the problem began.

To address this problem, we propose \textbf{\textsc{QMon}}, a
development-time method for recording selected intermediate measurement
outcomes while allowing the same circuit execution to continue.
\textsc{QMon} selects locations where a
numerical test indicates that the target qubit is not entangled with the
other qubits. At each selected mid-circuit location, \textsc{QMon} measures the qubit
to check for deviations from its expected one-qubit state, including
changes in relative phase. It then resets the qubit and restores its
state by repeating the relevant earlier gates that prepared it, using
newly initialized extra qubits in place of the other qubits involved
in those operations, before resuming the remaining gates of the circuit. We call the repetition of these earlier gates \emph{causal-cone replay}.

The key idea is simple. In an ideal noise-free setting, measuring one qubit
does not disturb the other qubits when it is not entangled with them. If
\textsc{QMon} then reconstructs the measured qubit exactly, the complete
circuit state can be restored before execution continues. \textbf{This guarantee
assumes that the reconstruction repeats the same preparation operations,
that newly initialized extra qubits are used, and that the recorded
measurement result is not used to control later operations.} If the monitored
qubit is entangled with the other qubits, the same procedure changes the
full state. Section~\ref{sec:no-disturbance} states the guarantee, with proofs in Appendix~\ref{app:no-disturbance-proof} and the disturbance law in Appendix~\ref{app:tightlaw}.

\textsc{QMon} is intended for debugging and
pre-deployment validation of circuit logic rather than for characterizing
hardware noise. In contrast to approaches that mainly characterize hardware
noise~\cite{muqeet2024mitigating} or study quantum fault
tolerance~\cite{willsch2018testing}, \textsc{QMon} focuses on logic bugs in
circuit programs. We evaluate the complete method in noise-free simulation~(RQ1--RQ3)
and under calibration-informed simulated noise (RQ4).
The RQ4 noise model uses gate error rates and qubit relaxation times
from device calibration data.
The shot-based experiments still repeat each circuit a finite number of
times, so they include ordinary sampling variation.

We implement and evaluate \textsc{QMon} on quantum benchmark circuits. The
study examines whether monitoring changes the circuit state, how much of a
circuit can be monitored under a limit on the number of qubits used in one
run, how effectively faults can be detected and located, how many
instrumented runs are required, and how the selection process scales to
larger circuits. We organize the evaluation around the following research
questions.

\begin{itemize}[leftmargin=*]
    \item \textbf{RQ1: When Does \textsc{QMon} Preserve the Full Circuit State?}
    We prove that \textsc{QMon} leaves the full circuit state unchanged when
    the monitored qubit is not entangled with the other qubits and its
    preparation is reconstructed exactly under the stated conditions. Across
    310 circuits, the numerical test selects 22,332 of 56,386 gate--qubit
    locations, and an independent numerical calculation reproduces the same
    selected set. We also test all 56,386 locations separately, inserting one monitor
    per evaluation, including at locations that fail the selection test. The results match
    the theoretical boundary: monitoring leaves the full state unchanged only
    when the target qubit is not entangled with the others, and we derive an
    exact expression for the amount of state change otherwise.

    \item \textbf{RQ2: How much of a circuit can \textsc{QMon} monitor?}
    We measure the fraction of gates with observable points on at least one operand,
    the fraction of circuit qubits monitored at least once, and how
    far through the circuit the observable points reach, including eligible final readouts. When each
    instrumented run is limited to 24 qubits, \textsc{QMon} monitors
    44.54\% of gates on average, monitors at least one
    location on 89.75\% of the qubits on average, and has its deepest
    observable point at 48.57\% of the circuit depth on average. We also study how
    monitoring coverage varies with circuit structure and entanglement.

    \item \textbf{RQ3: How effectively does \textsc{QMon} detect and locate
    programming errors?}
    On 238 circuits and five mutation seeds, we compare \textsc{QMon} with
    three assertion-based methods and a simulation-only debugger. All methods return
    a result for 2,860 faulty circuit variants whose final output distribution
    differs from that of the original circuit. \textsc{QMon} detects
    $71.1\pm4.5\%$, compared with $67.4\pm4.4\%$ to $71.3\pm4.4\%$ for
    the three assertion methods. For the 2,217 faults whose changed gate can
    affect at least one location selected by \textsc{QMon}'s numerical test,
    the detection rate is $91.4\pm0.8\%$. Because each instrumented run is limited to 24 qubits, instrumenting all checkpoints of a circuit at once may exceed this limit. We therefore partition the checkpoints into batches, each executed in a separate instrumented run. When the assertion methods use the same checkpoints and batches as
    \textsc{QMon}, their observed detection rates are similar. Batched
    statistical assertions have a false-alarm rate of $52.9\%$, compared with
    $0.8\%$ for \textsc{QMon}. Moreover, 125 of the 238 \textsc{QMon} plans require only a single batch. 

    \item \textbf{RQ4: How does noise affect detection and false alarms?}
    We evaluate 100 circuits under five scales of calibration-informed
    simulated noise. For 762 of the 1,373 faults tested, the changed gate
    can affect at least one location selected by \textsc{QMon}'s numerical
    test. \textsc{QMon} detects $98.0\%$ of these faults at the lowest
    noise scale and $96.5\%$ at the highest.
\end{itemize}

This work makes the following contributions.

\begin{itemize}[leftmargin=*]
    \item We introduce \textsc{QMon}, a measure--reset--replay monitoring method
    that records selected intermediate 0/1 outcomes, reconstructs the monitored
    qubit, and allows the same circuit execution to continue. We provide an exact
    state-level characterization of this process by proving the preservation
    conditions for one monitor and composed monitor groups and by quantifying the
    state change caused by applying one monitor to an entangled qubit.

    \item We evaluate where \textsc{QMon} can select and deploy monitors, how
    its coverage varies with circuit structure and entanglement, and how a
    limit on the number of qubits per run affects deployment. We also evaluate
    the selection process on larger circuits using a matrix-product-state
    (MPS) simulator.

    \item We compare \textsc{QMon} with three published assertion methods and a simulation-only debugger. The comparison covers fault detection, false alarms, identification of suspect circuit regions, categories of missed faults, and multiple simultaneous faults. We also evaluate \textsc{QMon} and the three assertion methods under calibration-informed simulated noise.
\end{itemize}

We provide a replication package for reproducing and extending this
work\footnote{\url{https://github.com/mooselab/quantum_monitoring}}.

\textbf{Paper organization.}
Section~\ref{sec:related-work} discusses related research.
Section~\ref{sec:background} introduces the quantum-computing concepts needed
for this work. Section~\ref{sec:methodology} presents the \textsc{QMon}
methodology. Section~\ref{sec:setup} describes the experimental setup, and
Section~\ref{sec:result} presents the results. Section~\ref{sec:discussion}
discusses their implications, Section~\ref{sec:threats-to-validity} describes
the threats to validity, and Section~\ref{sec:conclusion} concludes the paper.

\section{Related Work}
\label{sec:related-work}
Prior work on quantum assertions, debugging, testing, monitoring, and
mid-circuit measurement differs in the intermediate information it provides
and in how the check affects the execution being examined.
Table~\ref{tab:related-compare} compares the closest approaches in terms of
the information returned, separability requirements, explicit state
restoration, the effect on the checked execution, localization, and
end-to-end evaluation on physical quantum hardware. The following five
subsections place these approaches in the broader literature.

\begin{table*}[!t]
\centering
\caption{Comparison with the closest approaches for observing or checking
intermediate quantum-circuit behavior.}
\label{tab:related-compare}
\scriptsize
\setlength{\tabcolsep}{2pt}
\renewcommand{\arraystretch}{1}
\begin{tabular}{@{}
>{\raggedright\arraybackslash}p{0.17\textwidth}
>{\raggedright\arraybackslash}p{0.16\textwidth}
>{\raggedright\arraybackslash}p{0.08\textwidth}
>{\raggedright\arraybackslash}p{0.16\textwidth}
>{\raggedright\arraybackslash}p{0.18\textwidth}
>{\raggedright\arraybackslash}p{0.12\textwidth}
>{\centering\arraybackslash}p{0.06\textwidth}
@{}}
\toprule
Approach
& \shortstack[l]{Information returned}
& \shortstack[l]{Separability\\required?}
& \shortstack[l]{Explicit state restoration}
& \shortstack[l]{Effect on checked execution}
& Localization
& \shortstack{End-to-end\\hardware} \\
\midrule
Statistical assertions~\cite{10.1145/3307650.3322213}
& 0/1 counts and test result & No & None
& Run ends at checkpoint & Assertion site & No \\
Proq~\cite{10.1145/3428218}
& Pass/fail result & No & No reset or replay
& Depends on checked state & Assertion site & No \\
Dynamic assertions~\cite{10.1145/3373376.3378488,9407042,li2022exploiting}
& Ancilla outcome & No & No reset or replay
& Depends on checked state & Assertion site & Yes (IBM) \\
Test points~\cite{acharya2021test}
& Test-circuit outcomes & No & Not applicable
& Separate test run & Coarse gate set & Yes (IBM) \\
Circuit slicer~\cite{metwalli2022tool}
& Slice outcomes & No & Not applicable
& Separate slice run & Slice/source gates & No \\
MQT debugger~\cite{rovara2024debugger}
& Simulator state comparison & Yes & Not applicable
& Simulation only & Assertion cone & No \\
\textsc{QMon}
& Recorded 0/1 outcome & Yes & Reset + causal-cone replay
& Continues after reset + replay & Flagged causal cone & No \\
\bottomrule
\end{tabular}
\par\smallskip
\begin{minipage}{\textwidth}
\footnotesize
\raggedright
\textit{Note.} Here, separability means that the checked qubit or
subsystem is not entangled with the rest of the circuit.
``Explicit state restoration'' indicates whether an approach reconstructs
a directly measured program qubit before the checked execution continues.
A satisfied projection- or ancilla-based assertion may preserve the
checked state without explicit restoration. The final column indicates
whether the complete cited approach was evaluated end to end on physical
quantum hardware.
The MQT separability requirement applies only to the one-qubit
equality-assertion configuration evaluated here.
\end{minipage}
\end{table*}

\subsection{Classical software monitoring}

Software monitoring observes a running system and collects data about its
behavior and performance. These data help developers check whether the system
behaves as expected, identify problems, and improve the software over
time~\cite{cassar2017survey,9373305,bass2015devops,10.1145/3168505}. In classical software engineering, logs are a primary source of runtime data
and may be the only available record of how software behaved during an
execution~\cite{he2021survey,chen2021survey}. Logs record important states and
activities of operating systems and applications on common computing
platforms~\cite{9548437}. Developers insert log statements into source code
to capture this information and store it, for example in files or databases,
for later analysis~\cite{10.1007/s10664-024-10452-w}. These records support
tasks such as diagnosing failures~\cite{8719429}, predicting
performance~\cite{8526889}, and detecting anomalies
~\cite{hashemi2022sialog}.

Quantum circuits cannot be monitored in the same way by freely copying their
internal states, and measurement may change the computation. Quantum
log-like observations must therefore respect the no-cloning theorem and
manage measurement back-action, meaning the change caused by the measurement
itself. Classical monitoring techniques that assume access to program state
without changing it therefore require substantial adaptation for quantum
programs.

\subsection{Quantum monitoring}

Monitoring a quantum system is difficult because an arbitrary unknown state
cannot be copied and measurement generally changes the state being observed.
A theoretical debugging protocol uses projection measurements to test whether
a system has departed from its expected state without disturbing the expected
evolution~\cite{li2014debugging}. For noisy intermediate-scale quantum (NISQ)
devices, a continuous-monitoring approach estimates gate and readout noise
from routinely executed circuits and their measurement outcomes using a
tensor-network model~\cite{zolotarev2023continuous}. QuEST estimates
confidence in a circuit's output by comparing its measured distribution with
a pretrained reference distribution modeled using Gaussian functions. It can
terminate a computation when the estimated confidence falls below a chosen
threshold~\cite{kundu2024quest}.

Physical quantum devices can also drift after calibration, motivating methods
that assess their reliability during use. One approach inserts classical,
superposition, and uncompute test points and executes the resulting test
circuits, whose outputs are known in advance, to detect changes in
physical-qubit error rates~\cite{acharya2021test}. A later data-driven approach
combines backend and runtime error information through machine learning,
predicts the Probability of Successful Trials (PST), and uses this prediction
to assess circuit-output reliability~\cite{saravanan2021test}.

Although the theoretical projection-based protocol can preserve the expected
evolution, the practical systems above mainly characterize processor noise,
estimate confidence in the final circuit output, or execute separate test
circuits. None provides the combination targeted in this work: observing an
eligible intermediate value, restoring the measured state, and continuing
the same monitored execution. \textsc{QMon} addresses this gap through
mid-circuit measurement, reset, and reconstruction of the operations that
prepared the monitored qubit.

\subsection{Quantum assertion}
Quantum assertions check whether an intermediate quantum state satisfies a
property specified by the programmer. A common taxonomy distinguishes
measurement-based assertions from quantum-based assertions. The latter
typically use ancillas, which are auxiliary extra qubits, to infer properties
of the data state~\cite{ramalho2024testingdebuggingquantumprograms}. Because
an assertion may involve measurement, its effect on the current execution
depends on how the assertion is constructed and on whether the actual state
satisfies the asserted property.

Statistical assertions repeatedly execute the part of a circuit up to the
assertion point, measure the relevant qubits, and apply a statistical test to
determine whether the observed outcomes agree with an expected pattern
~\cite{10.1145/3307650.3322213}. Because the checked qubits are measured, each
such execution ends at the assertion point instead of continuing through the
rest of the circuit.

Projection-based assertions represent the acceptable states as a subspace and
use a projective measurement to test whether the current state lies in that
subspace~\cite{10.1145/3428218}. In the representative projection-based
formulation, the checked part of the circuit need not be unentangled from the
remaining qubits. A state that lies entirely in the asserted subspace passes
without being changed. Other states may be changed by the projection, and a
failed assertion terminates the execution
~\cite{10.1145/3428218,10.1145/3643667.3648226,
ramalho2024testingdebuggingquantumprograms}.

Ancilla-assisted assertions couple data qubits to extra qubits and then
measure the extra qubits to infer selected properties of the data state
~\cite{10.1145/3373376.3378488,9407042}. Although these methods avoid directly
measuring the data qubits, they do not always leave the data state unchanged.
When the actual state does not satisfy the asserted property, an extra qubit
may remain entangled with the data. Measuring it can then project or otherwise
change the data state
~\cite{ramalho2024testingdebuggingquantumprograms,li2022exploiting}.
Selected stabilizer-based assertion configurations have also been evaluated
on IBM quantum devices~\cite{li2022exploiting}.

Related debugging frameworks also use different execution models. A
circuit-slicing approach executes separately prepared stand-alone or
accumulated circuit slices, either in simulation or on a quantum device, and
maps the observed behavior back to the corresponding gates in the original
circuit~\cite{metwalli2022tool}. A simulation-only debugger instead accesses
intermediate states directly through the simulator, compares them with
user-specified expectations, and applies cone-of-influence analysis after a
failure. This analysis traces the instructions that could have affected the
failed assertion~\cite{rovara2024debugger}. These frameworks therefore obtain
diagnostic evidence from separate slice executions or from states available
inside a simulator, rather than restoring a measured state within the same
monitored execution.

Continuation without disturbance is not unique to \textsc{QMon}: a satisfied
projection- or ancilla-based assertion may also preserve the checked state. Although those assertion-based approaches can continue execution when the asserted condition holds, they may collapse/change states when the asserted conditions are violated. \textsc{QMon} selects locations where a numerical test indicates that
the target qubit is not entangled with the rest of the circuit. It combines
this selection with two intermediate reads and reset-and-replay
reconstruction. Under the ideal conditions in Section~\ref{sec:no-disturbance},
replay restores the state of the original circuit qubits before execution
continues.

\subsection{Quantum software testing}

The set of possible quantum-program inputs is effectively infinite because
qubits can be prepared in arbitrary superpositions, making exhaustive testing
impractical~\cite{paltenghi2024surveytestinganalysisquantum}. Several automated
testing methods have been proposed to address this problem. QuSBT (Quantum
Search-Based Testing) uses genetic algorithms to search for test cases that
expose failures in quantum programs
~\cite{10.1145/3510454.3516839}. QuCAT (Quantum Combinatorial Testing) applies
combinatorial testing~\cite{10.1145/1883612.1883618} to construct efficient
test suites and detects more faults than random testing in its evaluation
~\cite{wang2023qucatcombinatorialtestingtool}. Programs containing multiple
quantum subroutines create additional testing challenges. Recent work
addresses these challenges through frameworks that combine unit and
integration testing, thereby supporting a broader range of quantum programs
~\cite{long2023testingquantumprogramsmultiple}.

Benchmarks and mutation-analysis tools also support the evaluation and
improvement of quantum testing methods. Muskit provides mutation analysis for
IBM Qiskit programs by systematically generating faulty circuit variants,
called mutants, to evaluate test quality~\cite{9678563}. A large empirical
study of more than 700,000 mutants from real-world quantum circuits examines
how circuit characteristics, algorithm types, and mutation operators affect
testing results and provides recommendations for more targeted testing
~\cite{mendiluze2025quantum}.

\textsc{QMon} complements test generation and mutation analysis by providing
evidence from inside an execution, similar to checking logged values during
classical testing~\cite{1214327}. With expected one-qubit states derived from an original circuit,
these outcomes can support tests that check expected relations
between related executions. They can also narrow a detected deviation
to the earlier gates that may have caused it.

\subsection{Mid-circuit measurement and reset}

Mid-circuit measurement and reset allow a qubit to be measured and
reinitialized before the circuit finishes. Prior work has demonstrated these
operations on physical quantum hardware, enabling applications such as
quantum error correction and dynamic circuit execution
~\cite{nation2021measure, cj89-4h5t}. Complementary work developed a benchmarking suite based on
randomized benchmarking to measure the errors caused by mid-circuit
measurements on both the measured qubits and their unmeasured neighbors,
called spectator qubits~\cite{govia2023randomized}.

Researchers have also used these capabilities to reduce the number of qubits
required by a circuit. One compiler-assisted approach uses mid-circuit
measurement to enable qubit reuse~\cite{hua2022exploiting}. Later work
formalized qubit-reuse compilation through both exact and heuristic
algorithms~\cite{decross2023qubit}.

These studies establish hardware support for the individual mid-circuit
measurement and reset operations and use them for dynamic execution or
resource reduction. They do not evaluate the complete
measure--reset--replay process used by \textsc{QMon}. In this work, these
operations support development-time observation of selected intermediate
values followed by reconstruction and continued test execution.

\section{Background}
\label{sec:background}

This section introduces the quantum-computing concepts needed to understand
\textsc{QMon}, with a focus on the properties that affect circuit monitoring
and debugging.

\subsection{Quantum States, Superposition, and Entanglement}

A quantum bit (qubit) can exist in a superposition of $|0\rangle$ and
$|1\rangle$, described as
$
|\psi\rangle=\alpha|0\rangle+\beta|1\rangle,
$
where $\alpha,\beta\in\mathbb{C}$ and
$|\alpha|^2+|\beta|^2=1$~\cite{nielsen2010quantum}. For a pure multi-qubit state, a chosen subsystem is \emph{separable} from the
rest if the full state can be written as a product of states for the two
parts. In this case, the subsystem can be described independently of the
others. Otherwise, the two parts are entangled. For example, the Bell state
$
|\Phi^+\rangle=\frac{1}{\sqrt{2}}(|00\rangle+|11\rangle)
$
is maximally entangled. If both qubits are measured in the computational
basis, their outcomes are perfectly correlated: they are either both $0$ or
both $1$. For \textsc{QMon}, the key question is whether the state of one
qubit can be described independently of all other qubits at a selected point
in the circuit.

\subsection{No-Cloning Theorem}

The no-cloning theorem~\cite{nielsen2010quantum} states that an exact copy of
an arbitrary unknown quantum state cannot be created. This differs from
classical computing, where program data can usually be copied before it is
inspected or modified.

The no-cloning theorem does not mean that no information can ever
be obtained from a quantum state. However, measurement may change
the state being observed. Together, these two properties create important constraints for
quantum software testing and monitoring
~\cite{nielsen2010quantum}:

\begin{itemize}
    \item \textbf{Inspection:} A debugger cannot create an exact copy of an
    arbitrary unknown state for later inspection.

    \item \textbf{Testing:} Classical testing techniques that depend on
    copying intermediate program states cannot be transferred directly to
    quantum programs.

    \item \textbf{Monitoring:} An intermediate measurement may disturb the
    computation, so a monitoring method must state the conditions under which
    the execution is preserved.
\end{itemize}


\subsection{Quantum Circuits}

\begin{figure}[h]
    \centering
    \includegraphics[width=0.8\linewidth]{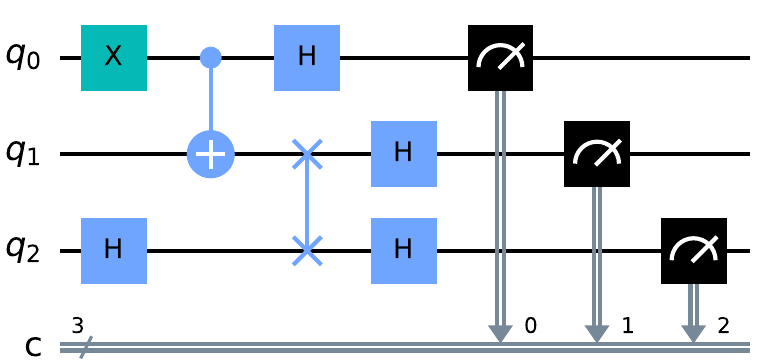}
    \caption{An example of a quantum circuit}
    \label{fig:An example of a Quantum Circuit}
    \vspace{-2mm}
\end{figure}

A quantum circuit applies a sequence of gates to qubit wires and stores
measurement outcomes in classical bits.
Fig.~\ref{fig:An example of a Quantum Circuit} shows a simple quantum circuit
developed with Qiskit~\cite{javadi2024quantum}. In this example,
$q_0,q_1,q_2$ are three qubits, and \texttt{c} denotes three classical bits. The circuit contains a NOT gate $X$ on $q_0$, shown as a green square labeled
``X'', and a Hadamard gate $H$ on $q_2$, shown as a blue square labeled
``H''. It also contains a controlled-NOT (CNOT) gate with control $q_0$ and
target $q_1$, shown as a filled control dot on $q_0$ connected to a
$\oplus$ symbol on $q_1$. A SWAP gate exchanges the states of $q_1$ and
$q_2$ and is shown as two $\times$ symbols connected by a vertical line. The
circuit also applies Hadamard gates to $q_0$, $q_1$, and $q_2$. Finally, it
measures the three qubits, shown by black meter icons, and stores their
outcomes in \texttt{c[0]}, \texttt{c[1]}, and \texttt{c[2]}, respectively.

In the following sections, a \emph{gate--qubit location} pairs a gate with
one of the qubits on which it acts. A one-qubit gate therefore defines one
such location, while a two-qubit gate defines two. Gate indices refer to the
order of gates in the circuit representation rather than to visual columns in
the circuit drawing.

\subsection{Phase Information and Measurement}

Quantum information is encoded not only in measurement probabilities but
also in the relative phases of the state amplitudes. Two states can therefore
produce the same measurement statistics in the computational basis, the
standard $\{|0\rangle,|1\rangle\}$ basis, while having different relative
phases. For example,
$
|\psi_1\rangle=\frac{1}{\sqrt{2}}(|0\rangle+|1\rangle)
$
and
$
|\psi_2\rangle=\frac{1}{\sqrt{2}}(|0\rangle+e^{i\phi}|1\rangle)
$
both produce outcomes $0$ and $1$ with probability $\frac{1}{2}$ when
measured in the computational basis, even though their relative phases can
differ. A computational-basis measurement alone does not reveal this phase
difference. Debugging a quantum circuit may therefore require checking both
measurement probabilities and phase relationships. For \textsc{QMon}, this
also means that reconstructing only the 0/1 probabilities is not sufficient;
the reconstruction must restore the monitored qubit's amplitudes and relative
phase.

\subsection{Density Matrix and Partial Trace}

The density matrix formalism~\cite{nielsen2010quantum} provides a general way
to describe both pure and mixed quantum states. A pure state $|\psi\rangle$
has density matrix
$
\rho=|\psi\rangle\langle\psi|.
$
Density matrices are especially useful for describing part of a larger
quantum system.

Given a composite system $AB$ in state $\rho_{AB}$, the state of subsystem
$A$ alone is obtained by taking the partial trace over subsystem $B$. This
operation is written as $\operatorname{Tr}_B$. For example, let $AB$ be
prepared in the Bell state
$
|\Phi^+\rangle=\frac{1}{\sqrt{2}}(|00\rangle+|11\rangle).
$
Its density matrix is
\[
\rho_{AB}
=\ket{\Phi^+}\!\bra{\Phi^+}
=\tfrac12
\begin{pmatrix}
1&0&0&1\\
0&0&0&0\\
0&0&0&0\\
1&0&0&1
\end{pmatrix}.
\]

The reduced state of qubit $A$ is obtained by tracing out $B$:

$$
\rho_A
=\operatorname{Tr}_B(\rho_{AB})
=\tfrac12\bigl(\ket0\!\bra0+\ket1\!\bra1\bigr)
=\tfrac12
\begin{pmatrix}
1&0\\[2pt]
0&1
\end{pmatrix}.
$$

Thus, measuring $A$ in the computational basis yields $\ket0$ and $\ket1$
with equal probability. The quantity $\operatorname{Tr}(\rho^2)$ is called
the purity of a density operator $\rho$. A state is pure if its purity is
one and mixed otherwise. Here, $\operatorname{Tr}(\rho_A^2)=1/2$, so
$\rho_A$ is mixed even though the global Bell state is pure. For a pure
bipartite state, a mixed reduced state implies entanglement between the two
subsystems~\cite{nielsen2010quantum}.

\subsection{Schmidt Decomposition and the Separability Certificate}
\label{sec:schmidt-certificate}

Schmidt decomposition is a mathematical criterion for \textsc{QMon} to decide whether one qubit
is separable from all other qubits. For a bipartite, or two-subsystem, pure state $\ket{\psi}_{XY}$, the Schmidt
decomposition is
$
\ket{\psi}_{XY}
=\sum_{k=1}^{r}a_k\,\ket{u_k}_X\ket{v_k}_Y,
$
where $\{\ket{u_k}_X\}_{k=1}^{r}$ and
$\{\ket{v_k}_Y\}_{k=1}^{r}$ are paired orthonormal Schmidt vectors,
$r$ is the Schmidt rank, and $a_k$ are the Schmidt amplitudes. The
amplitudes are ordered as
$
a_1\geq\cdots\geq a_r>0
$
and satisfy
$
\sum_{k=1}^{r}a_k^2=1.
$
Define the Schmidt probabilities as
$
\lambda_k=a_k^2.
$
These probabilities are the nonzero eigenvalues of either reduced state. In
particular,
$
\begin{aligned}
\rho_X=\operatorname{Tr}_Y\!\left(\ket{\psi}\!\bra{\psi}\right)
=\sum_{k=1}^{r}a_k^2\,\ket{u_k}\!\bra{u_k}
=\sum_{k=1}^{r}\lambda_k\,\ket{u_k}\!\bra{u_k}.
\end{aligned}
$

A bipartite pure state is separable exactly when its Schmidt decomposition
contains only one term, that is, when $r=1$. To use the same notation in this
case, we set $a_2=\lambda_2=0$. Separability is therefore equivalent to
$
r=1,a_2=0,\text{or }\lambda_2=0
$
~\cite{ekert1995entangled,nielsen2010quantum}. We now divide the circuit state into the monitored qubit $q$ and all other
qubits $R$. This is the one-qubit-versus-rest split $q\mid R$ immediately
after gate $i$, where
$
[n]=\{0,\ldots,n-1\},
R=[n]\setminus\{q\}.
$
Let
$
M_{i,q}\in\mathbb{C}^{2\times 2^{n-1}}
$
be the coefficient matrix obtained by reshaping the amplitudes of
$\ket{\psi_i}$. Its rows correspond to the two computational-basis states of
$q$, and its columns correspond to the computational-basis states of $R$.
The two nonnegative values returned by the singular-value decomposition of
$M_{i,q}$ are exactly the Schmidt amplitudes $a_1(i,q)$ and $a_2(i,q)$ across
this split. We order them from largest to smallest and set $a_2(i,q)=0$ when
the Schmidt rank is one.

The reduced state of the monitored qubit is
$
\rho_q(i)=M_{i,q}M_{i,q}^{\dagger}.
$
Consequently, the eigenvalues of $\rho_q(i)$ are
$a_1(i,q)^2$ and $a_2(i,q)^2$, or equivalently
$\lambda_1(i,q)$ and $\lambda_2(i,q)$. Hence,
$
\det\rho_q(i)
=a_1(i,q)^2a_2(i,q)^2
=\lambda_1(i,q)\lambda_2(i,q).
$

Because $\operatorname{Tr}\rho_q(i)=1$, we have
$
a_1(i,q)^2+a_2(i,q)^2=1,
$
and therefore $a_1(i,q)>0$. Thus,
\[
\det\rho_q(i)=0
\quad\Longleftrightarrow\quad
a_2(i,q)=0
\quad\Longleftrightarrow\quad
\lambda_2(i,q)=0.
\]
Since the global state $\ket{\psi_i}$ is pure, these conditions are also
equivalent to the separability of $q$ from $R$. In other words, the monitored
qubit is not entangled with the other qubits exactly when any of the three
quantities above is zero. The exact condition
$
\det\rho_q(i)=0
$
is the separability certificate assumed by
Theorem~\ref{thm:nodist} (see Section~\ref{sec:no-disturbance}). The implementation later approximates this exact
condition using a finite-precision numerical test.

\section{Methodology}
\label{sec:methodology}

\begin{figure*}[!t]
\centering
\includegraphics[width=0.9\linewidth]{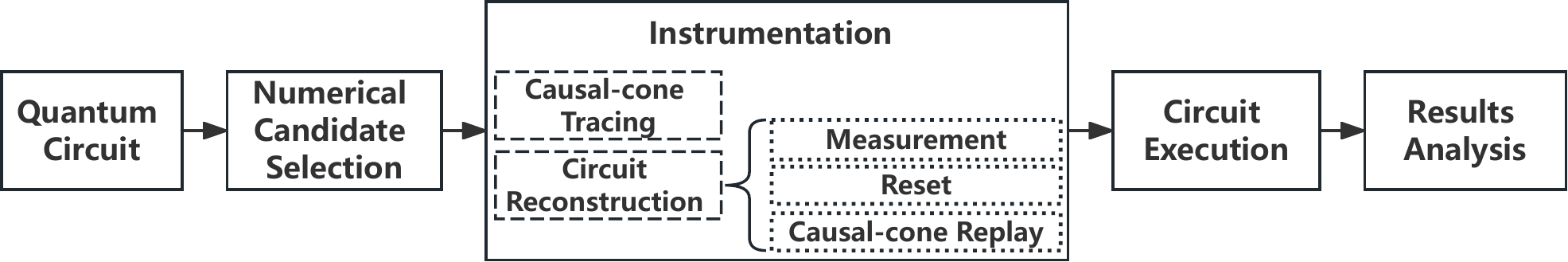}
\caption{Overview of the \textsc{QMon} workflow. Each deployed
mid-circuit checkpoint performs a computational-basis read and an
expected-state read. Each read is followed by reset and causal-cone
replay.}
\label{fig:methodology}
\vspace{-2mm}
\end{figure*}

In this section, we describe the \textsc{QMon} methodology for monitoring
quantum circuits using mid-circuit measurement, reset, and reconstruction.
The overall workflow is illustrated in Fig.~\ref{fig:methodology}. We first
provide an overview of \textsc{QMon} and then describe each step in detail.

\subsection{Overview}
\label{sec:overview}

Given a circuit, \textsc{QMon} first uses a numerical separability
test to select gate--qubit locations where the target qubit appears to be
separable from all other qubits
(Section~\ref{sec:node_selection}). It then traces the backward causal cone of
each selected location (Section~\ref{sec:dfs}) and reconstructs the
circuit under test (CUT) by inserting measurement, reset, and
causal-cone replay operations (Section~\ref{sec:reconstruction}).
The recorded outcomes are used to check consistency with the expected
one-qubit states (Section~\ref{sec:circuit_execution}).

Theorem~\ref{thm:nodist} proves state preservation for one exactly separable
monitored qubit. In Appendix~\ref{app:no-disturbance-proof}, Proposition~\ref{prop:grouped-monitor} states the condition
for monitoring several qubits after the same gate, and
Corollary~\ref{cor:compose} gives sufficient conditions for composing such
monitor groups in increasing gate order. These results assume exact causal-cone
replay and no use of the recorded
measurement outcomes to control later operations.

\subsection{Monitoring Location Selection Based on the No-Disturbance Theorem}
\label{sec:node_selection}
\label{sec:no-disturbance}

We analyze the target circuit after decomposing it into one- and two-qubit
gates. As defined in
Section~\ref{sec:background}, a \emph{gate--qubit location} pairs a gate
with one of the qubits on which it acts. Gate index $i$ follows zero-based
program order, so the first gate has index~0; it does not refer to a visual
column in the circuit drawing. We first state the condition that guarantees
no disturbance, then describe the numerical selection procedure and
illustrate it with an example.

\paragraph{\textbf{Exact condition and no-disturbance guarantee}}

Consider a circuit $C=U_{m-1}\cdots U_0$ consisting of $m$ unitary gates
on $n$ qubits initialized in $\ket{0^n}$. Let
$[n]=\{0,\ldots,n-1\}$ index these qubits. For
$i\in\{0,\ldots,m-1\}$, the state immediately after gate $i$, before
any monitor is inserted, is
$
\ket{\psi_i}=U_i\cdots U_0\ket{0^n}.
$
For a gate--qubit location $(i,q)$, let $R=[n]\setminus\{q\}$ denote
the other circuit qubits. The reduced density matrix of $q$ is
$
\rho_q(i)=\operatorname{Tr}_R
\left(\ket{\psi_i}\!\bra{\psi_i}\right).
$
We call $(i,q)$ \emph{monitorable} if $\det\rho_q(i)=0$.
Because $\ket{\psi_i}$ is pure, this condition holds exactly when $q$
is separable from all other qubits
(Section~\ref{sec:schmidt-certificate}).

At such a location, \textsc{QMon} measures $q$ in the computational
basis, records the outcome, resets $q$ to $\ket{0}$, and performs
replay of its backward causal cone on $q$ and extra qubits initialized
in $\ket{0}$, called fresh ancillas. The replay repeats the CUT's cone
gates with the same parameters, operand roles, and gate order, while
mapping each non-target wire to an ancilla. We call this \emph{matched
replay}. The other original circuit qubits are left untouched during replay. Let
$
V_{\mathrm{tail}}=U_{m-1}\cdots U_{i+1}
$
denote the remaining gates of the original circuit, with
$V_{\mathrm{tail}}=I$ when $i=m-1$.

\begin{theorem}[No disturbance]
\label{thm:nodist}
Assume ideal, noise-free execution of the original closed, unitary
circuit. Suppose that matched causal-cone replay is exact, uses fresh
ancillas, and does not use the recorded measurement outcome to choose
or modify later operations. If one \textsc{QMon} monitor is inserted
at a monitorable location $(i,q)$, then, after measurement, reset,
and replay, the reduced density operator of the original $n$ circuit
qubits is $\ketbra{\psi_i}{\psi_i}$. Here, the original qubits are
those of $C$, excluding the replay ancillas.

After the remaining original gates $V_{\mathrm{tail}}$ are executed,
the density operator of these qubits, before final measurement, is
$
V_{\mathrm{tail}}
\ketbra{\psi_i}{\psi_i}
V_{\mathrm{tail}}^\dagger.
$
This is exactly the final quantum state of the uninstrumented circuit.
The mid-circuit outcome remains available as a separate classical
record, while the final output distribution is unchanged.
\end{theorem}

To see why, the separability condition allows us to write
$
\ket{\psi_i}=\ket{u}_q\otimes\ket{\chi}_R,
$
where $\ket{u}$ is the state of $q$ and $\ket{\chi}$ is the joint
state of the remaining circuit qubits. Measuring $q$ may change
$\ket{u}$ but leaves $\ket{\chi}$ unchanged. Resetting $q$ then
produces $\ket{0}_q\otimes\ket{\chi}_R$ for either recorded outcome.
Exact matched replay reconstructs the complete state $\ket{u}$,
including its relative phase, and leaves the replay ancillas separate
from the original qubits. The original circuit therefore resumes from
$\ket{\psi_i}$, the same state it had immediately before the monitor.
This argument uses the standard measurement and partial-trace properties
of a product state~\cite{nielsen2010quantum}.

Appendix~\ref{app:no-disturbance-proof} provides the full proof of
Theorem~\ref{thm:nodist}, the converse for one forced monitor, and
sufficient conditions for composing multiple monitor groups. The converse shows that applying the same procedure to an entangled
qubit changes the joint state of the original qubits by a strictly positive
amount, although a particular final measurement may not reveal that change.
Appendix~\ref{app:tightlaw} derives the exact disturbance caused by
one forced monitor.

\paragraph{\textbf{Numerical selection rule}}

Theorem~\ref{thm:nodist} requires exact separability, whereas the
implementation uses floating-point arithmetic. Let $a_2(i,q)$ be the
second Schmidt amplitude across the split between $q$ and all other
qubits, or $0$ if the Schmidt rank is one. Its square is the smallest
eigenvalue of $\rho_q(i)$, denoted by $\lambda_{\min}(\rho_q(i))$.
As shown in Section~\ref{sec:schmidt-certificate},
\[
\det\rho_q(i)=0
\quad\Longleftrightarrow\quad
a_2(i,q)=0.
\]
The implementation selects $(i,q)$ when
$
a_2(i,q)\leq\varepsilon_{\mathrm{sv}},
\varepsilon_{\mathrm{sv}}=10^{-12}.
$
A location passing this finite-precision test is a
\emph{numerical candidate}. Passing the test provides numerical
evidence of separability, not a proof of exact equality to zero.
A \emph{deployed checkpoint} is a numerical candidate later included
in an instrumented run under the qubit budget.

We test $a_2(i,q)$ directly because the statevector implementation
obtains it through singular-value decomposition. Near a rank-one
state, computing $\det\rho_q(i)$ can subtract nearly equal
floating-point values, and the determinant decreases quadratically
with the small amplitude $a_2(i,q)$. It is therefore less suitable
as the selection quantity.

Algorithm~\ref{alg:screen} simulates the original circuit once using
a statevector, which stores its complete vector of amplitudes.
After each gate $i$, it tests every qubit acted on by that gate and
records the accepted gate indices in a separate list for each qubit.
For a two-qubit gate, the two qubits are tested separately, each
against all other circuit qubits. Untouched qubits need not be
retested because a unitary operation acting only on other qubits
cannot change their reduced density matrices. A qubit that becomes
entangled and later becomes separable again can therefore be selected
at a later location. All reported analyses use this per-gate rule.

\begin{algorithm}[t]
\begingroup
\makeatletter
\renewcommand{\fnum@algorithm}{\fname@algorithm~\thealgorithm}
\makeatother
\caption[Monitoring location selection]{Monitoring location selection}
\label{alg:screen}
\endgroup
\begin{algorithmic}[1]
\Require Circuit $qc$ with $n$ qubits; statevector tolerance
$\varepsilon_{\mathrm{sv}}$
\Ensure Per-qubit lists of numerical-candidate gate indices
\texttt{m\_lists}
\State Initialize $\ket{\psi}\gets\ket{0^n}$
\State Initialize $\texttt{m\_lists}[q]\gets[\,]$ for every qubit $q$
\For{each unitary gate index $i$ in execution order}
  \State Apply gate $qc.\texttt{data}[i]$ to $\ket{\psi}$
  \State Let \texttt{q\_list} be the qubits acted on by the gate
  \For{each $q\in\texttt{q\_list}$}
    \State $\rho_q(i)\gets\operatorname{Tr}_{[n]\setminus\{q\}}
      \left(\ket{\psi}\!\bra{\psi}\right)$
    \State Let $a_2(i,q)$ be the second Schmidt amplitude across
      $q\mid([n]\setminus\{q\})$, or $0$ if the Schmidt rank is one,
      so that $a_2(i,q)^2=\lambda_{\min}(\rho_q(i))$
    \If{$a_2(i,q)\le\varepsilon_{\mathrm{sv}}$}
      \State Append $i$ to $\texttt{m\_lists}[q]$
    \EndIf
  \EndFor
\EndFor
\State \Return \texttt{m\_lists}
\end{algorithmic}
\end{algorithm}

RQ1 reports the largest $a_2(i,q)$ among the numerical candidates
and an independently computed determinant residual. The residual
is only a cross-check, not part of the selection rule.
Appendix~\ref{app:proof} further distinguishes this numerical test
from the exact condition in Theorem~\ref{thm:nodist}, and
Appendix~\ref{sec:scalability} evaluates the same criterion using a
matrix product state (MPS) representation.

Terminal measurements are excluded from selection. A numerical
candidate immediately followed on that qubit by its ordinary final
computational-basis measurement does not receive an extra mid-circuit
monitor, because the final measurement already provides that 0/1 outcome.
These locations receive no second read.

Appendix~\ref{app:implementation-walkthroughs} illustrates the selection pass
using the circuit in Figure~\ref{fig:An example of a Quantum Circuit}.

\subsection{Tracing the Backward Causal Cone}
\label{sec:dfs}

A computational-basis measurement returns $0$ or $1$ and leaves the measured
qubit in the corresponding basis state. A subsequent reset prepares the qubit
in $\ket{0}$. These operations do not, by themselves, restore the amplitudes
and relative phase of the qubit before measurement. To reconstruct this
information, \textsc{QMon} traces the earlier gates that could have
contributed to preparing the monitored qubit and later replays those gates.

Starting from a numerical candidate, Algorithm~\ref{alg:dfs} scans backward
along the monitored qubit's wire. A one-qubit gate is added directly to
\texttt{path}. At a two-qubit gate, the search also follows the other input
wire because the state on that wire may have affected the monitored qubit
through the gate. The recursion continues until every branch reaches circuit
initialization.

The result is not a single path. It is the union of all backward paths that
could affect the monitored qubit, which forms its backward causal cone. The
\texttt{visited} set prevents the same gate--wire pair from being explored
more than once. Because the same gate may still be reached through different
wires, the final step removes duplicate gate indices and sorts the remaining
gates in increasing program order. This produces the ordered backward causal
cone used for replay. The tracing step uses only the circuit structure and
does not simulate a quantum state.

\begin{algorithm}[t]
\caption{Depth-first tracing of a backward causal cone}
\label{alg:dfs}
\begin{algorithmic}[1]
\Require Circuit $qc$; target qubit $q^\star$; starting gate index $i_0$
\Ensure Ordered gate list \texttt{path} for the backward causal cone
\State \texttt{visited}$\gets\{\,\}$; \texttt{path}$\gets[\,]$
\Function{Trace}{$q^\star,i_0$}
  \For{$i=i_0$ down to $0$}
    \If{$(i,q^\star)\in\texttt{visited}$ or gate $i$ does not act on $q^\star$}
      \State \textbf{continue}
    \EndIf
    \State add $(i,q^\star)$ to \texttt{visited}
    \If{gate $i$ acts on one qubit}
      \State append gate $i$ to \texttt{path}
    \Else
      \For{each other input qubit $q$ of gate $i$}
        \State \Call{Trace}{$q,i-1$}
      \EndFor
      \State append gate $i$ to \texttt{path}
    \EndIf
  \EndFor
\EndFunction
\State \Call{Trace}{$q^\star,i_0$}
\State deduplicate by gate index and sort in increasing program order
\color{black}
\State \Return \texttt{path}
\end{algorithmic}
\end{algorithm}

\subsection{Circuit Reconstruction}
\label{sec:reconstruction}
\label{sec:measurement-bases}

\textsc{QMon} uses two reads at each deployed checkpoint to obtain complementary information.
The first read measures the qubit in the computational basis. A computational-basis probability
can change linearly with a small parameter perturbation when its derivative is nonzero.
Statistical assertions can also detect this change. The second read checks the expected
one-qubit state, including its relative phase. A relative-phase change can leave computational-basis
probabilities unchanged while changing the overlap with the expected state.
Each read is followed by reset and causal-cone replay.

To construct an instrumented circuit, \textsc{QMon} copies the CUT gates in program order.
At each deployed checkpoint, it measures the target qubit in the computational basis,
stores the 0/1 outcome, and resets the qubit to $\ket{0}$. \textsc{QMon} then replays
the ordered backward causal cone returned by Algorithm~\ref{alg:dfs} using the CUT's
gates and parameters at the selected instruction indices. During the first replay,
the target wire remains the original monitored qubit. Each other wire in the causal cone
is mapped to a fresh ancilla. Here, a \emph{fresh ancilla} is an extra qubit initialized
in $\ket{0}$ and uncorrelated with the program qubits before the first replay.
The same ancillas are reset and reused for the second replay, as described below.
This mapping avoids applying the preparation gates a second time to the other original program qubits.

For the expected one-qubit state $\ket{\psi_q}$ at each checkpoint,
choose a unitary $U_q$ such that $U_q\ket{0}=\ket{\psi_q}$.
Applying $U_q^\dagger$ before a computational-basis measurement is equivalent to measuring
in the basis $\{U_q\ket{0},U_q\ket{1}\}$~\cite{nielsen2010quantum}.
The state $\ket{\psi_q}$ then produces outcome $0$ with certainty.
For a tested reduced state $\rho$, the probability of outcome $1$ is
$p_{\mathrm{adapt}}(1)=1-\bra{\psi_q}\rho\ket{\psi_q}$.
Appendix~\ref{app:rq3details:protocol} gives the numerical construction of $U_q$.
After the first replay, \textsc{QMon} applies $U_q^\dagger$, measures the target,
and records the second outcome. It then resets the target and all replay ancillas
to $\ket{0}$ and replays the same CUT cone before continuing the program.
The two replays reuse the same physical ancillas within the group.
Each replay starts with those ancillas in $\ket{0}$ and independent of the program.

The first read measures the incoming target state. For the second read,
$\rho$ is the target's reduced state after the first replay of the CUT,
before applying $U_q^\dagger$. The states before and after reconstruction agree
under the ideal preservation assumptions of Corollary~\ref{cor:two-read-preservation},
but may differ in a faulty or noisy execution.

We call both replays \emph{matched} when each applies the same causal-cone gates
with the same parameters and operand roles, in the same gate order as in the CUT,
while consistently replacing the non-target wires by their ancilla counterparts.
Under the separability and exact-replay conditions stated in Section~\ref{sec:no-disturbance},
matched replay reconstructs the monitored qubit's amplitudes and relative phase while
leaving the other original program qubits unchanged. After the second replay,
the remaining original circuit gates continue on the program qubits;
the replay ancillas are not used by those gates.

If several qubits acted on by the same gate are deployed as checkpoints,
\textsc{QMon} handles them together as one \emph{monitor group}. It measures
and resets all qubits in the group, forms the union of their backward causal cones,
and replays each gate in that union only once per replay.
This avoids repeating gates shared by the individual cones.
The monitored qubits remain on their original program wires, while every other wire
in the union cone is mapped to a fresh ancilla. The remaining program qubits are left untouched.
In Appendix~\ref{app:no-disturbance-proof}, Proposition~\ref{prop:grouped-monitor}
states the exact condition under which this grouped construction preserves the circuit state.

\subsection{Instrumentation Walkthrough}
\label{sec:example}

\begin{figure*}[t]
\centering
\newlength{\figThreeHeight}
\settoheight{\figThreeHeight}{%
    \includegraphics[width=0.69\textwidth]{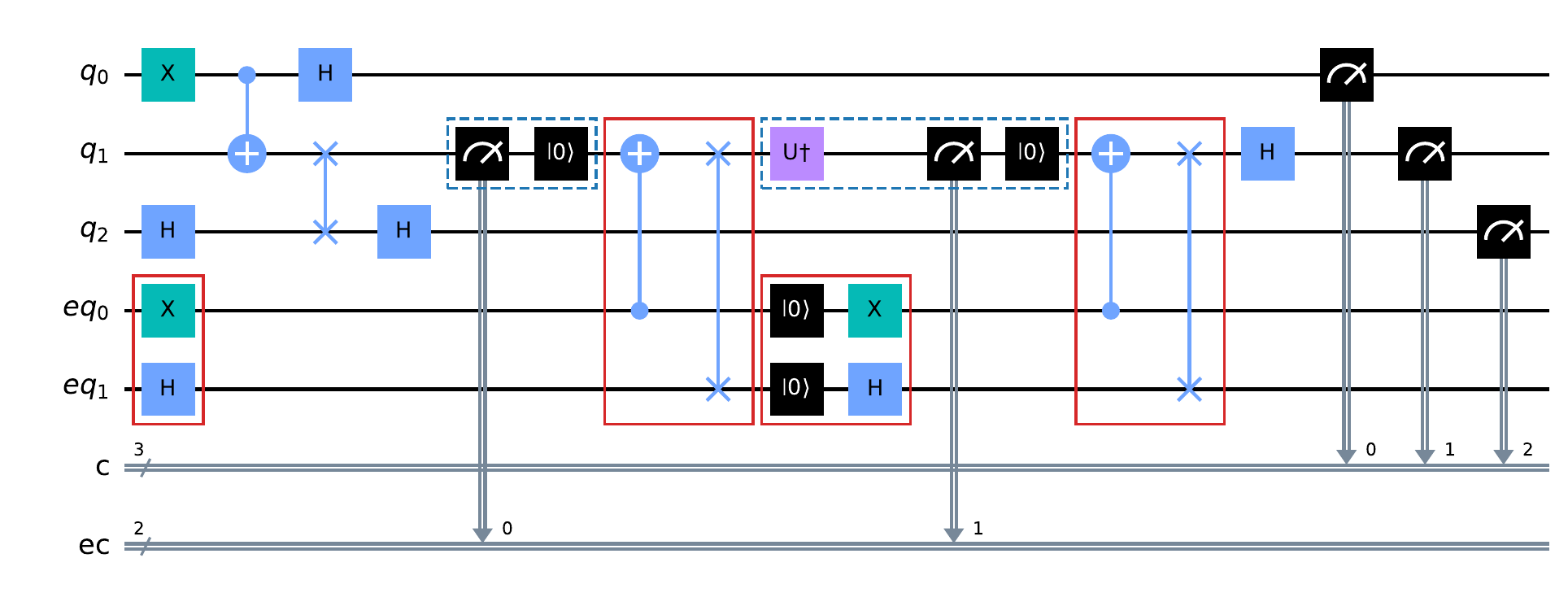}%
}
\subfloat[]{%
    \parbox[c][\figThreeHeight][c]{0.31\textwidth}{%
        \centering
        \includegraphics[width=\linewidth]{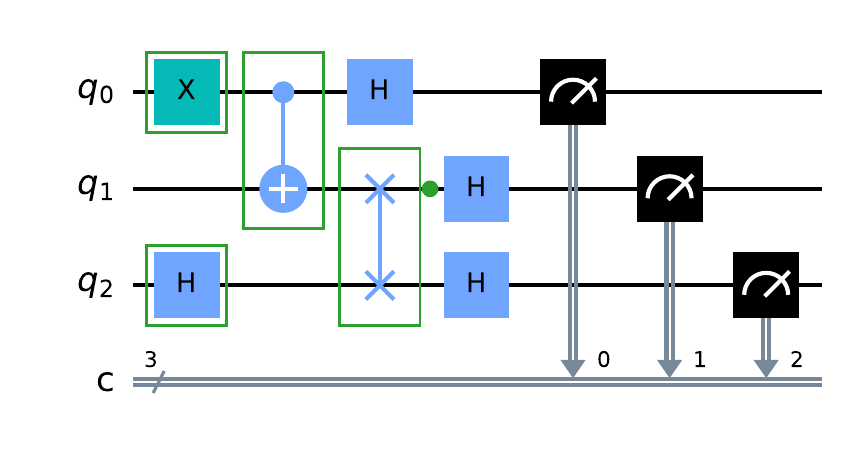}%
    }%
    \label{fig:dfs_a}%
}%
\subfloat[]{%
    \parbox[c][\figThreeHeight][c]{0.69\textwidth}{%
        \centering
        \includegraphics[width=\linewidth]{figures/fig3b.pdf}%
    }%
    \label{fig:dfs_b}%
}
\caption{Backward causal-cone tracing and circuit reconstruction in
\textsc{QMon}. (a)~Original circuit with the selected gate--qubit location
marked by a green dot and its full backward causal cone outlined in green.
(b)~The instrumented circuit, with two reads (dashed blue boxes) and
the causal cone replayed twice on the monitored
qubit and reset and reused ancillas. The replayed gates are outlined in red.}
\label{fig:dfs_pathtracing}
\end{figure*}

Figure~\ref{fig:dfs_pathtracing} illustrates the tracing and reconstruction
steps. The green dot in Fig.~\ref{fig:dfs_a} marks a selected gate--qubit
location on $q_1$, immediately after the SWAP gate between $q_1$ and $q_2$
and before the subsequent $H$ gate on $q_1$.

\paragraph{Backward causal-cone tracing}

Starting from the green dot, Algorithm~\ref{alg:dfs} first encounters the
SWAP gate between $q_1$ and $q_2$. Because SWAP has two inputs, the search
follows both input wires. The $q_2$ branch reaches and records the earlier
$H$ gate on $q_2$, while the $q_1$ branch continues to the CNOT gate.

At the CNOT gate, the search again follows both inputs. The $q_1$ input
reaches circuit initialization, whereas the $q_0$ input reaches and records
the initial $X$ gate. The later $H$ gate on $q_0$ is not included because it
occurs after the CNOT and therefore cannot affect the selected state of
$q_1$. Similarly, the later $H$ gates on $q_1$ and $q_2$ follow the selected
SWAP in the original program order and are not included.

After duplicate gates are removed and the remaining gates are sorted by gate
index, the backward causal cone contains four gates:
$X(q_0)$, CNOT$(q_0,q_1)$, $H(q_2)$, and SWAP$(q_1,q_2)$.
These gates are outlined in green in Fig.~\ref{fig:dfs_a}.

\paragraph{Circuit reconstruction}

Figure~\ref{fig:dfs_b} inserts a computational-basis measurement of $q_1$ at
the selected location, stores the 0/1 outcome in \texttt{ec[0]}, the first
bit of a new two-bit classical register \texttt{ec}, and resets $q_1$ to
$\ket{0}$. The final measurements still write to \texttt{c[0]},
\texttt{c[1]}, and \texttt{c[2]}, so they do not overwrite either
monitoring outcome stored in \texttt{ec}. Replaying the backward causal cone directly on
the original $q_0$ and $q_2$ wires would apply their preparation gates a
second time and change the program state. \textsc{QMon} therefore introduces $eq_0$ and $eq_1$ as fresh ancillas
representing the two non-target wires in the causal cone. The $X$ gate that
prepared $q_0$ is replayed on $eq_0$, and the $H$ gate that prepared $q_2$
is replayed on $eq_1$. The remaining replay operations are
CNOT$(eq_0,q_1)$ and SWAP$(q_1,eq_1)$.
At the selected location, $q_1$ is in $\ket{+}$.
The gate $U^\dagger$ then maps $\ket{+}$ to $\ket{0}$, so the second
read is an X-basis read, equivalently implemented by $H$ followed by Z
measurement. Its outcome is stored in \texttt{ec[1]}. For the ideal
correct circuit, the first read has probabilities $(1/2,1/2)$ and the
second read returns $0$ with certainty. After resetting $q_1$, $eq_0$,
and $eq_1$, \textsc{QMon} repeats the same cone replay.

The $X$ and $H$ gates acting only on the fresh ancillas appear early in the
drawing because they do not depend on the measurement or reset and may be
scheduled before them. This placement does not change the causal order of
the dependent replay operations. The ancilla resets and preparation for
the second replay likewise commute with the second read on $q_1$, provided
each reset precedes its ancilla's preparation and replay interactions.
The replayed gates are outlined in red.
The original $q_0$ and $q_2$ wires remain untouched by the replay operations. The original $H$ gates on $q_0$ and $q_2$ are also drawn before the
monitoring sequence because they act on different qubits and commute with
the measurement, reset, and replay operations. The $H$ gate on $q_1$ is
executed after the second replay as part of the original circuit tail, not as a replay
operation. The circuit then performs its original final measurements.

This reconstruction procedure also handles longer SWAP chains and
causal cones whose backward branches meet again. Under the assumptions
of Theorem~\ref{thm:nodist}, measuring, resetting, and replaying at an
exactly separable location restores the joint state of the original
circuit qubits before the remaining gates are executed. A causal cone
with more non-target wires requires more fresh ancillas, while one
with more gates requires more replay operations. The batching
procedure below manages these extra qubits under the available
qubit budget.

\subsection{Qubit-Budgeted Batching}
\label{sec:node_filtering}

A backward causal cone may require several replay ancillas, and placing many
monitor groups in one instrumented run may exceed the number of available
qubits. \textsc{QMon} therefore divides monitor groups among separate
instrumented runs, called \emph{batches}.

At gate index $i$, let $M_i$ contain all qubits whose gate--qubit locations
are numerical candidates considered for deployment after that gate. These
qubits form one monitor group
$
g_i=(i,M_i).
$
The qubits in a group are handled together and the group is not split across
batches.

Let $Q(i,q)$ be the set of circuit qubits touched by
$\Cone(i,q)$. The union
$
\bigcup_{q\in M_i}Q(i,q)
$
contains all wires needed to replay the causal cones of the group. The qubits
in $M_i$ remain on their original program wires, while every other wire in
the union is represented by a fresh ancilla. The ancilla demand of group
$g_i$ is therefore
$
d_i=
\left|
\left(
\bigcup_{q\in M_i}Q(i,q)
\right)
\setminus M_i
\right|.
$

Different monitor groups placed in the same batch receive disjoint sets of
fresh ancillas. The two replays within each group reuse its register
after reset. Their ancilla demands therefore add. We fix
$Q_{\max}=24$ for the main experiments, where $Q_{\max}$ is the maximum
total number of qubits in one instrumented run.
Appendix~\ref{sec:scalability} separately examines sensitivity to this
resource limit.

For a given circuit, let $Q_{\mathrm{base}}$ be the number of qubits in the
original uninstrumented circuit. The capacity available for replay ancillas
is
$
Q_{\mathrm{cap}}=Q_{\max}-Q_{\mathrm{base}}.
$
A group with
$
d_i>Q_{\mathrm{cap}}
$
cannot fit in one run even when deployed alone. It is therefore placed in the
cannot-fit-alone set. Because the group is not split, adding more batches
cannot make it fit.

All remaining groups are first packed using the deterministic
first-fit-decreasing (FFD) procedure in Algorithm~\ref{alg:batch}. The
procedure sorts the groups from largest to smallest ancilla demand, with ties
broken by gate index. It then places each complete group into the first batch
with enough remaining capacity.
Each batch's cumulative ancilla demand is stored and updated when a
group is added.

\begin{algorithm}[t]
\caption{First-fit-decreasing packing of monitor groups}
\label{alg:batch}
\begin{algorithmic}[1]
\Require $Q_{\mathrm{base}}$; selected sets $M_i$ and demands $d_i$;
         budget $Q_{\max}$
\Ensure FFD batches $\mathcal{B}_{\mathrm{FFD}}$;
        cannot-fit-alone groups $\mathcal{G}_{\mathrm{cannot}}$
\State $Q_{\mathrm{cap}}\gets Q_{\max}-Q_{\mathrm{base}}$
\State $\mathcal{G}\gets\{g_i=(i,M_i):M_i\ne\varnothing\}$
\State $\mathcal{G}_{\mathrm{cannot}}
       \gets\{g_i\in\mathcal{G}:d_i>Q_{\mathrm{cap}}\}$
\State $\mathcal{G}_{\mathrm{fit}}
       \gets\mathcal{G}\setminus\mathcal{G}_{\mathrm{cannot}}$
\State $\mathcal{B}_{\mathrm{FFD}}\gets[\,]$
\For{each $g_i\in\mathcal{G}_{\mathrm{fit}}$ from largest to smallest
      $d_i$, ties by $i$}
  \If{the first $b\in\mathcal{B}_{\mathrm{FFD}}$ with
      $\sum_{g_j\in b}d_j+d_i\le Q_{\mathrm{cap}}$ exists}
    \State append $g_i$ to $b$
  \Else
    \State append a new batch $\{g_i\}$ of cost $d_i$ to
           $\mathcal{B}_{\mathrm{FFD}}$
  \EndIf
\EndFor
\State \Return $\mathcal{B}_{\mathrm{FFD}}$ and sorted
       $\mathcal{G}_{\mathrm{cannot}}$
\end{algorithmic}
\end{algorithm}

\paragraph{Conditional Integer Linear Programming Refinement}

Let
$
N_g=|\mathcal{G}_{\mathrm{fit}}| \text{ and }
K_{\mathrm{FFD}}=|\mathcal{B}_{\mathrm{FFD}}|.
$
Here, $N_g$ is the number of groups that can fit individually, and
$K_{\mathrm{FFD}}$ is the number of batches produced by FFD.

When $K_{\mathrm{FFD}}>1$ and $N_g\leq120$, the implementation performs one
bounded integer linear programming (ILP) feasibility check. This check asks
only whether all feasible groups can be packed into
$K_{\mathrm{FFD}}-1$ batches. An assignment replaces the FFD plan only
when PuLP reports \texttt{Optimal} for this feasibility model. Otherwise,
including an unresolved timeout, the FFD plan is retained. The implementation
uses PuLP with Gurobi and a 600-second time limit for each refinement.

Table~\ref{tab:rq2-regression}, Table~\ref{tab:qmax-sweep}, and the RQ3
deployment plans all use this conditional FFD--ILP procedure. Because the ILP
checks only whether one fewer batch is possible, the reported batch counts
describe the plans found by this procedure. They are not claims that no plan
with still fewer batches exists.

Within each final batch, monitor groups are returned to increasing gate order
before instrumentation. Each batch is then executed as a separate
instrumented run, and the observations from all batches are combined using
their $(i,q)$ gate--qubit labels. If no monitor group fits individually, Algorithm~\ref{alg:batch}
returns no batches. If only eligible final readouts are available, we
execute the original circuit once without mid-circuit instrumentation,
provided $Q_{\mathrm{base}}\leq Q_{\max}$. This execution is not included
in the batch count returned by Algorithm~\ref{alg:batch}.
If neither mid-circuit checkpoints nor eligible final readouts are available,
there is no monitoring evidence, and no multiple-testing threshold is defined
for the empty test family. The RQ3 population requires a nonempty mid-circuit
deployment plan and excludes both boundary cases.

A batching walkthrough appears in Appendix~\ref{app:implementation-walkthroughs}.
Further implementation and complexity details appear in
Appendix~\ref{app:implementation-details}.

\subsection{Circuit Execution and Results Analysis}
\label{sec:circuit_execution}

After backward causal-cone tracing and circuit reconstruction, each
instrumented circuit contains mid-circuit measurements, resets, and
ordinary quantum gates. Physical execution requires a backend that
supports mid-circuit measurement and reset. In this paper, we
evaluate the complete instrumented circuits using noise-free simulation in
RQ3 and calibration-informed simulated noise in RQ4.
The shot-based analyses in both RQ3 and RQ4 collect outcomes from
repeated executions, so they retain finite-sampling variation.
RQ4 additionally includes simulated device noise.

\textsc{QMon} compares the observed outcome frequencies from mid-circuit measurements with the expected probabilities. A checkpoint is
flagged when this comparison satisfies the decision rule described in
Section~\ref{sec:rq3}, with full details in
Appendix~\ref{app:rq3details:protocol}. Otherwise, the observations are
treated as consistent with the expected probabilities under that rule.
Each batch uses 8,192 shots. Every shot records both reads at each
mid-circuit checkpoint. In RQ3, a statistical test compares the frequency of $1$ in the
first read with its expected probability. An outcome of $1$ in the
second read flags a deviation of the replayed target state from its
expected pure state. Localization uses the backward causal cone of the
earliest flagged checkpoint. In RQ4, the observed frequencies are
compared with calibration data collected separately for each method
under the same simulated noise conditions.
The absence of a flag does not prove that the circuit is correct.
Section~\ref{sec:result} presents the analyses used to answer the
research questions.

\section{Experiment Setup}
\label{sec:setup}

This section describes the circuits, qubit budget, and simulation backends
used to evaluate \textsc{QMon}.
\subsection{Quantum Circuit Selection}
\label{sec:prep:program}

We obtain our evaluation circuits from MQT Bench
~\cite{Quetschlich2023mqtbench}, which provides quantum circuits from
different applications and at several levels of abstraction. We use circuits
in the Qiskit format. Qiskit is an open-source software development kit (SDK)
for creating and manipulating quantum circuits and for executing them on
local simulators or compatible quantum backends. We use circuits at the target-independent abstraction level. These circuits
have not yet been mapped to a specific device topology or native gate set.
They therefore represent the circuit descriptions available before hardware-specific compilation. The main evaluation set contains 310 target-independent circuits with
2--14 qubits. All 310 circuits complete the RQ1 and RQ2 analyses without compilation failures or timeouts. RQ3 uses the 238 circuits with at least one deployed mid-circuit checkpoint and enough eligible gate locations for its five-seed mutation experiment (Section~\ref{sec:rq3}). RQ4 uses the 100 circuits whose deployment plan consists of one batch
and uses at most 12 qubits in total (Section~\ref{sec:rq4}).
The single-batch restriction avoids combining checkpoint results across
multiple batches when evaluating the effect of noise. The 12-qubit limit
includes both program qubits and replay ancillas and keeps the repeated
noisy simulations computationally manageable.

Appendix~\ref{sec:scalability} presents a separate scalability study. It
evaluates MPS-based selection of numerical candidates and structural
deployment planning on 144 circuits from 27 families, with circuit widths
from 15 to 50 qubits. These 144 circuits form a separate scalability set and
are not part of the 310-circuit main evaluation set.

At the target-independent abstraction level, circuits assume all-to-all
connectivity, meaning that any pair of qubits can interact directly. Mapping
a circuit to a device with restricted connectivity may require
additional routing SWAP gates, including SWAPs for interactions involving
replay ancillas. We do not model this hardware-dependent overhead.

\subsection{Simulation Backends}
\label{sec:prep:backend}

We evaluate \textsc{QMon} and all baselines in noise-free simulation in RQ1--RQ3 to study circuit logic bugs without the effects of hardware noise. Experiments using a finite number of shots still include sampling variation in the measurement outcomes. RQ4 adds calibration-informed simulated noise.

We use Qiskit Aer%
\footnote{\url{https://qiskit.github.io/qiskit-aer/}}
to simulate \textsc{QMon} executions and the statistical and dynamic-ancilla assertion baselines. Depending on the experiment, we use statevector or density-matrix simulation to obtain quantum states and probabilities, or collect measurement outcomes from repeated shots. We also use matrix-product state simulation for the scalability study. The projection assertion baseline is implemented in pytket, but its circuits are executed by Aer through pytket's Aer interface. The MQT debugger instead runs on its own decision-diagram backend. We analyze the resulting states, measurement counts, and assertion outcomes to assess state preservation, fault detection, and execution cost.

\subsubsection{Statevector Simulator}

We use statevector simulation to compute reduced one-qubit states,
separability scores, expected checkpoint probabilities, state fidelity, and
final measured-output $L_1$ distances. An $L_1$ distance is the sum of the
absolute differences between corresponding output probabilities. These
calculations may contain floating-point rounding error, but they do not
contain shot-sampling error.

\subsubsection{Shot-Based Simulator}

The noise-free detection and false-alarm experiments use
$
S=8{,}192
$
noise-free shots per explicitly executed run. Each shot is an independent
execution of the circuit. This setting retains the randomness of measurement
outcomes while excluding device noise. It also allows \textsc{QMon} and the
three finite-shot assertion baselines to be compared using the same number of
shots per run.

\subsubsection{Calibration-Informed Simulated Noise}
\label{sec:prep:noise}

We evaluate how noise affects detection and false alarms using a
simulation model based on the \texttt{FakeMontrealV2} calibration snapshot of
15 March 2021. We use median calibration parameters uniformly across
device locations, rather than modeling differences between physical
qubits. The model includes depolarizing gate errors, reset and measurement
errors, and thermal relaxation during gates, idle periods, and measurements.

We test five noise scales: $0$, $0.25$, $0.5$, $1$, and $2$.
Scale $0$ introduces no simulated device noise, while scale $1$ uses
the calibration-derived parameters. The other scales adjust error
parameters and relaxation rates without changing gate durations.
The simulation assumes all-to-all connectivity and does not model
routing, crosstalk, or leakage. \textsc{QMon} and the three
assertion baselines execute different gate sequences. Running each
circuit the same number of times therefore does not equalize the number
of gates executed or the time spent executing them. Appendix~\ref{app:rq4details} gives the full noise model and compilation
details.

\section{Experiment Results}
\label{sec:result}

\subsection{RQ1: When Does \textsc{QMon} Preserve the Full Circuit State?}
\label{sec:rq1}

\subsubsection{Motivation}

\textsc{QMon} is useful only if instrumenting a circuit does not change what
the circuit computes. For one monitor, Theorem~\ref{thm:nodist} gives the
forward guarantee: if the monitored qubit is separable from the other qubits
and its preparation is replayed exactly, the full circuit state is preserved.
Proposition~\ref{prop:converse} (see Appendix~\ref{app:proof}) gives the converse for one forced monitor: if
the monitored qubit is entangled with the other qubits, the full state
changes.

RQ1 examines the single-monitor boundary in two ways. First, we check whether
an independent numerical calculation reproduces the locations selected by
Algorithm~\ref{alg:screen} and whether their values are numerically consistent
with the exact separability condition. Second, we deliberately insert one
monitor at every gate--qubit location, including locations rejected by the
selector, and measure the resulting change in the full circuit state.
Each inserted monitor includes the first read and the second read,
each followed by reset and replay. We evaluate the state change after
the second replay.
RQ1 does not use a tool-to-tool baseline because it evaluates the state
change caused specifically by \textsc{QMon}'s measure--reset--replay
instrumentation. The direct reference is therefore the same circuit
executed without that instrumentation.

\subsubsection{Approach}

\paragraph{Independent numerical check}

For each numerical candidate $(i,q)$ selected by
Algorithm~\ref{alg:screen}, we simulate the circuit through gate
$U_i$ and record the second Schmidt amplitude $a_2(i,q)$, which is the
quantity used by the selector. As established in
Section~\ref{sec:schmidt-certificate}, the exact conditions
$
a_2(i,q)=0
\text{ and }
\det\rho_q(i)=0
$
are equivalent. The implementation applies the finite-precision rule
$
a_2(i,q)\leq\varepsilon_{\mathrm{sv}}=10^{-12}.
$
As a separate numerical check, we also construct $\rho_q(i)$ and compute its
determinant. RQ1 reports the largest selected value of $a_2(i,q)$ and the
largest absolute determinant residual.

\paragraph{Forced-monitor disturbance evaluation}

We deliberately insert one monitor at each of the 56,386 gate--qubit
locations in the 310-circuit evaluation set, including locations not selected
by Algorithm~\ref{alg:screen}. Each location pairs a gate with one of the
qubits on which it acts.

Let $\rho_{\mathrm{orig}}$ be the final density operator of the
uninstrumented circuit, and let $\rho_{\textsc{QMon}}$ be the final
density operator of the instrumented circuit after the measurement record and
replay ancillas have been traced out. We compare the complete final states
rather than only their output probabilities because a chosen measurement
basis can hide changes in phase or correlation.

We compute the squared Uhlmann fidelity
\begin{equation}
F=
\Bigl(
\operatorname{Tr}
\sqrt{
\sqrt{\rho_{\mathrm{orig}}}\,
\rho_{\textsc{QMon}}\,
\sqrt{\rho_{\mathrm{orig}}}
}
\Bigr)^2,
\label{eq:uhlmann}
\end{equation}
which, because $\rho_{\mathrm{orig}}$ is pure, reduces to
$
F=
\langle\psi_{\mathrm{orig}}\rvert
\rho_{\textsc{QMon}}
\lvert\psi_{\mathrm{orig}}\rangle.
$
We report the state infidelity $1-F$. A value of zero means that the full
final state is unchanged, while a positive value indicates a full-state
difference.

\subsubsection{Results}

\paragraph{Numerical selection check}

Of the 56,386 gate--qubit locations in the 310-circuit evaluation set,
Algorithm~\ref{alg:screen} selects 22,332, or 39.6\%. An independently
implemented singular-value-decomposition calculation reproduces both the
second Schmidt amplitudes and the complete selected set. The largest selected
amplitude is
$
5.57\times10^{-14},
$
which is below the selection tolerance of $10^{-12}$. The independent
determinant calculation has a maximum absolute residual of
$
4.61\times10^{-15}.
$

This check confirms that the implementation applies its numerical selection
rule as intended. It does not prove that the selected floating-point values
are mathematically equal to zero. The results instead show that the selected
locations are numerically consistent with the exact separability condition.
The forced-monitor evaluation below provides separate evidence about the
state-preservation boundary because it includes all locations, not only those
accepted by the selector.

The selected locations occur in circuits with 2 to 14 qubits, including
Greenberger--Horne--Zeilinger (GHZ), entangled quantum Fourier transform
(QFT), and graph-state circuits. A circuit may therefore create entanglement
during its execution while still containing particular locations at which
one qubit is separable from all other qubits.

\paragraph{For one forced monitor, exact state preservation requires
separability and follows an exact law}

\begin{figure}[t]
\centering
\includegraphics[width=\linewidth]{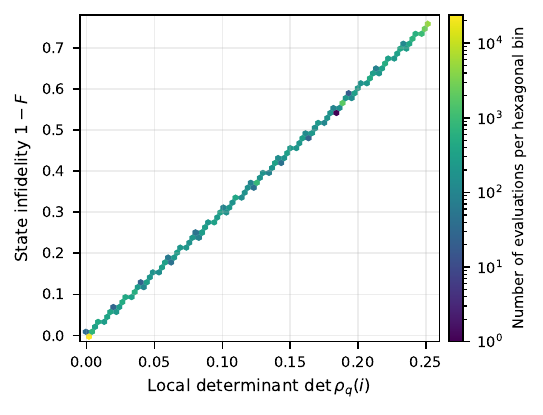}
\caption{State infidelity versus the local determinant for all 56,386
single-monitor evaluations. Color indicates the number of evaluations per hexagonal bin.
The computed values agree with the proved identity
$1-F=3\det\rho_q(i)$, with a maximum absolute residual of
$3.17\times10^{-13}$.
Under the stated ideal conditions, exact full-state preservation
requires $\det\rho_q(i)=0$.}
\label{fig:ent-dist}
\end{figure}

Figure~\ref{fig:ent-dist} shows the relationship between state infidelity and
the local determinant for all 56,386 single-monitor evaluations. The
evaluated values follow the exact closed-form law
$
1-F=3\det\rho_q(i),
$
derived as Eq.~\eqref{eq:tightlaw} in
Appendix~\ref{app:tightlaw}. Under the ideal single-monitor conditions of
Proposition~\ref{prop:converse}, a zero determinant gives zero full-state
disturbance:
$
\det\rho_q(i)=0
\Longrightarrow
1-F=0.
$
Any positive determinant instead gives a strictly positive full-state
change:
$
\det\rho_q(i)>0
\Longrightarrow
1-F>0.
$
Exact separability is therefore the precise boundary for preserving the full
state when one monitor is inserted under these conditions.

This is a statement about the complete quantum state. A particular final
measurement distribution may remain unchanged even when the underlying state
has changed, because that measurement may not reveal a phase or correlation
difference.

\paragraph{Replay depth affects cost, not the theorem}

Under the stated ideal conditions, the state-preservation
guarantees of Theorem~1 and Corollary~2 do not depend on
the depth of the replayed causal cone, the number of SWAP
gates, or the number of two-qubit interactions
(Appendix~\ref{app:no-disturbance-proof}).
A deeper causal cone may nevertheless require more replay
ancillas and more replayed gates, increasing the resource
cost of instrumentation (Appendix~\ref{sec:scalability}).

\begin{tcolorbox}[
    enhanced jigsaw,
    breakable,
    colback=white,
    colframe=black,
    boxrule=0.4pt,
    arc=0pt,
    boxsep=0pt,
    left=3pt,
    right=3pt,
    top=3pt,
    bottom=3pt,
    before skip=6pt,
    after skip=6pt
]
\textbf{Summary of RQ1.}
\textsc{QMon} selects 22,332 of 56,386 gate--qubit locations
(39.6\%) as numerical candidates for monitoring. For a single monitor under
the stated ideal conditions, the full circuit state is preserved only when
the monitored qubit is not entangled at all with the other qubits at the monitored location; any
non-zero entanglement changes the full state.
\end{tcolorbox}

\subsection{RQ2: How Much of a Circuit Can \textsc{QMon} Monitor?}
\label{sec:rq2}

\subsubsection{Motivation}

RQ2 measures how much of a
circuit remains available for monitoring after applying these conditions and
the qubit budget. It also examines which circuit properties are associated
with higher or lower monitoring coverage. RQ2 does not present a raw tool-to-tool coverage comparison because its coverage measures are defined by \textsc{QMon}'s own selection and deployment rules. A \emph{numerical candidate} is a gate--qubit location that passes \textsc{QMon}'s finite-precision separability test, and a \emph{deployed mid-circuit checkpoint} is a numerical candidate included in an instrumented run under the $Q_{\max}=24$ qubit budget. The assertion and debugging baselines do not define comparable checkpoint eligibility. 

\subsubsection{Approach}

\paragraph{Coverage metrics}

For a circuit $C$, let
$
\mathcal{P}_C
=
\{(i,q):q\text{ is acted on by gate }i\}
$
denote its set of gate--qubit locations. A one-qubit gate contributes one
location, while a two-qubit gate contributes two because
Algorithm~\ref{alg:screen} tests both qubits separately. We distinguish the numerical candidates selected by
Algorithm~\ref{alg:screen} from the deployed mid-circuit checkpoints that can be
included in an instrumented run under the budget $Q_{\max}=24$.
For the coverage metrics, an \emph{observable point} is a deployed mid-circuit
checkpoint or an eligible final readout. For a qubit measured at the end of the circuit, we check whether the location after its last gate is a numerical candidate. If so, we use the existing final measurement to observe that location and assign the readout to that gate. No mid-circuit monitor is inserted at that location. For each circuit, we report three coverage metrics:

\begin{itemize}
    \item \emph{Deployed gate coverage} is the number of source gates with
    an observable point on at least one operand divided by the total number of gates.
    Each gate is counted once.

    \item \emph{Qubit coverage} is the number of circuit qubits with at least
    one observable point divided by the total number of qubits.

    \item \emph{Normalized depth reach} is the deepest source-gate layer
    containing an observable point divided by the total source-circuit depth.
\end{itemize}

Each metric is averaged equally over all 310 circuits.

We also report the pooled numerical-candidate rate, which divides the
total number of numerical candidates by the total number of gate--qubit
locations across all 310 circuits.

\paragraph{Circuit properties}
We examine five circuit properties: the mean and peak local entanglement scores, source-circuit depth, two-qubit-gate density, and qubit count.
The entanglement scores are relevant because exact state preservation requires the monitored qubit to be separable from the rest of the circuit.
Depth and two-qubit-gate density describe how many layers the computation spans and how frequently gates act on two qubits, respectively. These properties may be associated with where separable qubits occur and how
many gates and extra qubits are needed for replay. Qubit count refers to the number of qubits in the source circuit and determines how many extra qubits remain available for replay under the fixed $Q_{\max}=24$ budget.

To quantify local entanglement, we define the one-qubit-versus-rest score
$
e_{i,q}=\det\rho_q(i)
$
for every $(i,q)\in\mathcal{P}_C$. Because the circuit state immediately
after each gate is pure,
$
e_{i,q}
=
\tfrac12\bigl(1-\Tr[\rho_q(i)^2]\bigr)
\in[0,\tfrac14].
$
The score is zero exactly when qubit $q$ is separable from all other qubits.
In this pure-state setting, a larger value means that the reduced state of
$q$ is more mixed and that $q$ is more entangled with the other qubits.

We summarize these local scores for each circuit as
\begin{equation}
\begin{aligned}
E_{\mathrm{mean}}(C)
&:=
\frac{1}{|\mathcal{P}_C|}
\sum_{(i,q)\in\mathcal{P}_C}e_{i,q},\\
E_{\mathrm{peak}}(C)
&:=
\max_{(i,q)\in\mathcal{P}_C}e_{i,q}.
\end{aligned}
\label{eq:rq2-ent-summary}
\end{equation}
Here, $E_{\mathrm{mean}}$ is the average local score across the circuit, while
$E_{\mathrm{peak}}$ is the largest local score encountered. These quantities
summarize one-qubit-versus-rest entanglement across circuit locations; they
are not measures of entanglement in the full circuit state as a whole.

We compute source-circuit depth using as-soon-as-possible (ASAP) layering,
which places each gate in the earliest layer allowed by its dependencies. The
calculation uses the target-independent circuit after its existing
decomposition into one- and two-qubit gates and before \textsc{QMon}
instrumentation or hardware-specific compilation.

We initialize $L_q=0$ for every qubit $q$. For each included gate $U_i$,
processed in program order, let $Q_i$ be the set of qubits on which the gate
acts. Its layer is
$
\ell_i
=
1+\max_{q\in Q_i}L_q.
$
After assigning $\ell_i$, we set $L_q=\ell_i$ for every $q\in Q_i$. The
circuit depth is then
$
D(C)=\max_i\ell_i.
$
Measurements, barriers, resets, delays, snapshots, and simulator-only save
instructions are excluded from this calculation.

We define two-qubit-gate density as
$
\delta_2(C)
=
\frac{N_2(C)}{N_{\mathrm{gate}}(C)}
=
\frac{N_2(C)}{N_1(C)+N_2(C)},
$
where $N_1(C)$ and $N_2(C)$ are the numbers of included one- and two-qubit
gates, respectively. Each two-qubit gate is counted once, rather than once
for each of its two qubits. The non-gate instructions excluded from the depth
calculation are also excluded from these counts.

\paragraph{Statistical analysis}

We use regression models to examine whether the circuit properties defined above are
associated with deployed gate coverage or normalized depth reach. Both
outcomes are proportions between zero and one, so we fit a separate
fractional-logit model for each outcome. Fractional-logit regression is
designed for proportion-valued outcomes.

Each model includes five circuit properties:
$E_{\mathrm{mean}}$, $E_{\mathrm{peak}}$, source-circuit depth,
two-qubit-gate density, and qubit count. Before fitting the models, we
standardize each property by subtracting its mean and dividing by its sample
standard deviation across all 310 circuits. This places properties measured
on different scales onto a comparable scale. The sample standard deviation
uses denominator $310-1$, and standardization is performed over the complete
sample rather than separately within each circuit family.

Let $\mu$ denote the modeled mean coverage or depth reach. The fractional-logit
model uses the transformed value
$
\log\left(\frac{\mu}{1-\mu}\right).
$
In Table~\ref{tab:rq2-regression}, $\hat{\beta}$ is the estimated change on
this transformed scale for a one-standard-deviation increase in one circuit
property when the other four properties are held fixed. The sign of
$\hat{\beta}$ gives the direction of the association. Its magnitude is not a
direct percentage-point change in coverage.

Circuits from the same family are generated from related templates and may
therefore be more similar to one another than to circuits from other
families. We account for this dependence by clustering the standard errors
by circuit family.

We compute each 95\% confidence interval (CI) using 2,000 family-level
bootstrap samples. For each sample, we draw 42 families with replacement
from the original 42 families. Whenever a family is drawn, all circuits in
that family are included. The reported CI is defined by the 2.5th and 97.5th
percentiles of the 2,000 coefficient estimates.

For each coefficient, we test whether it is zero using the estimate divided
by its family-clustered standard error. Raw two-sided $p$-values use a
Student $t$ distribution with $42-1=41$ degrees of freedom.

The two models contain ten predictor--outcome tests in total. Testing many
associations increases the chance of reporting a false positive, so we apply
the Benjamini--Hochberg (BH) correction to the ten $p$-values. We denote the adjusted
values by $p_{BH}$ and treat $p_{BH}<0.05$ as statistically significant. The reported
95\% CIs are not adjusted for these ten tests. These models identify statistical associations and do not establish that a
circuit property causes a change in coverage.

\subsubsection{Results}

Table~\ref{tab:rq2-regression} reports coverage under the default
$Q_{\max}=24$ budget and the associations between circuit properties and
deployed gate coverage or normalized depth reach.

\begin{table*}[!t]
\centering
\caption{Coverage achieved by \textsc{QMon} and circuit properties associated
with coverage at $Q_{\max}=24$.}
\label{tab:rq2-regression}
\scriptsize
\setlength{\tabcolsep}{3.5pt}
\renewcommand{\arraystretch}{1.02}

\begin{tabular}{@{}p{0.27\textwidth}p{0.50\textwidth}
>{\raggedleft\arraybackslash}p{0.18\textwidth}@{}}
\multicolumn{3}{@{}l}{\textbf{Panel A: Coverage summary}} \\
\toprule
Metric & Calculation & Result \\
\midrule

Pooled numerical-candidate rate
& Numerical candidates / all gate--qubit locations, pooled across all
  circuits
& 22,332 / 56,386 (39.61\%) \\

Mean deployed gate coverage
& Source gates with an observable point on at least one operand / source gates, computed per circuit and
  then averaged
& 44.54\% \\

Mean qubit coverage
& Qubits with at least one observable point / circuit qubits, computed per
  circuit and then averaged
& 89.75\% \\

Mean normalized depth reach
& Deepest source-gate layer with an observable point / source-circuit depth, computed per
  circuit and then averaged
& 48.57\% \\

Circuits with full qubit coverage
& Circuits in which every qubit has at least one observable point
& 254 / 310 (81.94\%) \\
\bottomrule
\end{tabular}

\vspace{0.6em}

\begin{tabular}{@{}lccc ccc@{}}
\multicolumn{7}{@{}l}
{\textbf{Panel B: Circuit-level associations with coverage}} \\
\toprule
& \multicolumn{3}{c}{Deployed gate coverage}
& \multicolumn{3}{c}{Normalized depth reach} \\
\cmidrule(lr){2-4}\cmidrule(l){5-7}
Predictor & $\hat{\beta}$ & 95\% CI & $p_{\mathrm{BH}}$
& $\hat{\beta}$ & 95\% CI & $p_{\mathrm{BH}}$ \\
\midrule

$E_{\mathrm{mean}}$
& $-0.4407$ & $[-0.9994,\ 0.2054]$ & $0.1818$
& $-1.4111$ & $[-2.6322,\ -0.0371]$ & $0.0517$ \\

$E_{\mathrm{peak}}$
& $-1.3810$ & $[-2.1298,\ -0.8910]$ & $\mathbf{0.0002}$
& $-1.1126$ & $[-2.6286,\ -0.2231]$ & $0.0739$ \\

Source-circuit depth
& $-0.0129$ & $[-0.2581,\ 0.2550]$ & $0.9084$
& $1.4558$ & $[0.6923,\ 4.0510]$ & $\mathbf{0.0002}$ \\

Two-qubit-gate density
& $-0.3507$ & $[-0.6509,\ -0.0597]$ & $0.0517$
& $0.2864$ & $[-0.2673,\ 0.9721]$ & $0.3847$ \\

Qubit count
& $-0.0810$ & $[-0.3263,\ 0.1051]$ & $0.4685$
& $-0.3546$ & $[-0.8231,\ -0.0287]$ & $0.0837$ \\
\bottomrule
\end{tabular}
\end{table*}

\paragraph{Monitoring coverage}

Across all 310 circuits, Algorithm~\ref{alg:screen} selects 22,332 of 56,386
gate--qubit locations. The pooled numerical-candidate rate is therefore
39.61\%. Under the $Q_{\max}=24$ budget, the mean deployed gate coverage
is 44.54\%. The difference arises from the averaging schemes, counting units, and selection and deployment scopes. The 39.61\% rate pools all locations across all circuits before applying the deployment budget, so circuits with more locations receive more weight. The 44.54\% value first
computes deployed gate coverage within each circuit and then gives every
circuit equal weight when averaging the 310 values.

Although fewer than half of the source gates have an observable point on at least one operand on
average, the mean qubit coverage is 89.75\%. Thus, most qubits are monitored
at least once even though many individual circuit locations are not
monitored. Moreover, 254 of the 310 circuits, or 81.94\%, have at least one observable point on every qubit. The mean normalized depth reach is 48.57\%, so the deepest observable point lies near the midpoint of the circuit on average. A possible explanation is that qubits may remain entangled at later locations, while backward causal cones involving more qubits may require more extra qubits for replay than the budget allows. The separability test and the qubit budget can therefore exclude later mid-circuit checkpoints.

\paragraph{Circuit properties associated with coverage}
Panel B in Table~\ref{tab:rq2-regression} reports ten associations between five circuit properties and two outcomes. After the BH correction, two associations satisfy $p_{BH}<0.05$. Each association is estimated with the other four circuit properties held fixed.

First, a larger $E_{\mathrm{peak}}$ is associated with lower deployed gate coverage. This negative association is consistent with the separability requirement: exact state preservation requires the monitored qubit to be separable from the rest of the circuit. However, $E_{\mathrm{peak}}$ records only the largest local entanglement score, not the number of locations that satisfy this requirement. It therefore does not determine coverage by itself. Second, greater source-circuit depth is associated with larger normalized depth reach. Thus, in this circuit set, the deepest observable point tends to occur later as a fraction of the total depth, not merely at a larger absolute layer index. The mean and peak entanglement scores do not describe where separable locations occur along the circuit. One possible explanation is that some deeper circuits contain late sequences of gates
that leave qubits separable or disentangle them. Such sequences can provide monitoring locations closer to the end of the computation, through eligible final readouts or mid-circuit checkpoints whose replay fits within the qubit budget. The regression does not directly test this explanation. No other association satisfies the $p_{BH}<0.05$ criterion. Some unadjusted 95\% CIs exclude zero, but their corresponding BH adjusted $p$-values remain above 0.05. We therefore do not classify these associations as statistically significant.

\paragraph{Sensitivity to the qubit budget}

Appendix~\ref{sec:scalability}, Table~\ref{tab:qmax-sweep}, reports coverage
for several values of $Q_{\max}$. Increasing the budget from 24 to 28 raises
mean deployed gate coverage from 44.54\% to 45.89\%. At
$Q_{\max}=28$, every numerical candidate is observable in every circuit.
None of the coverage metrics changes further as $Q_{\max}$ increases from 28
to 40. We do not fit the same five-property model for qubit coverage.
Qubit coverage is 100\% in 254 of the 310 circuits and varies within
only 4 of the 42 families. We therefore report it descriptively.

\begin{tcolorbox}[
    enhanced jigsaw,
    breakable,
    colback=white,
    colframe=black,
    boxrule=0.4pt,
    arc=0pt,
    boxsep=0pt,
    left=3pt,
    right=3pt,
    top=3pt,
    bottom=3pt,
    before skip=6pt,
    after skip=6pt
]
\textbf{Summary of RQ2.}
With a 24-qubit limit per instrumented run, \textsc{QMon} covers
44.54\% of source gates on at least one operand on average and monitors 89.75\% of circuit
qubits at least once. The deepest observable point lies at 48.57\% of the
circuit depth on average, and every qubit is monitored in 254 of the 310
circuits. Higher peak one-qubit-versus-rest entanglement is associated with
lower gate coverage. At $Q_{\max}=28$, all numerical candidates are observable.
\end{tcolorbox}

\subsection{RQ3: How Effectively Does \textsc{QMon} Detect and Locate Programming Errors?}
\label{sec:rq3}

\subsubsection{Motivation}

RQ3 asks whether the resulting intermediate observations can be used to detect gate-level logic errors and narrow their possible causes. We compare \textsc{QMon} with three quantum assertion methods and a simulation-only debugger.

\subsubsection{Mutation}
\paragraph{Rationale for mutation-based evaluation}

We use mutation testing to construct controlled circuit variants
with known injected fault locations, enabling reproducible
comparisons of fault detection and localization. Its classical
rationale includes the competent-programmer hypothesis, which
motivates studying small deviations from an intended implementation,
and the coupling-effect hypothesis, which suggests that tests
exposing simple faults may also expose more complex
faults~\cite{demillo1978hints}. These hypotheses provide
methodological motivation rather than guarantees for quantum
programs. Quantum mutation tools such as Muskit~\cite{9678563}
and QMutPy~\cite{10.1145/3533767.3543296} similarly use predefined
gate-level changes to assess testing effectiveness.
Our gate-replacement operator models local mistakes in gate
choice or rotation parameters while preserving the instruction's
qubit operands. Randomness samples locations and replacements
within this predefined fault model, and all methods are evaluated
on the same mutants. The injected locations are withheld from
the detectors and used to assess whether the reported suspect
regions contain the faults. Thus, the experiment measures
effectiveness against the specified gate-level faults, rather
than estimating detection rates for all real-world programming
errors.

\subsubsection{Approach}
\paragraph{Mutation protocol}

The RQ3 subset contains 238 circuits with at least one deployed mid-circuit \textsc{QMon} checkpoint and
at least three eligible one- or two-qubit unitary instructions. An instruction need not be within the backward causal cone of any candidate. For each circuit and each of five mutation seeds, we sample three distinct
eligible gate locations without replacement. We construct three separate
single-fault circuit variants, called mutants, by replacing the gate at one
sampled location in each variant. The replacement has the same number of
operands as the original gate. The injected fault location is not given to
any detector or localizer.
We classify a mutant as \emph{output-equivalent} when the $L_1$ distance
between its final measured-output distribution and that of the original
circuit is below $10^{-9}$.
Of the 3,570 mutation attempts, 710 are output-equivalent and 2,860 are
non-output-equivalent. All five methods return a decision for every one of
the 2,860 non-output-equivalent mutants.
Appendix~\ref{app:rq3details:protocol} provides the complete mutation and
evaluation protocol.

\paragraph{Baselines and checkpoint scope}

We compare \textsc{QMon} with four published methods: statistical
assertions~\cite{10.1145/3307650.3322213}, projection assertions
(Proq)~\cite{10.1145/3428218}, dynamic ancilla assertions
~\cite{10.1145/3373376.3378488,li2022exploiting}, and the MQT
debugger~\cite{rovara2024debugger}.
Every method is evaluated on the same sets of mutants.
We conduct two comparisons, described below.

\textsc{QMon} evaluates the deployed checkpoints under the constraint of
$Q_{\max}=24$.
Each mutant is evaluated using a monitoring plan derived only from the
original circuit.
The checkpoint locations, expected states and batch
assignments are fixed from the original circuit; replay uses the circuit
under test's gates and parameters at the planned instruction indices.
The preservation guarantee does not extend automatically to faulty
executions that violate its separability or matched-replay assumptions.
Each mid-circuit checkpoint undergoes the first read and the second
read within the same batch. Eligible final readouts are tested once in
the computational basis, using the existing final measurements.

In the first comparison, \textsc{QMon} uses its deployed checkpoints
and batches, whereas the three assertion methods use the complete set
of numerical candidates in the original circuit. Each assertion
checkpoint is evaluated in a separate circuit prefix ending at that
checkpoint. Thus, this comparison does not use identical checkpoint
sets or batch assignments across methods.
MQT starts from the complete candidate set,
subject to its additional applicability rule.
An MQT assertion is considered applicable only if it can be evaluated on
the original, unmodified circuit and the evaluation returns a pass.

In the second comparison, the three assertion methods use exactly
the same deployed checkpoints and batch assignments as \textsc{QMon}.
The assertions assigned to a batch are inserted in gate order and
executed together, using 8,192 shots per batch. This comparison therefore
holds the checkpoint sets, batch counts, and test-shot budgets equal
across \textsc{QMon} and the three assertion methods.
MQT is not included in this batched comparison because it uses a
simulation-based debugging procedure rather than shot-based execution.
\paragraph{Checkpoint decision rules}

In our experiment, \textsc{QMon} and the three finite-shot assertion methods
use $S=8{,}192$ shots per explicitly executed run. MQT is deterministic
and simulation-only. For a nondeterministic Z-basis reference distribution, \textsc{QMon} uses a
two-sided exact binomial test. Its second read has expected outcome~0 and flags any observed~1 in this noise-free experiment. The statistical assertion method uses
Pearson's $\chi^2$ goodness-of-fit test when both expected outcome counts
are at least five; otherwise, it uses the same exact binomial test.

For \textsc{QMon} and the separate-run statistical assertion method, a false alarm
occurs when finite-shot sampling variation causes a correct circuit
to be flagged at one or more checkpoints. We set $\alpha=0.01$
to target a probability of at most 1\% for this circuit-level event.
Bonferroni correction divides this budget equally among all tested reads
of the circuit. Let $K_{\mathrm{mid}}$ count deployed mid-circuit
checkpoints and $K_{\mathrm{final}}$ count eligible final readouts.
\textsc{QMon}'s tested-read family has size
$K_{\mathrm{QMon}}=2K_{\mathrm{mid}}+K_{\mathrm{final}}$, giving a
per-read threshold of $\alpha/K_{\mathrm{QMon}}$.
Statistical assertions use $\alpha/K_{\mathrm{stat}}$, where
$K_{\mathrm{stat}}$ is the declared checkpoint count.
For example, a method with 100 tested reads uses a threshold of
$0.01/100=0.0001$ at each read: its statistical test flags a
checkpoint when the resulting $p$-value is below this threshold.
Within one circuit evaluation, this single budget covers the complete
checkpoint set across all batches; each batch does not receive a
separate 1\% budget.
The two \textsc{QMon} reads share the same 8,192 shots, with both
reads performed in each shot.

Both methods handle reference distributions treated as deterministic
using a direct rule. If the expected outcome is 0, any observed 1
flags the checkpoint; if it is 1, any observed 0 flags the checkpoint. Projection and dynamic ancilla assertions retain their direct any-failure
rules in our noise-free evaluation: neither computes a statistical
$p$-value nor applies Bonferroni correction. On a mutant, either an evaluated state mismatch or an unevaluable
result counts as an MQT combined flag. An unevaluable result means that MQT cannot extract the pure one-qubit
state required for the comparison; it is not an evaluated state mismatch.
Appendices~\ref{app:rq3details:baselines} and~\ref{app:rq3details:protocol} provide the complete baseline implementations and checkpoint decision rules.

\paragraph{Evaluation metrics}
A method flags the circuit if at least one checkpoint is flagged.
We call a mutant \emph{candidate-covered} when its changed gate lies in the
backward causal cone of at least one numerical \textsc{QMon} candidate,
whether or not that candidate is deployed.
Of the 2,860 mutants, 2,217 are candidate-covered. For each method, the
candidate-covered detection rate is the fraction of these mutants that the
method flags.
Checkpoints are ordered by gate index.
For a detected mutant, the earliest flagged checkpoint defines the suspect
region as its backward causal cone. The first and second reads at one mid-circuit checkpoint share the same source-gate cone.
We define localization precision as
$\frac{1}{|\mathrm{Cone}|}$ when the injected gate belongs to the earliest
flagged cone, and as zero otherwise. Here, $|\mathrm{Cone}|$ is the number
of unique source-gate indices in that cone, including the gate at the
flagged checkpoint.
Localization precision is averaged over the mutants detected by the method,
so the methods use different localization denominators.
The false-alarm experiment performs one correct-circuit trial for each of
the 238 circuits and each finite-shot method, using a fixed shot seed.
Appendix~\ref{app:rq3details:protocol} provides the detailed metric
definitions and simulation procedure.

\subsubsection{Results}
In Table~\ref{tab:rq3-results}, detection and candidate-covered detection are the unweighted mean $\pm$ sample standard deviation of the five per-seed rates, with pooled counts in parentheses. Candidate-covered detection uses the 2,217 candidate-covered mutants. Localization precision is conditional on each configuration's detected mutants, with the mean and sample standard deviation computed across the five per-seed means. False alarms use all 238 correct circuits. The MQT row reports combined flags consisting of evaluated state mismatches and unevaluable outcomes.

\begin{table*}[!t]
\centering
\caption{RQ3 results for the two comparisons on 2,860 non-output-equivalent mutants.}
\label{tab:rq3-results}
\begingroup\scriptsize
\setlength{\tabcolsep}{2pt}
\renewcommand{\arraystretch}{1.04}
\begin{tabular*}{\textwidth}{@{\extracolsep{\fill}}lcccc@{}}
\toprule
\multicolumn{5}{@{}l}{\textbf{(a) First comparison: QMon batches and separate-checkpoint assertions}} \\
\midrule
Configuration & Detection & Candidate-covered detection & Localization precision & False alarms \\
\midrule
\textsc{QMon} & $71.1\pm4.5\%$ (2,031/2,860) & $91.4\pm0.8\%$ (2,026/2,217) & $0.502\pm0.131$ ($n=2{,}031$) & 0.8\% (2/238) \\
\midrule
Statistical assertion & $67.4\pm4.4\%$ (1,928/2,860) & $86.7\pm1.6\%$ (1,924/2,217) & $0.494\pm0.126$ ($n=1{,}928$) & 0.0\% (0/238) \\
Projection assertion & $71.1\pm4.2\%$ (2,031/2,860) & $91.6\pm1.0\%$ (2,031/2,217) & $0.500\pm0.130$ ($n=2{,}031$) & 0.0\% (0/238) \\
Dynamic ancilla & $71.3\pm4.4\%$ (2,036/2,860) & $91.9\pm0.8\%$ (2,036/2,217) & $0.501\pm0.131$ ($n=2{,}036$) & 0.0\% (0/238) \\
MQT debugger & $59.2\pm6.2\%$ (1,688/2,860) & $76.1\pm4.0\%$ (1,688/2,217) & $0.476\pm0.122$ ($n=1{,}688$) & -- \\
\bottomrule
\end{tabular*}
\par\vspace{5pt}
\begin{tabular*}{\textwidth}{@{\extracolsep{\fill}}lcccc@{}}
\toprule
\multicolumn{5}{@{}l}{\textbf{(b) Second comparison: identical deployed checkpoints and batches}} \\
\midrule
Configuration & Detection & Candidate-covered detection & Localization precision & False alarms \\
\midrule
Batched statistical & $70.9\pm4.8\%$ (2,027/2,860) & $90.7\pm1.6\%$ (2,012/2,217) & $0.429\pm0.123$ ($n=2{,}027$) & 52.9\% (126/238) \\
Batched projection & $70.7\pm4.2\%$ (2,019/2,860) & $91.1\pm1.2\%$ (2,019/2,217) & $0.503\pm0.131$ ($n=2{,}019$) & 0.0\% (0/238) \\
Batched dynamic ancilla & $70.8\pm4.4\%$ (2,024/2,860) & $91.3\pm0.8\%$ (2,024/2,217) & $0.502\pm0.131$ ($n=2{,}024$) & 0.0\% (0/238) \\
\bottomrule
\end{tabular*}
\endgroup\end{table*}

\paragraph{Detection and localization results}
Table~\ref{tab:rq3-results}(a) shows the three assertion methods have observed detection rates of $71.3\pm4.4\%$ for dynamic ancilla assertions, $71.1\pm4.2\%$ for projection assertions, and $67.4\pm4.4\%$ for statistical assertions. \textsc{QMon} detects $71.1\pm4.5\%$ of the same 2,860 mutants. MQT produces a combined flag for $59.2\pm6.2\%$. The assertion baselines evaluate checkpoints on separately executed mutant prefixes, so their readings are not affected by earlier \textsc{QMon} measure--reset--replay operations. Among the 2,217 candidate-covered mutants, \textsc{QMon} detects $91.4\pm0.8\%$. The three assertion methods detect $86.7\pm1.6\%$ to $91.9\pm0.8\%$ of this population, while MQT produces a combined flag for $76.1\pm4.0\%$. Among the mutants detected by each method, \textsc{QMon} has an observed mean localization precision of $0.502\pm0.131$. The localization-precision values in Table~\ref{tab:rq3-results}(a) are conditional on different detected sets and do not form a common-subset ranking. In the correct-circuit experiment, \textsc{QMon} flags 2 of the 238 circuits, giving a false-alarm rate of 0.8\%. The three finite-shot assertion methods produce no false alarms in this noise-free experiment. MQT's correct-circuit pass is used to establish whether its assertions are applicable, so we do not report a separate MQT false-alarm rate. Appendix~\ref{app:rq3details:mqt} provides the detailed MQT flag composition.

\paragraph{Testing overhead}
Table~\ref{tab:rq3-fullscan-counts} compares the planned numbers of circuit runs needed for a complete checkpoint scan on the 238 studied circuits in the first comparison. Each run uses $S=8{,}192$ shots.
The median plan requires only one run for batched \textsc{QMon}, compared with $21$ runs for the separate-run assertion methods.
Specifically, $125$ of the $238$ \textsc{QMon} plans $(52.5\%)$ fit into a single batch. The 75th-percentile counts are $14$ and $89$ runs, respectively.
The distributions are right-skewed: their means are $30.0$ runs for batched \textsc{QMon} and $92.1$ for the assertion methods. The ratio of these means is $3.07$ ($92.1/30.0$).
\textsc{QMon} detects $71.1\%$ of the mutants, compared with $67.4$--$71.3\%$ for the separate-run assertion methods, and requires fewer planned circuit runs in this comparison.
The two dimensions should therefore be considered together rather than treated as a single ranking.

\begin{table*}[t]
\centering
\caption{Planned runs and extra qubits on 238 circuits, with 8,192 shots per run.}
\label{tab:rq3-fullscan-counts}
\scriptsize
\setlength{\tabcolsep}{3pt}
\renewcommand{\arraystretch}{1.04}
\begin{tabular*}{\textwidth}{@{\extracolsep{\fill}}lrrrrrrrrr@{}}
\toprule
Method
& \multicolumn{3}{c}{Planned runs (first comparison)}
& \multicolumn{3}{c}{Planned runs (second comparison)}
& \multicolumn{3}{c}{Extra qubits (both comparisons)} \\
\cmidrule(lr){2-4}
\cmidrule(lr){5-7}
\cmidrule(l){8-10}
& Median & P75 & Mean
& Median & P75 & Mean
& Median & P75 & Maximum \\
\midrule
\textsc{QMon} & 1 & 14 & 30.0 & 1 & 14 & 30.0 & 10 & 15 & 21 \\
Statistical assertion & 21 & 89 & 92.1 & 1 & 14 & 30.0 & 0 & 0 & 0 \\
Projection assertion & 21 & 89 & 92.1 & 1 & 14 & 30.0 & 0 & 0 & 0 \\
Dynamic ancilla & 21 & 89 & 92.1 & 1 & 14 & 30.0 & 1 & 1 & 1 \\
\bottomrule
\end{tabular*}
\par\smallskip
\begin{minipage}{\textwidth}\footnotesize
P75 denotes the 75th percentile. Planned full-scan test shots equal
$8{,}192\times$ planned runs. In the second comparison, all four methods
use a median of 8,192 and a P75 of 114,688 total test shots per circuit.
\end{minipage}
\end{table*}

\paragraph{Miss analysis}
On the 2,860 non-output-equivalent mutants, \textsc{QMon} detects 2,031 and misses 829. Among the 643 mutants that are not candidate-covered, \textsc{QMon} misses 638. A further 180 misses occur even though the changed gate lies in the backward causal cone of at least one deployed checkpoint. Structural coverage does not guarantee an observable or statistically detectable checkpoint change. The remaining 11 misses are deployment gaps: a numerical candidate can be affected by the changed gate, but that candidate is not included in the $Q_{\max}=24$ deployment plan. Appendix~\ref{app:rq3-misses} provides the complete miss analysis, including the circuit-family and gate-replacement results in Table~\ref{tab:rq3-misses}.
\paragraph{Comparison at identical checkpoints and batches}
In the second comparison, the three assertion methods use exactly the same deployed checkpoints and batch assignments as \textsc{QMon}. Table~\ref{tab:rq3-results}(b) reports the results. Batched projection and dynamic-ancilla assertions detect $70.7\pm4.2\%$ and $70.8\pm4.4\%$ of the mutants, respectively, compared with $71.1\pm4.5\%$ for \textsc{QMon}. Batched statistical assertions detect $70.9\pm4.8\%$, but flag 126 of the 238 correct circuits, giving a false-alarm rate of $52.9\%$. Direct measurements at earlier checkpoints can disturb correct superpositions and affect later checks.

Localization precision is $0.502\pm0.131$ for \textsc{QMon}, $0.429\pm0.123$ for batched statistical assertions, $0.503\pm0.131$ for batched projection assertions, and $0.502\pm0.131$ for batched dynamic-ancilla assertions. Each value is averaged over the mutants detected by that method,
so the methods are compared on different sets of mutants.

In the second comparison, all four methods execute every batch in
the \textsc{QMon} plan with 8,192 shots per batch. Their planned run
counts and test-shot costs are therefore identical, as shown in
Table~\ref{tab:rq3-fullscan-counts}.
The qubit costs differ: \textsc{QMon} uses a median of 10 replay ancillas,
compared with no extra qubits for statistical and projection assertions
and one reused ancilla for dynamic assertions.

For miss analysis, \textsc{QMon} misses 829 of the 2,860 mutants, compared with 833 for batched statistical assertions, 841 for batched projection assertions, and 836 for batched dynamic-ancilla assertions. Among the 2,217 candidate-covered mutants, the corresponding miss counts are 191, 205, 198, and 193. These miss counts must be considered together with the false-alarm results, especially for batched statistical assertions.

\begin{tcolorbox}[enhanced jigsaw,breakable,colback=white,colframe=black,boxrule=0.4pt,arc=0pt,boxsep=0pt,left=3pt,right=3pt,top=3pt,bottom=3pt,before skip=6pt,after skip=6pt]
\textbf{Summary of RQ3.} \textsc{QMon} detects $71.1\pm4.5\%$ of the 2,860 mutants with changed final outputs and $91.4\pm0.8\%$ of the 2,217 candidate-covered mutants.
With different checkpoint sets, the three assertion methods detect $67.4$--$71.3\%$, while \textsc{QMon} requires fewer planned runs: the median is one batch per circuit, compared with 21 circuit runs for the separate-run assertion methods.
At identical checkpoints and batches, projection and dynamic-ancilla
assertions detect $70.7$--$70.8\%$, close to \textsc{QMon}'s observed
detection rate. Batched statistical assertions detect $70.9\%$ but
have a false-alarm rate of $52.9\%$, compared with $0.8\%$ for \textsc{QMon}.
\end{tcolorbox}

\subsection{RQ4: How Does Noise Affect Detection and False Alarms?}
\label{sec:rq4}

\subsubsection{Motivation}
The ideal preservation guarantee does not cover hardware noise. RQ4 examines whether noise obscures software faults or produces false alarms on correct circuits, using the calibration-informed simulated noise described in Section~\ref{sec:prep:noise}.

\subsubsection{Approach}
We evaluate the 100-circuit subset described in Section~\ref{sec:prep:program}. The five-seed population contains 1,373 non-output-equivalent mutants, including 762 candidate-covered mutants.

\textsc{QMon}, projection assertions, and dynamic-ancilla assertions collect checkpoint outcomes from 8,192 test shots per batch. All shots execute the complete batched circuit.
For projection and dynamic-ancilla assertions, a failure in one shot does not by itself indicate a logic bug. We compare each checkpoint's failure rate with the correct-circuit calibration data. At each checkpoint, these two methods use only shots with no earlier recorded assertion failure. For example, if exactly one shot fails the first assertion, the second checkpoint has 8,191 usable samples, although all 8,192 shots were executed. No replacement shots are collected.
\textsc{QMon} uses all 8,192 shots at every checkpoint. Statistical assertions test each checkpoint in a separate execution of the prefix of the circuit under test ending at that checkpoint, followed by measurement. Each prefix starts from the circuit's initial state and contains no earlier assertion measurements. Each checkpoint test uses 8,192 shots.

Separately, we collect 32,768 calibration shots per batch or checkpoint-prefix circuit from the correct circuit under the same noise conditions. These data describe the measurement outcomes expected in the presence of noise but without logic faults, and provide the comparison baseline for RQ4's detection and false-alarm tests.
The calibration data are collected for each circuit, method configuration, and noise scale, then reused across fault tests under the same conditions, rather than recollected for each fault. This calibration cost applies to all methods and is reported separately from the test-shot cost.

At each noise scale, every method tests all 1,373 mutants. To measure false alarms, each method also repeats the complete detection test 20 times on each unmodified circuit, collecting fresh test shots for every repetition. Each repetition checks whether the method incorrectly reports a fault in that correct circuit. This gives $100\times20=2{,}000$ correct-circuit tests per method and noise scale.
All five scales, including scale~0, use the same calibration-based decision rules. \textsc{QMon} tests for changes in the outcome frequencies of both the first read and the second read; statistical assertions test for changes in computational-basis frequencies; projection and dynamic-ancilla assertions test for increased assertion-failure rates. We use a circuit-level significance target of $\alpha=0.01$, with Bonferroni correction across checkpoint reads. These calibration-based rules are evaluation adaptations for the noisy comparison. Appendix~\ref{app:rq4details} gives the full protocol.

\subsubsection{Results}
Table~\ref{tab:rq4-noise} reports candidate-covered detection as the mean and sample standard deviation of the five seed-specific rates. For false alarms, we calculate a rate across all 100 correct circuits in each of the 20 repetitions and report the mean and sample standard deviation of these 20 rates.
These false-alarm statistics describe variation between repetitions using the same circuits and fixed calibration data. All evaluated cases remain in the denominators, and only an explicit fault flag contributes to the numerator.

\textsc{QMon}'s mean candidate-covered detection decreases from $98.1\pm1.1\%$ at scale~0 to $96.4\pm1.6\%$ at scale~2, while the mean false-alarm rate increases from $0.20\pm0.41\%$ to $1.20\pm0.95\%$.
Across all five seeds, the corresponding detection counts are 747/762 and 735/762; across all 20 repetitions, the false-alarm counts are 4/2,000 and 24/2,000.

The higher detection rate than the 91.4\% reported in RQ3 is measured on a different population: RQ4 uses 762 candidate-covered mutants from 100 single-batch circuits with at most 12 qubits, whereas RQ3 uses 2,217 candidate-covered mutants from 238 circuits. RQ4 also uses calibration-based decision rules even at scale~0. The percentages therefore summarize different populations and testing procedures.

From scale~0 to scale~2, mean candidate-covered detection decreases from 93.5\% to 93.0\% for statistical assertions, from 98.1\% to 95.0\% for projection assertions, and from 98.3\% to 90.8\% for dynamic-ancilla assertions. At scale~2, the corresponding mean false-alarm rates are 0.75\%, 0.95\%, and 0.65\%, respectively.

\begin{table*}[t]
\centering
\caption{Candidate-covered detection / correct-circuit false-alarm rates (\%) under simulated noise.}
\label{tab:rq4-noise}
\scriptsize
\setlength{\tabcolsep}{3pt}
\renewcommand{\arraystretch}{1.04}
\begin{tabular*}{\textwidth}{@{\extracolsep{\fill}}lcccc@{}}
\toprule
Scale & \textsc{QMon} & Statistical & Projection & Dynamic ancilla \\
\midrule
0 & $98.1\pm1.1\,/\,0.20\pm0.41$ & $93.5\pm3.2\,/\,1.20\pm1.20$ & $98.1\pm1.1\,/\,0.00\pm0.00$ & $98.3\pm1.2\,/\,0.00\pm0.00$ \\
0.25 & $97.3\pm1.1\,/\,0.60\pm0.75$ & $93.4\pm3.3\,/\,1.00\pm1.12$ & $97.1\pm1.1\,/\,0.70\pm0.80$ & $95.5\pm1.7\,/\,1.05\pm1.28$ \\
0.5 & $97.3\pm1.3\,/\,0.90\pm1.25$ & $93.2\pm3.2\,/\,1.30\pm1.17$ & $96.6\pm1.4\,/\,0.65\pm0.67$ & $94.7\pm3.2\,/\,1.05\pm1.15$ \\
1 & $96.9\pm1.2\,/\,1.05\pm0.89$ & $93.0\pm2.9\,/\,0.70\pm0.92$ & $96.0\pm1.3\,/\,0.75\pm0.64$ & $94.1\pm1.4\,/\,0.75\pm0.85$ \\
2 & $96.4\pm1.6\,/\,1.20\pm0.95$ & $93.0\pm3.1\,/\,0.75\pm0.79$ & $95.0\pm1.8\,/\,0.95\pm1.28$ & $90.8\pm3.2\,/\,0.65\pm0.59$ \\
\bottomrule
\end{tabular*}
\par\smallskip
\begin{minipage}{\textwidth}
\footnotesize
\raggedright
Each entry reports detection / false alarms, both as mean $\pm$ sample standard deviation. Detection rates are summarized over five mutation seeds. False-alarm rates are summarized over 20 repetitions, each testing all 100 correct circuits.
\end{minipage}
\end{table*}

We compare candidate-covered detection using paired exact McNemar tests at the nominal 0.01 level. At this threshold, no comparison between \textsc{QMon} and projection assertions is statistically significant. This does not establish equivalence.
\textsc{QMon} has significantly higher candidate-covered detection than statistical assertions at all five scales and dynamic-ancilla assertions at all four positive scales. The $p$-values range from $1.46\times10^{-11}$ to $7.92\times10^{-9}$ for statistical assertions, and from $1.82\times10^{-12}$ to 0.0034 for dynamic-ancilla assertions. These comparisons are exploratory. Appendix~\ref{app:rq4details} provides the paired counts, full results, and limitations of the tests.

\begin{tcolorbox}[enhanced jigsaw,breakable,colback=white,colframe=black,boxrule=0.4pt,arc=0pt,boxsep=0pt,left=3pt,right=3pt,top=3pt,bottom=3pt,before skip=6pt,after skip=6pt]
\textbf{Summary of RQ4.} Under the specified simulated noise and calibration procedure, \textsc{QMon} retains a mean candidate-covered detection rate of 96.4\% at noise scale~2, with a mean false-alarm rate of 1.2\%.
\end{tcolorbox}

\section{Discussion}
\label{sec:discussion}

Appendix~\ref{app:discussion} provides a more detailed discussion. The main
findings are summarized below.

\paragraph{Scalability}
MPS-based selection of numerical candidates completes for all 144 circuits
in the 15--50-qubit scalability set.

At $Q_{\max}=100$, every selected monitor group in the 144 circuits can be
placed in a deployment plan without exceeding the qubit limit. This does not
make every plan practical. The largest case still requires 134,893 batches,
and its planned count of original-circuit, replay, basis-rotation, measurement, and reset
operations is $229{,}306.30$ times the original gate count
(Appendix~\ref{sec:scalability}). Having enough qubits to fit a deployment
plan therefore does not guarantee a practical instrumentation cost. These
counts describe the circuit structure and are not estimates of hardware
runtime.

\paragraph{Multiple faults}
The additional finite-shot experiment evaluates one, two, or three
simultaneous faults~(Appendix~\ref{sec:multi-error}, Table~\ref{tab:multi-error}). The one-fault condition uses only the first of the three
locations sampled for each circuit--seed pair. It is therefore a different
set of mutants from the primary single-fault experiment, which evaluates all
three sampled locations separately. Among the non-output-equivalent cases completed by all methods in each
condition, the observed \textsc{QMon} detection rate increases from
$74.0\pm7.4\%$ with one fault to $86.5\pm7.4\%$ with three faults. Because
the mutant populations differ across the three conditions, this result does
not show the effect of adding faults to the same mutants.

Among detected cases in which at least one injected fault can affect a
deployed \textsc{QMon} checkpoint, the earliest flagged causal cone contains
an injected fault in more than 99\% of cases for all three fault counts.
Among detected cases in which every injected fault can affect a deployed
checkpoint, the union of all flagged cones contains every fault in
$99.9\pm0.3\%$ of the one-fault cases, $94.4\pm2.2\%$ of the two-fault
cases, and $94.8\pm1.4\%$ of the three-fault cases. Thus, the larger-fault
populations have higher observed detection rates.

\paragraph{Use in software testing}
\textsc{QMon} compares checkpoint observations with expected one-qubit
states and probabilities computed from an original circuit assumed to be
correct. In the mutation experiments, each mutant is tested against the
expectations obtained from its unmodified circuit. The recorded outcomes
can also be inspected as an execution trace without assigning a pass/fail
verdict.
\label{sec:discussion:simulation-limit}
These uses require a platform that supports compatible mid-circuit
measurement and reset. Appendix~\ref{sec:se-workflow} provides further discussion of these uses in the quantum software workflow.

\section{Threats to Validity}
\label{sec:threats-to-validity}

\noindent\textbf{Internal Validity.}
RQ3 replaces eligible one- and two-qubit gates without changing
their qubit operands. This model does
not cover all bug types or
how source-level errors appear after compilation. In particular,
an error in a source-level multi-qubit gate may change several
one- and two-qubit gates on different qubits after compilation;
our experiment does not model this transformation.

\noindent\textbf{Construct Validity.}
Candidate selection uses numerical tolerances and does not establish
the exact separability assumed by the preservation theorem.
The first and second reads check selected one-qubit distributions,
not arbitrary joint states. Finite-shot tests may miss faults that
leave these distributions unchanged or change them too little to detect.
Output equivalence refers only to the final measured-output distribution.
Excluding cases where any method returns no decision may favor
easier cases in the multiple-fault comparison.

\noindent\textbf{External Validity.}
RQ1 and RQ2 use 310 circuits with 2--14 qubits, RQ3 uses 238 of these
circuits, RQ4 uses a 100-circuit subset, and the MPS-based scalability analysis uses 144 circuits with 15--50 qubits. These
datasets may not represent other circuit families, input states,
architectures, or entanglement structures. All experiments are simulated.
RQ4 uses homogeneous medians from a historical
device-calibration snapshot, with all-to-all connectivity. These simulations do not establish
performance on a current physical device.

\noindent\textbf{Conclusion Validity.}
The planned run and test-shot counts in
Table~\ref{tab:rq3-fullscan-counts} describe a complete checkpoint scan.
They do not account for early stopping or the number and duration of
operations within each run. Appendix~\ref{sec:scalability} separately
counts original-circuit, replay, basis-rotation, measurement, and reset operations;
these structural counts are not hardware-runtime measurements. The RQ4 paired comparisons are exploratory. Their $p$-values are
not adjusted for the comparisons across the three assertion methods and
five noise scales. Test results may be correlated when mutants originate from the
same circuit or tests share calibration data.

\section{Conclusion}
\label{sec:conclusion}

\textsc{QMon} records selected intermediate outcomes through
mid-circuit measurement, reset, and causal-cone replay, preserving the
full circuit state under the stated ideal conditions. We evaluate
\textsc{QMon}'s usefulness in detecting programming errors in noise-free
and noisy simulations.
At matched checkpoints and batches, \textsc{QMon}, projection assertions,
and dynamic-ancilla assertions have similar observed noise-free detection
rates for mutants with changed final outputs and identical planned run
and test-shot counts.
At the largest tested noise scale, \textsc{QMon} achieves a mean
detection rate of 96.4\% on candidate-covered mutants, with a mean
false-alarm rate of 1.2\%.
Evaluation of the method on physical quantum hardware remains
future work.

\section*{Acknowledgments}
We used OpenAI Codex to assist
with implementing and running experiment scripts, generating tables
from experiment outputs, polishing
the main text and appendices, and reviewing the proofs. The authors are responsible for the accuracy and
integrity of all AI-assisted content.

\balance
\bibliographystyle{IEEEtran}
\bibliography{quantum-monitoring_copy}

\newpage
\appendices
\section{THEORETICAL AND IMPLEMENTATION ANALYSIS}
\label{app:proof}

\subsection{Proof of the No-Disturbance Theorem}
\label{app:no-disturbance-proof}

This appendix proves Theorem~\ref{thm:nodist} from
Section~\ref{sec:no-disturbance}. It identifies the exact single-monitor
condition under which measure--reset--replay instrumentation preserves the
full circuit state. It also explains how Algorithm~\ref{alg:screen} tests
this condition numerically and shows that the state-preservation guarantee
does not depend on the depth or internal gate structure of the replayed
backward causal cone.

Corollary~\ref{cor:two-read-preservation} extends the single-monitor
guarantee to the complete two-read checkpoint. The arguments specialize
standard product-state measurement, partial-trace, and fidelity
facts~\cite{nielsen2010quantum,jozsa1994fidelity}.

\paragraph{Setup}

Consider a circuit
\[
C=U_{m-1}\cdots U_0
\]
consisting of $m$ one- and two-qubit unitary gates. The circuit acts on
\[
\HS=(\mathbb{C}^2)^{\otimes n}
\]
and starts in the all-zero state $\zeron$. Define the prefix unitary and its
output state as
\[
V_i:=U_i\cdots U_0,
\qquad
\ket{\psi_i}:=V_i\zeron,
\qquad
i\in\{0,\ldots,m-1\}.
\]

Fix a monitored gate--qubit location $(i,q)$, where $i$ is a gate index and
$q$ is one of the qubits acted on by gate $i$. Let
\[
[n]:=\{0,\ldots,n-1\},
\qquad
R:=[n]\setminus\{q\},
\]
and write the full Hilbert space as
\[
\HS=\HS_q\otimes\HS_R.
\]
The reduced state of $q$ immediately after gate $i$ is
\[
\rho_q(i):=
\Tr_R\ketbra{\psi_i}{\psi_i}.
\]

Because $q$ is one qubit, $\rho_q(i)$ has two eigenvalues, which we denote by
$\lambda_1$ and $\lambda_2$. Its determinant is therefore
\[
\det\rho_q(i)=\lambda_1\lambda_2.
\]

To describe the effect of measuring $q$, write $\ket{\psi_i}$ in the
computational basis of $q$ as
\[
\ket{\psi_i}
=
\ket{0}_q\ket{\widetilde{s}_0}_R
+
\ket{1}_q\ket{\widetilde{s}_1}_R,
\text{ }
p_b:=
\langle\widetilde{s}_b
\vert
\widetilde{s}_b\rangle,
\text{ }b \in\{0,1\}
\]
The vectors $\ket{\widetilde{s}_0}_R$ and
$\ket{\widetilde{s}_1}_R$ are generally unnormalized states of $R$
associated with the two possible measurement outcomes of $q$. The value
$p_b$ is the probability of observing outcome $b$ when $q$ is measured in
the computational basis.

Taking the partial trace over $R$ gives
\[
\begin{aligned}
\rho_q(i)
&=
\operatorname{Tr}_R
\left(
\ket{\psi_i}\!\bra{\psi_i}
\right)\\
&=
\sum_{b,b'\in\{0,1\}}
\langle\widetilde{s}_{b'}|
\widetilde{s}_{b}\rangle
\ket b\!\bra{b'}\\
&=
\begin{pmatrix}
p_0 &
\langle\widetilde{s}_1|\widetilde{s}_0\rangle\\
\langle\widetilde{s}_0|\widetilde{s}_1\rangle &
p_1
\end{pmatrix}.
\end{aligned}
\]

It follows that
\[
\begin{aligned}
\det\rho_q(i)
&=
p_0p_1
-
\langle\widetilde{s}_1|\widetilde{s}_0\rangle
\langle\widetilde{s}_0|\widetilde{s}_1\rangle\\
&=
p_0p_1
-
\left|
\langle\widetilde{s}_0|\widetilde{s}_1\rangle
\right|^2.
\end{aligned}
\]

When $p_0p_1>0$, define the normalized conditional states
\[
\ket{s_b}
:=
\frac{\ket{\widetilde{s}_b}}{\sqrt{p_b}}.
\]
Because
\[
\langle\widetilde{s}_0|\widetilde{s}_1\rangle
=
\sqrt{p_0p_1}\,
\langle s_0|s_1\rangle,
\]
the determinant can also be written as
\[
\det\rho_q(i)
=
p_0p_1
\left(
1-
\left|
\langle s_0|s_1\rangle
\right|^2
\right).
\]

The following definition and proposition connect this determinant to
separability.

\begin{definition}[Monitorable location]
A gate--qubit location $(i,q)$ is \emph{monitorable} if and only if
\[
\det\rho_q(i)=0.
\]
\end{definition}

\begin{proposition}[Separability]
\label{prop:schmidt}

For a pure state $\ket{\psi_i}$, the following conditions are equivalent:

\begin{enumerate}
    \item $\det\rho_q(i)=0$;
    \item $\rho_q(i)=\ketbra{u}{u}$ is a rank-one projector;
    \item
    $
    \ket{\psi_i}
    =
    \ket{u}_q\otimes\ket{\chi}_R
    $
    for unit vectors
    $\ket{u}\in\HS_q$ and $\ket{\chi}\in\HS_R$.
\end{enumerate}
\end{proposition}

\begin{proof}

Write the Schmidt decomposition of $\ket{\psi_i}$ across the split
$q\mid R$ as
\[
\ket{\psi_i}
=
\sum_{k=1}^{r}
\sqrt{\lambda_k}\,
\ket{u_k}_q\ket{v_k}_R,
\]
where
\[
\lambda_1\geq\cdots\geq\lambda_r>0,
\]
$\{\ket{u_k}\}$ and $\{\ket{v_k}\}$ are orthonormal sets, and
$r\leq2$ is the Schmidt rank. The values $\lambda_k$ are the Schmidt
probabilities, while $\sqrt{\lambda_k}$ are the Schmidt amplitudes.

The reduced state of $q$ is
\[
\rho_q(i)
=
\sum_k
\lambda_k\ketbra{u_k}{u_k}.
\]
Its two eigenvalues are $\lambda_1$ and $\lambda_2$, with the convention
$\lambda_2=0$ when $r=1$. Therefore,
\[
\det\rho_q(i)
=
\lambda_1\lambda_2.
\]

The determinant is zero if and only if $r=1$. Schmidt rank one is equivalent
to $\rho_q(i)$ being a pure rank-one state and to the factorization
\[
\ket{\psi_i}
=
\ket{u}_q\otimes\ket{\chi}_R.
\]
The three conditions are therefore equivalent.
\end{proof}

Thus,
\[
\det\rho_q(i)=0
\]
is necessary and sufficient for $q$ to be separable from the other qubits at
location $(i,q)$. This is the exact mathematical condition used by
Theorem~\ref{thm:nodist}.

Algorithm~\ref{alg:screen} cannot test exact equality to zero using
floating-point arithmetic. It instead identifies $(i,q)$ as a numerical
candidate when
\[
a_2(i,q)\leq\varepsilon_{\mathrm{sv}}.
\]
Passing this finite-precision test provides numerical evidence that the exact
condition holds, but it does not prove that $a_2(i,q)$ or
$\det\rho_q(i)$ is mathematically equal to zero.

\paragraph{Backward causal cone}

Define $\Cone(i,q)$ as the full backward causal cone of location $(i,q)$. It
contains gate $U_i$ and, recursively, the immediately preceding gate on every
input wire of each gate already included in the cone. Equivalently,
$\Cone(i,q)$ is the union of all backward causal paths from $(i,q)$, as
computed by Algorithm~\ref{alg:dfs}.

Let
\[
Q\subseteq[n]
\]
be the set of qubits touched by the cone. If
\[
j_1<\cdots<j_\ell
\]
are the gate indices in the cone, define
\[
V_{\mathrm{cone}}
:=
U_{j_\ell}\cdots U_{j_1},
\]
so that gates with smaller indices act first. The operator
$V_{\mathrm{cone}}$ is a unitary on $\HS_Q$. Write
\[
\HS_Q=\HS_q\otimes\HS_A,
\qquad
A:=Q\setminus\{q\}.
\]
We use $\ket{0}_Q$ to denote the all-zero state of the qubits in $Q$.

\begin{proposition}[Cone marginal]
\label{prop:cone}

The reduced state of $q$ immediately after gate $i$ is determined entirely
by its backward causal cone:
\begin{equation}
\rho_q(i)
=
\Tr_A
\left(
V_{\mathrm{cone}}
\ketbra{0}{0}_Q
V_{\mathrm{cone}}^\dagger
\right).
\label{eq:cone}
\end{equation}
\end{proposition}

\begin{proof}

We use induction on the number of gates in $V_i$ that do not belong to
$\Cone(i,q)$.

If there are no such gates, then the part of $V_i$ that can affect $q$ is
exactly $V_{\mathrm{cone}}$. All qubits outside $Q$ remain in their initial
state and can be traced out without changing the reduced state of $q$.
Equation~\eqref{eq:cone} therefore follows directly.

Now suppose that $V_i$ contains at least one gate outside the cone. Let
$U_j\notin\Cone(i,q)$ be the latest such gate, and let $S$ be the set of
qubits on which $U_j$ acts. Write
\[
V_i=W_2U_jW_1,
\]
where
\[
W_2:=U_i\cdots U_{j+1},
\qquad
W_1:=U_{j-1}\cdots U_0,
\]
with an empty product interpreted as the identity.

Because $U_j$ is the latest gate outside the cone, every gate in $W_2$
belongs to $\Cone(i,q)$. No gate in $W_2$ can act on a qubit in $S$. If a
later cone gate acted on one of these qubits, tracing that gate backward along
the corresponding input wire would include $U_j$ in the cone, contradicting
$U_j\notin\Cone(i,q)$. Thus, $W_2$ and $U_j$ act on disjoint sets of qubits
and commute:
\[
W_2U_j=U_jW_2.
\]

In addition, $q\notin S$, because every gate acting on $q$ before or at gate
$i$ belongs to the backward causal cone of $(i,q)$.

Let
\[
\rho_0=\ketbra{0^n}{0^n}.
\]
Using $W_2U_j=U_jW_2$, we obtain
\begin{align}
\rho_q(i)
&=
\Tr_R
\left(
W_2U_jW_1
\rho_0
W_1^\dagger U_j^\dagger W_2^\dagger
\right)
\nonumber\\
&=
\Tr_R
\left(
U_jW_2W_1
\rho_0
W_1^\dagger W_2^\dagger U_j^\dagger
\right)
\nonumber\\
&=
\Tr_R
\left(
W_2W_1
\rho_0
W_1^\dagger W_2^\dagger
\right).
\end{align}
The last equality holds because $U_j$ acts only on
$S\subseteq R$, and the partial trace over $R$ is unchanged by a unitary
acting only on the subsystem being traced out. Removing $U_j$ therefore does
not change $\rho_q(i)$.

Repeating this argument removes every gate outside $\Cone(i,q)$ without
changing the reduced state of $q$. The remaining circuit is precisely
$V_{\mathrm{cone}}$ acting on the cone qubits $Q$. All qubits outside $Q$
remain in $\ket{0}$ and factor from the cone state. Tracing out
$A=Q\setminus\{q\}$ gives Eq.~\eqref{eq:cone}.
\end{proof}

Combining Propositions~\ref{prop:schmidt} and~\ref{prop:cone} gives the
property needed for reconstruction. If $(i,q)$ is monitorable, then the
reduced state of $q$ in
\[
V_{\mathrm{cone}}\ket{0}_Q
\]
is the pure state $\ketbra{u}{u}$. Applying
Proposition~\ref{prop:schmidt} to
\[
\ket{\Phi}
:=
V_{\mathrm{cone}}\ket{0}_Q
\in
\HS_q\otimes\HS_A
\]
therefore gives
\begin{equation}
V_{\mathrm{cone}}\ket{0}_Q
=
\ket{u}_q\otimes\ket{a}_A
\label{eq:cone-product}
\end{equation}
for some unit vector $\ket{a}\in\HS_A$.

Equation~\eqref{eq:cone-product} means that replay reproduces the complete
state $\ket{u}$ of the monitored qubit, including its relative phase. The
monitored qubit is in a product state with the other cone wires, so the replay
ancillas are not entangled with it after exact reconstruction.

\paragraph{Instrumentation}

At a monitored location $(i,q)$, the monitor performs the following
operations. An arbitrary read basis is implemented by a local
unitary $W$ before computational-basis measurement; $W=I$ for the first read
and $W=U^\dagger$ for the second read.

\begin{enumerate}
    \item it applies $W$, then measures $q$ in the computational basis and records the outcome;
    \item it resets $q$ to $\ket{0}$;
    \item it introduces a fresh ancilla register $A'$ with
    $\HS_{A'}\cong\HS_A$, initialized in its all-zero state;
    \item it applies $V_{\mathrm{cone}}$ to
    $\HS_q\otimes\HS_{A'}$, with every non-target cone wire mapped to its
    corresponding fresh ancilla.
\end{enumerate}

The original qubits in
\[
R=[n]\setminus\{q\}
\]
are left untouched during replay. Let
\[
V_{\mathrm{tail}}
:=
U_{m-1}\cdots U_{i+1}
\]
be the remaining part of the original circuit, with
$V_{\mathrm{tail}}=I$ when $i=m-1$.

\medskip
\noindent\textbf{Theorem~\ref{thm:nodist} (No disturbance).}
\emph{
Assume an ideal setting in which the original circuit is closed and unitary,
the matched causal-cone replay is exact, the replay uses fresh ancillas, and
the recorded measurement outcome is not used to choose or modify later
operations. If $(i,q)$ is monitorable, then the final density operator of the
instrumented circuit on the original $n$ qubits is
\[
V_{\mathrm{tail}}
\ketbra{\psi_i}{\psi_i}
V_{\mathrm{tail}}^\dagger,
\]
which is exactly the final state of the uninstrumented circuit. The
mid-circuit outcome is still recorded, but it does not change the continued
computation. The final output distribution is therefore identical, and the
state disturbance is exactly zero.
}
\medskip

\begin{proof}[Proof of Theorem~\ref{thm:nodist}]

By Proposition~\ref{prop:schmidt},
\[
\ket{\psi_i}
=
\ket{u}_q\otimes\ket{\chi}_R.
\]
Thus, the monitored qubit is separable from all other original qubits.

Applying $W$ and measuring $q$ in the computational basis gives outcome
$b\in\{0,1\}$ with probability
\[
p_b^{(W)}=
\left|
\langle b|W|u\rangle
\right|^2.
\]
The corresponding post-measurement state is
\[
\ket{b}_q\otimes\ket{\chi}_R.
\]
The state $\ket{\chi}_R$ is unchanged because it was already in a product
state with $q$.

This is the step that fails when $\det\rho_q(i)>0$. Conditioning on the measurement outcome can change the state of $R$,
although its outcome-averaged reduced state remains unchanged.
Proposition~\ref{prop:tightlaw} quantifies the disturbance to the joint state of the
original circuit qubits, and
Fig.~\ref{fig:ent-dist} illustrates the resulting disturbance.

Reset maps
\[
\ket{b}_q
\longmapsto
\ket{0}_q
\]
for either measurement outcome. The post-reset state is therefore
\[
\ket{0}_q\otimes\ket{\chi}_R,
\]
which is independent of the recorded value $b$.

The fresh ancilla register $A'$ is initialized in the all-zero state. Exact
matched replay applies $V_{\mathrm{cone}}$ to $q$ and $A'$. By
Eq.~\eqref{eq:cone-product}, this gives
\[
\ket{u}_q
\otimes
\ket{\chi}_R
\otimes
\ket{a}_{A'}
=
\ket{\psi_i}_{qR}
\otimes
\ket{a}_{A'}.
\]
Thus, immediately before the circuit tail begins, the original qubits have
been restored to their uninstrumented state $\ket{\psi_i}$.

Let $B$ be the classical register that stores the measurement outcome.
Because the outcome is recorded but never used to control later gates, the
joint classical--quantum state after the tail is
\[
\Omega_{BqRA'}
=
\left(
\sum_b
p_b^{(W)}\ketbra{b}{b}_B
\right)
\otimes
V_{\mathrm{tail}}
\ketbra{\psi_i}{\psi_i}
V_{\mathrm{tail}}^\dagger
\otimes
\ketbra{a}{a}_{A'}.
\]

Tracing out the classical record and replay ancillas gives
\[
\Tr_{BA'}\Omega_{BqRA'}
=
V_{\mathrm{tail}}
\ketbra{\psi_i}{\psi_i}
V_{\mathrm{tail}}^\dagger
\]
on the original qubits. This is exactly the final state of the
uninstrumented circuit. Recording the outcome therefore leaves no residual
effect on the continued computation.
\end{proof}

\begin{corollary}[Depth independence]
\label{cor:depth}

Theorem~\ref{thm:nodist} holds for any backward causal cone
$\Cone(i,q)$, regardless of its depth, number of SWAP gates, or number of
repeated two-qubit interactions.
\end{corollary}

\begin{proof}

The proof of Theorem~\ref{thm:nodist} uses only two properties:

\begin{enumerate}
    \item the exact separability condition
    $\det\rho_q(i)=0$, stated in
    Proposition~\ref{prop:schmidt}; and
    \item the product-state reconstruction in
    Eq.~\eqref{eq:cone-product}, obtained from
    Propositions~\ref{prop:schmidt} and~\ref{prop:cone}.
\end{enumerate}

Neither property places a limit on the depth or branching structure of
$V_{\mathrm{cone}}$. The proof of Proposition~\ref{prop:cone} removes gates
outside the cone without making assumptions about the gates inside it.
Equation~\eqref{eq:cone-product} then follows from the purity of the
monitored qubit's reduced state.

A SWAP is an ordinary two-qubit unitary inside
$V_{\mathrm{cone}}$. A chain of SWAP gates or repeated CNOT gates only
changes the size and content of $V_{\mathrm{cone}}$. It does not change the
separability condition used by the theorem. The structure of the cone
therefore affects the number of replay gates and fresh ancillas, but not the
conditional state-preservation guarantee
(Appendix~\ref{sec:scalability}).
\end{proof}

\begin{proposition}[Converse: separability is necessary for full-state
preservation]
\label{prop:converse}

Under the same ideal single-monitor setting, if
\[
\det\rho_q(i)>0,
\]
then the instrumentation does not preserve the full circuit state. Let
\[
\ket{\phi}
:=
V_{\mathrm{tail}}\ket{\psi_i}
\]
be the uninstrumented final state. The final state of the instrumented circuit
on the original qubits differs from $\ketbra{\phi}{\phi}$, and its
full-state infidelity is strictly positive.
\end{proposition}

\begin{proof}

By Proposition~\ref{prop:schmidt},
\[
\det\rho_q(i)>0
\]
means that $\ket{\psi_i}$ has Schmidt rank two across the split $q\mid R$.
We can therefore write
\[
\ket{\psi_i}
=
\sqrt{\lambda_1}
\ket{u_1}_q\ket{v_1}_R
+
\sqrt{\lambda_2}
\ket{u_2}_q\ket{v_2}_R,
\]
where
\[
\lambda_1,\lambda_2>0,
\qquad
\lambda_1+\lambda_2=1,
\]
and $\{\ket{u_1},\ket{u_2}\}$ and
$\{\ket{v_1},\ket{v_2}\}$ are orthonormal sets.

Measuring $q$ and discarding the measurement outcome leaves $R$ in the
reduced state
\[
\rho_R
=
\lambda_1\ketbra{v_1}{v_1}
+
\lambda_2\ketbra{v_2}{v_2}.
\]
The reset and replay operations act only on $q$ and the fresh ancillas. They
do not act on $R$. After the replay ancillas are traced out, the state of the
original qubits immediately before the tail therefore has the product form
\[
\sigma_q\otimes\rho_R
\]
for some one-qubit density operator $\sigma_q$.

Let $F$ be the fidelity between the instrumented and uninstrumented final
states. Because $V_{\mathrm{tail}}$ is unitary, applying it to both states
does not change their fidelity. Hence,
\[
\begin{aligned}
F
&=
\bra{\psi_i}
\left(
\sigma_q\otimes\rho_R
\right)
\ket{\psi_i}\\
&=
\sum_{k=1}^{2}
\lambda_k^2 w_k,
\end{aligned}
\]
where
\[
w_k
:=
\bra{u_k}\sigma_q\ket{u_k}.
\]

Because $\sigma_q$ is a density operator and
$\{\ket{u_1},\ket{u_2}\}$ is an orthonormal basis of the one-qubit space,
\[
w_k\geq0,
\qquad
\sum_k w_k
=
\Tr\sigma_q
=
1.
\]
Moreover, Schmidt rank two implies
\[
0<\lambda_k<1.
\]
Therefore,
\[
F
\leq
\max_k\lambda_k^2
\sum_k w_k
=
\max_k\lambda_k^2
<
1.
\]
The full-state infidelity $1-F$ is thus strictly positive.
\end{proof}

Together, Theorem~\ref{thm:nodist} and
Proposition~\ref{prop:converse} establish the exact single-monitor boundary
used in RQ1. Under the stated ideal conditions, the instrumentation preserves
the full circuit state if and only if the monitored qubit is separable from
the other qubits:
\[
\det\rho_q(i)=0.
\]

The forward direction also guarantees that every final measurement
distribution is unchanged. The converse is a statement about the complete
quantum state. It does not claim that every particular final measurement
distribution must change, because a chosen measurement may not reveal a
difference in phase or correlation.

\begin{proposition}[Grouped monitors at one gate]
\label{prop:grouped-monitor}

Let
\[
M\neq\varnothing
\]
be the set of qubits monitored together immediately after gate $i$, and let
\[
R=[n]\setminus M.
\]
The grouped instrumentation measures and resets every qubit in $M$, forms the
union of their backward causal cones, and replays that union once. The cone
wires in $M$ remain on their original program wires, while every other cone
wire is mapped to a fresh ancilla.

Define
\[
\rho_M
=
\Tr_R\ketbra{\psi_i}{\psi_i},
\qquad
\rho_R
=
\Tr_M\ketbra{\psi_i}{\psi_i}.
\]
After the measurement records and replay ancillas are traced out, the state
of the original qubits immediately before the circuit tail is
\begin{equation}
\rho_M\otimes\rho_R.
\label{eq:grouped-channel}
\end{equation}

Consequently, the grouped monitor preserves $\ket{\psi_i}$ if and only if
$\rho_M$ has rank one. If $\{\lambda_k\}$ are the Schmidt probabilities
across the split $M\mid R$, then the fidelity between the grouped
instrumented state and the original state is
\[
F
=
\Tr(\rho_M^3)
=
\sum_k\lambda_k^3.
\]
\end{proposition}

\begin{proof}

When both operands of a two-qubit gate are monitored after the same gate,
their full backward causal cones coincide once gate $i$ is included, because
tracing backward from that gate follows every input wire recursively. More
generally, the grouped construction replays the union of the cones of the
qubits in $M$ and includes every shared gate only once.

Measuring and resetting all qubits in $M$, while discarding the outcomes and
using no outcome-dependent control, maps an arbitrary state $X_{MR}$ to
\[
X_{MR}
\longmapsto
\ketbra{0}{0}_M
\otimes
\Tr_M X_{MR}.
\]
The order of the measurements and resets on distinct qubits in $M$ does not
change this result.

The argument of Proposition~\ref{prop:cone} also applies when the
single-qubit subsystem $q$ is replaced by the block $M$. It shows that exact
replay of the union cone on $M$ and fresh ancillas reconstructs a joint state
whose reduced state on $M$ is $\rho_M$. The replay does not act on the
original qubits in $R$. After the replay ancillas are traced out, the state of
the original register is therefore
\[
\rho_M\otimes\rho_R,
\]
which proves Eq.~\eqref{eq:grouped-channel}.

Now write the Schmidt decomposition across $M\mid R$ as
\[
\ket{\psi_i}
=
\sum_k
\sqrt{\lambda_k}
\ket{u_k}_M\ket{v_k}_R.
\]
The fidelity between the original pure state and the product state in
Eq.~\eqref{eq:grouped-channel} is
\[
\begin{aligned}
F
&=
\bra{\psi_i}
\left(
\rho_M\otimes\rho_R
\right)
\ket{\psi_i}\\
&=
\sum_k\lambda_k^3\\
&=
\Tr(\rho_M^3).
\end{aligned}
\]
This fidelity equals one exactly when one Schmidt probability equals one and
all others are zero. Equivalently, the Schmidt rank is one and $\rho_M$ is
pure. Thus, the grouped monitor preserves the original state if and only if
$\rho_M$ has rank one.
\end{proof}

The idealized per-qubit rule requires $\det\rho_q(i)=0$ for every
$q\in M$. The implementation approximates this condition using
$a_2(i,q)\le\varepsilon_{\mathrm{sv}}$. Under the exact per-qubit condition, repeated application of
Proposition~\ref{prop:schmidt} then gives
\[
\ket{\psi_i}
=
\left(
\bigotimes_{q\in M}
\ket{u_q}
\right)
\otimes
\ket{\chi}_R.
\]
This factorization implies that $\rho_M$ is pure and therefore satisfies the
condition of Proposition~\ref{prop:grouped-monitor}.

The per-qubit rule is conservative. For example, a complete Bell-pair block
may be in a pure state and separable from $R$, even though each individual
qubit in that block has a mixed reduced state. Such a block satisfies the
joint condition in Proposition~\ref{prop:grouped-monitor}, but it does not
pass the implementation's per-qubit selection rule.

\begin{corollary}[Composition of monitor groups]
\label{cor:compose}

Group the selected gate--qubit locations by gate index. Suppose that:

\begin{enumerate}
    \item every qubit in every monitor group satisfies its exact per-qubit
    separability condition in the corresponding prefix of the uninstrumented
    circuit;
    \item every causal-cone replay is exact and matched to the original
    circuit;
    \item the recorded measurement outcomes are not used to choose or modify
    later operations; and
    \item different monitor groups use disjoint fresh ancilla registers.
\end{enumerate}

Then the monitor groups, inserted in increasing gate order, preserve the
final state and every final measurement distribution.
\end{corollary}

\begin{proof}

Order the monitor groups by increasing gate index. By
Proposition~\ref{prop:grouped-monitor}, the first group restores the original
prefix state on the program qubits and factors its measurement records and
replay ancillas from that state.

The original computation therefore continues from exactly the same prefix
state as the uninstrumented circuit. When execution reaches the next monitor
group, the exact separability conditions verified in the uninstrumented
circuit still hold because all earlier groups have restored the original
program state.

Applying the same argument to each group in sequence proves by induction that
all groups preserve the program state. The final state and every output
distribution are therefore identical to those of the uninstrumented circuit.
\end{proof}

\begin{corollary}[Preservation by the complete checkpoint]
\label{cor:two-read-preservation}
Assume ideal, noise-free execution of the original circuit, pure target marginals at
the selected locations, exact matched replay, and no outcome-dependent
control. Different groups use disjoint replay registers; each register is
initialized or reset to $\ket{0}$ before every replay. Then the computational
measurement, reset, and replay followed by the second measurement, reset, and replay
preserves the program state, and so does their composition in program order.
\end{corollary}

\begin{proof}
The first measurement, reset, and replay restore the original state by
Proposition~\ref{prop:grouped-monitor}. For the second read, local basis
rotations precede measurement. The targets and replay ancillas are then
reset before the same union cone is replayed, so
Proposition~\ref{prop:grouped-monitor} applies again.
Corollary~\ref{cor:compose} gives preservation across successive groups.
\end{proof}

For an isolated forced checkpoint on a pure circuit-under-test prefix
$\tau_{MR}$, exact replay reconstructs its marginal $\tau_M$. With the
measurement records discarded, each measurement--reset--replay step acts as
$\mathcal R_\tau(X)=\tau_M\otimes\operatorname{Tr}_M X$. Hence
$\mathcal R_\tau^2=\mathcal R_\tau$: resetting the reused ancillas before
the second replay gives the same final program state $\tau_M\otimes\tau_R$,
not a second independent fidelity loss. Its fidelity with the uninstrumented
pure prefix is $\operatorname{Tr}(\tau_M^3)$; for one target,
$1-F=3\det\tau_q$. For a general incoming faulty state $X$, the output is
instead $\tau_M\otimes\operatorname{Tr}_M X$, which need not equal $X$.

The exact premise of Theorem~\ref{thm:nodist} is
\[
\det\rho_q(i)=0.
\]
Across the 310 evaluation circuits, Algorithm~\ref{alg:screen} tests 56,386
gate--qubit locations and identifies 22,332 numerical candidates satisfying
\[
a_2(i,q)\leq10^{-12}.
\]
A separately implemented singular-value-decomposition calculation reproduces
the same selected set, and the largest selected value is
\[
5.57\times10^{-14}.
\]
An independent determinant calculation gives a maximum absolute residual of
\[
4.61\times10^{-15}.
\]

The determinant residual is used only as a numerical cross-check. Near a
rank-one state, its calculation subtracts nearly equal floating-point
quantities. It is therefore neither the selection threshold nor an estimate
of $a_2(i,q)^2$.

Theorem~\ref{thm:nodist},
Proposition~\ref{prop:grouped-monitor}, and Corollaries~\ref{cor:compose} and~\ref{cor:two-read-preservation} guarantee exact full-state preservation when the
exact separability condition is satisfied at every monitored location. The
benchmark calculation shows that the selected locations are numerically
consistent with this condition; it does not prove that their floating-point
values are mathematically equal to zero.

Proposition~\ref{prop:converse} establishes necessity for one forced monitor.
It does not establish a general necessity result for an arbitrary sequence of
monitors, and it does not claim that every chosen final measurement
distribution must change.

The forced-monitor evaluation covers all 56,386 gate--qubit locations shown
in Fig.~\ref{fig:ent-dist}. For one forced monitor under the stated ideal
conditions, the full-state disturbance $1-F$ is zero exactly when
\[
\det\rho_q(i)=0,
\]
and is positive whenever
\[
\det\rho_q(i)>0.
\]
The following subsection derives the exact relationship between this
determinant and the resulting full-state disturbance.

\subsection{An Exact Disturbance Law for Forced Monitors}
\label{app:tightlaw}

We now derive the full-state disturbance caused by a single
measure--reset--replay monitor, without requiring the monitored qubit to be
separable. We call this a \emph{forced monitor}. The circuit and replay
assumptions are those of Appendix~\ref{app:no-disturbance-proof}.
RQ1 evaluates the resulting identity at all 56,386 gate--qubit locations
in the 310 circuits (Fig.~\ref{fig:ent-dist}).

\paragraph{Setting}

As in Section~\ref{sec:no-disturbance}, consider a noise-free, closed circuit
\[
C=U_{m-1}\cdots U_0
\]
consisting of $m$ one- and two-qubit unitary gates and acting on $n$ qubits
initialized in $\zeron$. Every circuit prefix is therefore in a pure state.
Immediately after gate $i$,
\[
\ket{\psi_i}
=
U_i\cdots U_0\zeron,
\qquad
i\in\{0,\ldots,m-1\}.
\]

Fix a monitored gate--qubit location $(i,q)$, where gate $i$ acts on qubit
$q$. Let
\[
R=[n]\setminus\{q\},
\qquad
\HS=\HS_q\otimes\HS_R,
\]
and let
\[
\rho_q(i)
=
\Tr_R\ketbra{\psi_i}{\psi_i}
\]
be the reduced state of $q$.

Write the Schmidt decomposition across the split $q\mid R$ as
\[
\ket{\psi_i}
=
\sqrt{\lambda_1}\,
\ket{u_1}_q\ket{v_1}_R
+
\sqrt{\lambda_2}\,
\ket{u_2}_q\ket{v_2}_R,
\qquad
\lambda_1+\lambda_2=1,
\]
where $\{\ket{u_k}\}\subset\HS_q$ and
$\{\ket{v_k}\}\subset\HS_R$ are orthonormal sets. The values
$\lambda_k$ are the Schmidt probabilities and the eigenvalues of
$\rho_q(i)$, while $\sqrt{\lambda_k}$ are the Schmidt amplitudes. We use the
convention $\lambda_2=0$ when the Schmidt rank is one. Therefore,
\[
\det\rho_q(i)=\lambda_1\lambda_2.
\]

Exactly one monitor is inserted at $(i,q)$. The instrumented circuit performs
the following steps:

\begin{enumerate}
    \item It measures $q$ in an arbitrary orthonormal basis
    $\{\ket{\beta_0},\ket{\beta_1}\}$ of $\HS_q$.
    \textsc{QMon} uses the computational basis for the first read and
    the adapted basis for the second. The outcome is recorded but
    is not used to choose or modify later operations.

    \item It resets $q$ to $\ket{0}$.

    \item It performs exact matched replay of the backward causal cone on
    $q$ and a fresh ancilla register $A'$, initialized in the all-zero state.
    No later program gate acts on $A'$. Before another replay uses this register, it is reset to its all-zero state.

    \item It applies the remaining original circuit
    \[
    V_{\mathrm{tail}}
    =
    U_{m-1}\cdots U_{i+1},
    \]
    with $V_{\mathrm{tail}}=I$ when $i=m-1$.
\end{enumerate}

For the state comparison, the recorded outcome and the replay ancillas are
traced out. Let $\rho_{\mathrm{out}}$ be the final state of the original
$n$ qubits in the instrumented circuit, and let
\[
\ket{\phi}
=
V_{\mathrm{tail}}\ket{\psi_i}
\]
be the pure final state of the uninstrumented circuit. Their fidelity is
\[
F
=
\bra{\phi}
\rho_{\mathrm{out}}
\ket{\phi}.
\]
Because $\ket{\phi}$ is pure, this is the squared Uhlmann fidelity used in
RQ1.

\begin{proposition}[Exact disturbance law]
\label{prop:tightlaw}

Under the ideal pure-state, single-monitor setting above, with exact replay
and with the measurement record and replay ancillas traced out,
\[
F
=
\Tr\!\left[\rho_q(i)^3\right]
=
\lambda_1^3+\lambda_2^3.
\]
Equivalently,
\begin{equation}
1-F
=
3\lambda_1\lambda_2
=
3\det\rho_q(i).
\label{eq:tightlaw}
\end{equation}

This identity does not depend on the measurement basis, the depth or gate
content of the replayed backward causal cone, or the remaining circuit
$V_{\mathrm{tail}}$.
\end{proposition}

\begin{proof}

Expand $\ket{\psi_i}$ in the measurement basis of $q$:
\[
\ket{\psi_i}
=
\sum_{b\in\{0,1\}}
\ket{\beta_b}_q\otimes\ket{c_b}_R,
\]
where
\[
\ket{c_b}
:=
\left(
\bra{\beta_b}\otimes I_R
\right)
\ket{\psi_i},
\qquad
p_b
:=
\langle c_b|c_b\rangle,
\]
and $I_R$ is the identity operator on $\HS_R$. Let
\[
B_+
:=
\{b\in\{0,1\}:p_b>0\}.
\]

For $b\in B_+$, outcome $b$ occurs with probability $p_b$. Immediately
after measurement, the normalized state is
\[
\ket{\beta_b}_q
\otimes
\frac{\ket{c_b}_R}{\sqrt{p_b}}.
\]
Reset changes this state to
\[
\ket{0}_q
\otimes
\frac{\ket{c_b}_R}{\sqrt{p_b}}.
\]

The recorded outcome does not control replay, so the same replay operation
is applied for every outcome. Let
\[
\ket{\Phi'}_{qA'}
:=
V_{\mathrm{cone}}\ket{0}_{qA'}
\]
be the pure state prepared on the monitored qubit and replay ancillas. After
averaging over the discarded measurement outcome, the state entering the
circuit tail is
\[
\begin{aligned}
&\sum_{b\in B_+}
p_b\,
\ketbra{\Phi'}{\Phi'}_{qA'}
\otimes
\frac{\ketbra{c_b}{c_b}_R}{p_b}\\
&\qquad =
\ketbra{\Phi'}{\Phi'}_{qA'}
\otimes
\rho_R,
\end{aligned}
\]
where
\[
\rho_R
:=
\sum_{b\in B_+}
\ketbra{c_b}{c_b}
=
\Tr_q\ketbra{\psi_i}{\psi_i}.
\]

The last equality follows from the completeness relation
\[
\sum_b\ketbra{\beta_b}{\beta_b}=I_q,
\]
and therefore holds for any orthonormal measurement basis. Equivalently,
after the outcome is discarded, a local measurement and reset on $q$ do not
change the reduced state of $R$.

By Proposition~\ref{prop:cone}, exact replay prepares a pure joint state on
$qA'$ whose reduced state on $q$ is
\[
\Tr_{A'}
\ketbra{\Phi'}{\Phi'}
=
\rho_q(i).
\]
Since the circuit tail acts only on the original qubits and acts as the
identity on $A'$, tracing out $A'$ may be performed before the tail. The final
state on the original qubits is therefore
\[
\rho_{\mathrm{out}}
=
V_{\mathrm{tail}}
\left(
\rho_q(i)\otimes\rho_R
\right)
V_{\mathrm{tail}}^\dagger.
\]

Thus, after the record and replay ancillas are removed, the forced monitor
replaces the original joint state at $(i,q)$ by the product of its two
reduced states. Using
\[
\ket{\phi}
=
V_{\mathrm{tail}}\ket{\psi_i},
\]
we obtain
\[
\begin{aligned}
F
&=
\bra{\phi}
\rho_{\mathrm{out}}
\ket{\phi}\\
&=
\bra{\psi_i}
\left(
\rho_q(i)\otimes\rho_R
\right)
\ket{\psi_i}.
\end{aligned}
\]

In the Schmidt bases,
\[
\rho_q(i)
=
\sum_k
\lambda_k\ketbra{u_k}{u_k},
\qquad
\rho_R
=
\sum_k
\lambda_k\ketbra{v_k}{v_k}.
\]
Hence,
\[
\bra{u_j}\rho_q(i)\ket{u_k}
=
\lambda_k\delta_{jk},
\qquad
\bra{v_j}\rho_R\ket{v_k}
=
\lambda_k\delta_{jk}.
\]

Substituting the Schmidt decomposition of $\ket{\psi_i}$ gives
\[
\begin{aligned}
F
&=
\sum_{j,k}
\sqrt{\lambda_j\lambda_k}\,
\bra{u_j}\rho_q(i)\ket{u_k}
\bra{v_j}\rho_R\ket{v_k}\\
&=
\sum_{j,k}
\sqrt{\lambda_j\lambda_k}\,
\left(\lambda_k\delta_{jk}\right)
\left(\lambda_k\delta_{jk}\right)\\
&=
\lambda_1^3+\lambda_2^3\\
&=
\Tr\!\left[\rho_q(i)^3\right].
\end{aligned}
\]

Finally, because $\lambda_1+\lambda_2=1$,
\[
\begin{aligned}
1-F
&=
1-\lambda_1^3-\lambda_2^3\\
&=
(\lambda_1+\lambda_2)^3
-\lambda_1^3-\lambda_2^3\\
&=
3\lambda_1\lambda_2
(\lambda_1+\lambda_2)\\
&=
3\lambda_1\lambda_2\\
&=
3\det\rho_q(i).
\end{aligned}
\]
\end{proof}

Equation~\eqref{eq:tightlaw} has three direct consequences.

First, it gives the exact single-monitor boundary stated in
Theorem~\ref{thm:nodist}:
\[
F=1
\quad\Longleftrightarrow\quad
\det\rho_q(i)=0.
\]
Thus, under the stated ideal conditions, separability is necessary and
sufficient for one forced monitor to preserve the full circuit state exactly.
Algorithm~\ref{alg:screen} instead applies the finite-precision test
\[
a_2(i,q)\leq\varepsilon_{\mathrm{sv}}.
\]
Passing this numerical test is evidence that the exact condition holds, but
does not by itself prove zero disturbance. Moreover, separability is
sufficient but not necessary for preserving one particular output
distribution, because a fixed final measurement may not reveal a phase or
correlation change in the underlying state.

Second, the common unitary tail does not change the fidelity between the
instrumented and uninstrumented states. The full-state disturbance is
therefore determined by the state at the monitored location. Since
\[
0\leq\det\rho_q(i)\leq\frac14,
\]
the state infidelity satisfies
\[
0\leq1-F\leq\frac34.
\]
The upper bound is reached when
\[
\lambda_1=\lambda_2=\frac12,
\]
that is, when the monitored qubit is maximally entangled with the other
qubits across the split $q\mid R$.

Third, the result is independent of the measurement basis. The coefficient
$3$ follows from the relation $\lambda_1+\lambda_2=1$ and does not require
the measurement basis to match the Schmidt basis.

In RQ1, we compute the absolute residual between the two sides of
Eq.~\eqref{eq:tightlaw} at every gate--qubit location. At locations numerically consistent with
\[
\det\rho_q(i)=0,
\]
the computed state infidelity is also numerically consistent with zero under the same simulation setting. Across all the 56,386 evaluated locations, the maximum absolute residual $|(1-F_{i,q})-3\det\rho_q(i)|$ is
$3.17\times10^{-13}$.

\subsection{Selection and Batching Walkthroughs}
\label{app:implementation-walkthroughs}

\paragraph{\textbf{Selection walkthrough}}

We use the example circuit in
Figure~\ref{fig:An example of a Quantum Circuit} to illustrate the
complete selection pass. At the start, the selected-location list
for each of the three qubits is empty.

At index~0, the $X$ gate prepares $q_0$ in $\ket{1}$, so
$(0,q_0)$ is selected. At index~1, the CNOT control qubit is in
$\ket{1}$, and the gate maps
$
\ket{1}_{q_0}\ket{0}_{q_1}
\longmapsto
\ket{1}_{q_0}\ket{1}_{q_1}.
$
This remains a product state, so the CNOT creates no entanglement
in this execution. Both $(1,q_0)$ and $(1,q_1)$ are selected.

At index~2, the $H$ gate prepares the still-separable qubit $q_2$
in $\ket{+}=(\ket{0}+\ket{1})/\sqrt{2}$, so $(2,q_2)$ is
selected. At index~3, the SWAP gate exchanges the states of $q_1$
and $q_2$. The full state remains a product state, so both
$(3,q_1)$ and $(3,q_2)$ are selected.

The single-qubit $H$ gates at indices~4, 5, and~6 also leave
their respective qubits separable. The complete numerical-candidate
lists are therefore
$
\{
q_0:\{0,1,4\},
q_1:\{1,3,5\},
q_2:\{2,3,6\}
\}.
$
The ordinary final measurements read these qubits directly after
their respective $H$ gates. Removing
these three final-read locations leaves the lists used for
mid-circuit instrumentation:
$
\{
q_0:\{0,1\},
q_1:\{1,3\},
q_2:\{2,3\}
\}.
$

\paragraph{Batching Walkthrough}

Suppose a 10-qubit circuit is deployed with $Q_{\max}=24$. The capacity
available for replay ancillas is therefore
$
Q_{\mathrm{cap}}=24-10=14.
$
Consider four monitor groups with ancilla demands 9, 7, 5, and 3. Algorithm~\ref{alg:batch} processes them in the following order. The demand-9 group opens the first batch. The demand-7 group cannot join it
because
$
9+7>14,
$
so it opens a second batch. The demand-5 group then joins the first batch and
fills it exactly, while the demand-3 group joins the second batch. The two batch costs are therefore 14 and 10. Here, $K_{\mathrm{FFD}}=2$ and $N_g=4$, so the conditional ILP refinement
tests whether all four groups can fit into one batch. This is impossible
because their total demand is
$
9+7+5+3=24>14.
$
A two-batch FFD plan is therefore retained. Within each batch, the groups
are returned to gate order before instrumentation. For another 14-qubit circuit under the same $Q_{\max}=24$ budget,
a group requiring 11 extra qubits would be placed in the
cannot-fit-alone set because its demand exceeds the available
capacity of 10. Additional batches
cannot reduce the number of qubits required by that individual group.

\subsection{Implementation Complexity}
\label{app:implementation-details}

This subsection summarizes the worst-case time and space costs of
monitoring-location selection, backward causal-cone tracing, and batch
construction. These bounds describe the current implementation and do not
include the cost of executing the resulting instrumented circuits.

\paragraph{Monitoring-location selection complexity}

Consider a circuit with $m$ one- and two-qubit gates acting on $n$ qubits.
Updating the statevector for one gate takes $O(2^n)$ time. After each gate,
Algorithm~\ref{alg:screen} computes the reduced state of at most two acted-on
qubits, and each one-qubit reduction also takes $O(2^n)$ time. The complete
selection pass therefore takes
$
O(m2^n)
$
time.

The statevector requires $O(2^n)$ space. The lists of numerical candidates
contain at most two gate--qubit locations per gate and therefore require
$O(m)$ additional space. The total space requirement is
$
O(2^n+m).
$
The implementation never constructs the full
$2^n\times2^n$ density matrix.

Appendix~\ref{sec:scalability} separately evaluates candidate selection using
a matrix product state (MPS) representation, whose cost depends on the
observed bond dimension and therefore on the circuit's entanglement
structure.

\paragraph{Backward causal-cone tracing complexity}

For one monitor group, Algorithm~\ref{alg:dfs} records visited gate--wire
pairs so that the same pair is not recursively expanded more than once. This
prevents repeated expansion when different backward branches meet again.

However, in the current implementation, each recursive call may still scan
backward through up to $m$ gate positions. Since the number of distinct
gate--wire pairs is $O(m)$ for a circuit of one- and two-qubit gates, the
conservative worst-case running time is
$
O(m^2)
$
per monitor group. Removing duplicate gates and sorting the final cone takes
$O(m\log m)$ time, which is smaller than this worst-case bound.

The visited set and the collected gate list each contain at most $O(m)$
entries, so the space requirement is
$
O(m)
$
per monitor group. This step uses only the circuit structure and does not
simulate a quantum state.

\paragraph{Batch-construction complexity}

Let $N_g$ be the number of monitor groups that can each fit within the qubit
budget when deployed alone, and let $K_{\mathrm{FFD}}$ be the number of
batches returned by first-fit decreasing (FFD). Assume that the ancilla
demand of each group has already been computed.

FFD first sorts the $N_g$ groups by decreasing ancilla demand, which takes
$
O(N_g\log N_g)
$
time. It then considers each group and checks batches in order until it finds
one with sufficient remaining capacity. Each capacity check takes
$O(1)$ time using the stored cumulative batch demand. In the worst case, it checks all
$K_{\mathrm{FFD}}$ batches for each group. The total running time is
therefore
$
O\!\left(
N_g\log N_g+N_gK_{\mathrm{FFD}}
\right).
$
Because $K_{\mathrm{FFD}}\leq N_g$, this is at most $O(N_g^2)$. The group list and batch assignments require
$
O(N_g+K_{\mathrm{FFD}})
$
space.

\paragraph{Conditional ILP refinement}

All reported deployment plans first use FFD. When
$
K_{\mathrm{FFD}}>1
\text{ and }
N_g\leq120,
$
the implementation performs one bounded integer linear programming (ILP)
feasibility check. The check asks whether all individually feasible monitor
groups can be packed into
$
K_{\mathrm{FFD}}-1
$
batches.

An assignment replaces the FFD plan only when PuLP reports \texttt{Optimal}
for this feasibility model. Otherwise, including an unresolved timeout,
the FFD plan is retained. We do not give an asymptotic
time bound for this solver step; instead, its runtime is limited explicitly.

The analyses reported in Table~\ref{tab:rq2-regression},
Table~\ref{tab:qmax-sweep}, and the RQ3 deployment plans use PuLP with Gurobi
and a 600-second limit for each triggered ILP check. The 144-circuit
large-circuit planning sweep in Appendix~\ref{sec:scalability} uses the same
conditional FFD--ILP procedure with a 120-second limit.

The refinement checks only whether one fewer batch than the FFD plan is
possible. If such a plan is found, the implementation does not continue
testing still smaller batch counts. Therefore, the reported batch counts are
valid plans found by the stated procedure, not general claims of global
optimality.

\section{RQ3 Protocol, Baseline Details, and Detailed Results}
\label{app:rq3details}

This appendix provides the implementation details and detailed results for the RQ3 evaluation in
Section~\ref{sec:rq3}. It describes the published mechanisms of the four
baseline methods, the adaptations and decision rules used in our evaluation,
the complete mutant-evaluation protocol, and the applicability and flagging
rules of the MQT debugger, followed by the matched-batch comparison
and detailed miss analysis.

\subsection{Baseline Mechanisms and Decision Rules}
\label{app:rq3details:baselines}

\paragraph{Checkpoint sets and execution model}

The three finite-shot assertion methods in the primary comparison declare the complete set of numerical
candidates in the original circuit as their checkpoint set. These
checkpoints are considered in increasing gate order. Each checkpoint is checked using a separate circuit containing the
circuit-under-test prefix through that checkpoint and the corresponding
assertion operation. During mutant evaluation, this prefix comes from the
mutant, while the expected distribution or state and the assertion
parameters are derived from the original circuit.

The MQT debugger starts from the same set of numerical candidates, but it
also applies the method-specific applicability check described below. A
checkpoint is used for mutant evaluation only if the corresponding MQT
assertion can first be evaluated and passes on the original circuit.

During actual debugging, the location of a fault is unknown. A complete scan
of $K$ assertion checkpoints in these per-checkpoint configurations requires $K$ separate circuit
runs.
The planned full-scan counts for these per-checkpoint configurations
use this complete declared checkpoint set.

\paragraph{Evaluation-only no-effect rules}

The noise-free evaluator avoids simulating checkpoint states that the
injected gate cannot affect. Appendix~\ref{app:rq3details:protocol}
defines these method-specific shortcuts and explains why they do not
reduce the planned full-scan execution counts.

\paragraph{Shot budget}

Each explicitly executed checkpoint circuit of the three finite-shot
assertion methods uses
$
S=8{,}192
$
noise-free shots. The MQT debugger is deterministic and simulation-only and
does not use shots.

The four baselines are implemented as follows.

\begin{itemize}

    \item \textbf{Statistical assertions}
    ~\cite{10.1145/3307650.3322213}:

    The circuit prefix through the checkpoint is executed with a
    computational-basis measurement of the checked qubit. The observed 0/1
    counts are compared with the expected one-qubit distribution derived from
    the original circuit.

    We use Pearson's $\chi^2$ goodness-of-fit test when both expected outcome
    counts are at least five. Otherwise, we use a two-sided exact binomial
    test. When the original circuit predicts one outcome with probability
    one, observing the other outcome flags the checkpoint.

    We do not include the method's separate entanglement assertion because the checkpoints considered in this comparison are selected at
    locations where the checked qubit is numerically consistent with being
    separable from the remaining register.

    \item \textbf{Projection assertions (Proq)}
    ~\cite{10.1145/3428218}:

    The method tests the rank-one projector
    \[
    P
    =
    \ket{\psi_{\mathrm{exp}}}
    \bra{\psi_{\mathrm{exp}}},
    \]
    where $\ket{\psi_{\mathrm{exp}}}$ is the expected one-qubit state derived
    from the original circuit. The corresponding pass and fail projectors are $P$ and $I-P$,
    respectively, where $I$ is the identity operator on the monitored
    qubit. These projectors select the expected-state direction and
    its orthogonal complement, respectively. The two subspaces are
    orthogonal and together span the entire one-qubit state space.
    We use the authors' implementation and execute the resulting checkpoint
    circuit in noise-free simulation. Any projection failure among the
    $S$ shots flags the checkpoint. For the exact expected state,
    \[
    \operatorname{Tr}(P\rho)=1,
    \]
    so a correct noise-free execution has zero theoretical failure
    probability.

    \item \textbf{Dynamic ancilla assertions}
    ~\cite{10.1145/3373376.3378488,li2022exploiting}:

    The published ancilla-based assertion circuit is appended to the circuit
    prefix through the checkpoint. For a deterministic expected state, we use
    the published basis-state assertion construction. Otherwise, we use the
    construction for an arbitrary one-qubit superposition.

    The assertion couples the checked qubit to an extra qubit and then
    measures the extra qubit. Any nonzero extra-qubit outcome flags the
    checkpoint. The per-shot failure probability is
    \[
    1-
    \bra{\psi_{\mathrm{exp}}}
    \rho
    \ket{\psi_{\mathrm{exp}}}.
    \]
    This is the same one-qubit state-overlap signal tested by the projection
    assertion, although the two methods obtain the signal through different
    circuits.

   \item \textbf{MQT debugger}
      ~\cite{rovara2024debugger}:
      At each checkpoint, MQT attempts to extract the checked qubit as a
      normalized pure one-qubit state from the circuit prefix and compares it
      with the normalized expected state derived from the original circuit.
      Let $\mathbf{a}$ and $\mathbf{b}$ be the amplitude vectors of the
      extracted and expected states, respectively. MQT computes
      \[
      s(\mathbf{a},\mathbf{b})
      =
      \left|
      \sum_j \overline{b_j}a_j
      \right|
      =
      \left|
      \left\langle
      \psi_{\mathrm{exp}}
      \middle|
      \psi_{\mathrm{act}}
      \right\rangle
      \right|.
      \]
      Thus, $s$ is the absolute state-vector overlap, not its square. The
      equality assertion passes when $s\geq0.99$ and returns an evaluated
      state mismatch when $s<0.99$.

      MQT can perform this comparison only when the checked qubit can be
      represented as a pure one-qubit substate. If the checked qubit is
      entangled with the remaining register, its reduced state is mixed and
      the backend rejects the equality assertion instead of returning a
      similarity value. We record this result as an \emph{unevaluable}
      outcome: the backend cannot extract the pure one-qubit state needed
      for the comparison. This is an evaluation category used in this
      study, not a separate official MQT status.

      Before a checkpoint is used to evaluate a mutant, the same assertion
      must first be evaluable and must pass on the original circuit. Only
      checkpoints satisfying both conditions are considered
      \emph{applicable}. A checkpoint that fails or is unevaluable on the
      original circuit is excluded from mutant evaluation.

      On a mutant, either of the following outcomes counts as a combined
      flag:

      \begin{enumerate}
          \item the equality assertion is evaluated and
          $s<0.99$; or
          \item the result is recorded as an unevaluable outcome.
      \end{enumerate}

      Each original-circuit applicability invocation and each mutant
      assertion invocation has a hard timeout of 120 seconds. For each
      mutant, the complete MQT checkpoint scan is further limited to
      600 seconds of cumulative MQT invocation time and at most ten timed-out
      invocations. Reaching either limit produces a
      \emph{budget-exhausted} no-decision outcome. Other unsupported cases
      and tool errors are recorded as errors and also produce no decision.
      Timeout, error, and budget-exhausted outcomes do not count as flags. Because MQT directly accesses simulator-visible quantum states, it is
      simulation-only.
\end{itemize}

\subsection{Full Detection Protocol}
\label{app:rq3details:protocol}

\paragraph{Mutation generation}

The RQ3 subset contains 238 circuits with a \textsc{QMon} deployment plan
and at least three eligible one- or two-qubit unitary instructions. In this
experiment, a \emph{mutation location} is a gate instruction identified by
its program-order index. It is therefore different from a gate--qubit
location, which pairs a gate with one of the qubits on which it acts.

For each circuit and each of five seeds, we sample three distinct eligible
gate instructions without replacement. We then construct three separate
single-fault mutants, with one sampled gate replaced in each mutant. All
methods are evaluated on the same mutants.

The replacement has the same number of operands as the original gate. A
one-qubit gate is replaced by a gate drawn from
$
\{R_x,R_y,X,Y,H\},
$
and a two-qubit gate is replaced by a gate drawn from
$
\{CP,R_{ZX},R_{ZZ},CX,CZ\}.
$
Here, $CP$, $CX$, and $CZ$ denote the controlled-phase, controlled-NOT, and
controlled-Z gates, while $R_{ZX}$ and $R_{ZZ}$ denote two-qubit $ZX$ and
$ZZ$ rotations. Rotation angles are drawn uniformly from $[0,2\pi]$. If the
sampled gate and its parameters are identical to the original operation, we
sample again. This procedure produces
$
238\times5\times3=3{,}570
$
single-fault mutation attempts. The arity-preserving gate-replacement model
follows established practice in quantum mutation analysis
~\cite{9678792,9678563}.

\paragraph{Output-equivalent mutants}

We classify a mutant as \emph{output-equivalent} when the $L_1$ distance
between its final measured-output distribution and that of the original
circuit is below
$
10^{-9}.
$
Both distributions are computed from exact statevector probabilities and are
restricted to the qubits that are physically measured at the end of the
circuit. No shot sampling is used for this classification.

Output equivalence means only that the two circuits have the same final
measured-output distribution under this numerical criterion. It does not
mean that their complete quantum states, intermediate states, or quantum
processes are equal.

\paragraph{Original-circuit plan and checkpoint sets}

Each mutant is evaluated using a monitoring plan derived only from the original
circuit. For \textsc{QMon}, the following quantities remain fixed when the
mutant is executed:

\begin{itemize}
    \item the selected and deployed checkpoint locations;
    \item the expected one-qubit states, computational-basis probabilities, and basis rotations;
    \item the backward causal cones;
    \item the replay instruction indices and operand roles;
    \item the batch assignments.
\end{itemize}

The monitoring plan is not recomputed from the mutant. In particular,
causal-cone replay uses the gates and parameters from the circuit under test
at the fixed cone instruction indices. Each replacement preserves the original instruction's qubit operands,
so the precomputed mapping from those qubits to replay wires remains valid. This choice does not imply
that instrumenting a faulty execution leaves its state unchanged.

The \textsc{QMon} checkpoint set consists of the mid-circuit checkpoints in
its $Q_{\max}=24$ deployment plan and eligible final reads.
Each mid-circuit target is read first
in the computational basis and then, after reset and CUT replay, through its
basis rotation. The targets and the group's replay ancillas are
reset before a second CUT replay resumes the program. Final computational
reads are collected once, from the first batch.

The three assertion methods in the first comparison declare the complete set of numerical candidates
in the original circuit as their checkpoint set. MQT starts from the same
set but also applies its method-specific applicability rule.

The additional matched-batch comparison restricts the assertion methods to
the same deployed locations and batch membership as \textsc{QMon}.

\paragraph{Expected-state construction}
For each mid-circuit checkpoint, we compute the original circuit's
one-qubit reduced state $\rho_q$ and require
$\operatorname{Tr}(\rho_q^2)\geq1-10^{-9}$. The expected pure state is
the normalized eigenvector of $\rho_q$ with the largest eigenvalue.
Writing this vector as $(a,b)^{\mathsf T}$, we construct
$U_q=\left(\begin{smallmatrix}a&-b^*\\b&a^*\end{smallmatrix}\right)$
and apply $U_q^\dagger$ for the second read. We check that the transformed
expected state gives outcome 0 with probability at least $1-10^{-9}$;
the unitarity check uses absolute tolerance $10^{-9}$ and relative
tolerance $10^{-5}$. These checks do not replace the candidate-selection
tolerance or the statistical decision threshold.

\paragraph{Simulation backend}

RQ3 uses Qiskit Aer to execute the instrumented \textsc{QMon} circuits
and the three finite-shot assertion baselines, as described in
Section~\ref{sec:prep:backend}. All 2,860 mutants in the single-fault
comparison have a decision.

\paragraph{Checkpoint decision rules}

The finite-shot methods use
$
S=8{,}192
$
shots per explicitly executed run.

A false alarm occurs when a method flags a correct circuit.
For \textsc{QMon} and the statistical assertion method, we set
$\alpha=0.01$ as the nominal circuit-level false-alarm probability:
the target upper bound on the probability of at least one false flag
during one complete evaluation of that method's checkpoint set.
This budget accounts for random fluctuations in finite-shot measurements;
it is not an allowed fraction of incorrect measurement outcomes.
This budget applies to the complete checkpoint set
and is not restarted for each checkpoint, batch, or separate execution.

Let $K_{\mathrm{mid}}$ count deployed mid-circuit checkpoints and
$K_{\mathrm{final}}$ count eligible final readouts. The complete
\textsc{QMon} tested-read family has size
$K_{\mathrm{QMon}}=2K_{\mathrm{mid}}+K_{\mathrm{final}}$.
Let $K_{\mathrm{stat}}$ count the checkpoints declared by the statistical
assertion method. The corresponding Bonferroni thresholds
are
\[
\frac{\alpha}{K_{\mathrm{QMon}}}
\qquad\text{and}\qquad
\frac{\alpha}{K_{\mathrm{stat}}}.
\]
For valid checkpoint $p$-values, this correction bounds the probability
of at least one false rejection across the complete checkpoint set by
$\alpha$, without requiring independence among checkpoints.
False-alarm control for the $\chi^2$ branch is nominal, because its
reference distribution is an asymptotic approximation.

For either method, let $\pi_j$ denote the raw, unadjusted $p$-value
returned by its statistical test at checkpoint $j$. This is distinct
from the reference probability $p_j$ of measuring 1.
For the first read at a mid-circuit checkpoint, $p_j=\langle1|\rho_q|1\rangle$.

For a nondeterministic computational-basis reference distribution, \textsc{QMon}
uses a two-sided exact binomial test whose null hypothesis is that
the probability of measuring 1 equals $p_j$. Checkpoint $j$ is
flagged when
\[
\pi_j < \frac{\alpha}{K_{\mathrm{QMon}}}.
\]

For a nondeterministic reference distribution, the statistical assertion method uses
Pearson's $\chi^2$ goodness-of-fit test when both expected outcome
counts are at least five; otherwise, it uses a two-sided exact
binomial test. In either statistical-test branch, checkpoint $j$
is flagged when
\[
\pi_j < \frac{\alpha}{K_{\mathrm{stat}}}.
\]
Here, $\alpha/K_{\mathrm{QMon}}$ and $\alpha/K_{\mathrm{stat}}$ are the Bonferroni-adjusted
per-read significance thresholds against which the raw
$p$-values are compared.
The method flags the circuit if at least one checkpoint is flagged.

Both methods handle reference distributions treated as deterministic
using a direct rule rather than the $p$-value comparison above.
The implementation treats $p_j<10^{-12}$ as an expected outcome of 0
and $p_j>1-10^{-12}$ as an expected outcome of 1.
If the expected outcome is 0, any observed 1 flags the checkpoint;
if it is 1, any observed 0 flags the checkpoint.
Thus, a single contrary outcome among the $S$ shots is sufficient.
For exactly deterministic reference distributions, the contrary
outcome has zero probability under the null hypothesis; the
$10^{-12}$ cutoff is a numerical convention.

A second read has expected probability zero for outcome 1 and
flags on any 1 in the noise-free protocol. Both reads at a mid-circuit
location share the same $8{,}192$ shots; the second read adds one test
per mid-circuit checkpoint without increasing the number of batches or shots.
Eligible final readouts use the computational-basis rule once.

Projection and dynamic ancilla assertions use their published any-failure
rules and do not apply a Bonferroni-adjusted $p$-value. MQT uses the
deterministic mismatch and unevaluable rules described in
Appendix~\ref{app:rq3details:baselines}.

The values $K_{\mathrm{QMon}}$ and $K_{\mathrm{stat}}$ are fixed before execution and count the
complete tested-read families, not the numbers of batches or shots. They are not reduced
because checkpoints are placed in batches, handled by an evaluation-only
no-effect rule, or not reached after early stopping.

\paragraph{Evaluation-only no-effect rules}

The injected fault location is known to the evaluator because the experiment
created the mutant. It is not provided to any detector or localizer.

In the first comparison, when the mutated gate lies outside a checkpoint's backward causal cone, the
mutation cannot change the state at that checkpoint in the ideal noise-free
model. The evaluation handles this case as follows:

\begin{itemize}
    \item \textsc{QMon} executes its complete fixed set of batch circuits.
    It therefore retains the possibility of a finite-shot false flag at a
    checkpoint that the mutation cannot affect.

    \item The statistical assertion method samples $S$ outcomes from the
    unchanged original distribution and applies its usual statistical test.
    This preserves its finite-shot false-flag probability.

    \item Projection and dynamic ancilla assertions return a deterministic
    pass because their theoretical failure probability is zero when the
    asserted state is unchanged.

    \item MQT returns a pass without invoking the debugger on that unchanged
    checkpoint state.
\end{itemize}

These method-specific rules avoid unnecessary simulation while preserving
the implemented decisions under the noise-free model. Because they use the
known injected fault location, they are evaluation-time shortcuts rather than
practical debugging savings. They do not reduce the planned full-scan counts
for these per-checkpoint configurations.

\paragraph{Detection and localization}

All checkpoints are ordered by gate index. For \textsc{QMon}, the simulator
first generates the data for all batch circuits. The method then identifies
the earliest checkpoint that produces a flag. If multiple checkpoints after the same
gate flag, the one with the smaller qubit index is chosen. At the same location,
the first read precedes the second read. Data from later
checkpoints may therefore already have been generated even though they are
not needed to identify the earliest flag. The baseline scans stop at the first flag. For the three assertion methods,
this also stops further checkpoint-circuit execution. For MQT, no later
checkpoint is evaluated after the first flag. The matched-batch
assertion configurations instead execute all planned batches without early
stopping, then identify the earliest flag from the recorded outcomes.

For a detected mutant, the earliest flagged checkpoint defines the suspect
region as its backward causal cone in the original circuit. The injected
gate location is used only after the method has returned its result, to
evaluate whether the reported region contains the known fault. It is not
provided to the detector or localizer.

Localization precision is defined as
$
\frac{1}{|\mathrm{Cone}|}
$
when the injected gate belongs to the backward causal cone of the earliest
flagged checkpoint, and as zero otherwise. Here,
$|\mathrm{Cone}|$ is the number of unique source-gate indices in the cone,
including the gate at the flagged checkpoint.

A correct suspect region receives a higher precision value when it contains
fewer gates. Thus, the metric rewards a causal cone that contains the known
fault while narrowing the suspect region more precisely.

A finite-shot flag at a checkpoint that the mutation cannot affect still
counts as a detection under the method's decision rule, but its localization
precision is zero. Localization precision is averaged only over the mutants
detected by that method. The methods therefore use different localization
denominators, and the reported values are not a ranking on one common set of
detected mutants.

The single-fault comparison contains 2,860 non-output-equivalent
mutants. Every reported method returns a decision for every one of these
records. Detection is therefore evaluated on the same 2,860 mutants for all
methods, and no record is removed because of incomplete execution.

\paragraph{Candidate and deployment coverage}

We call a mutant \emph{candidate-covered} when its changed gate lies in the
backward causal cone of at least one numerical \textsc{QMon} candidate,
whether or not that candidate is deployed.

We call a mutant \emph{deployment-covered} when its changed gate lies in the
backward causal cone of at least one checkpoint included in the
$Q_{\max}=24$ deployment plan.

A mutant that is candidate-covered but not deployment-covered is called a
\emph{deployment gap}.

Among the 2,860 non-output-equivalent mutants, 2,217 are candidate-covered.
This population contains 2,206 deployment-covered mutants and 11 deployment
gaps.

Candidate-covered detection restricts the mutant population to the 2,217
candidate-covered mutants, while every method retains its own implemented
checkpoint set. It therefore measures performance on faults that can affect
at least one location selected by \textsc{QMon}'s numerical test. Candidate
coverage alone does not imply matched checkpoint sets.

\paragraph{False alarms}

The false-alarm experiment performs one correct-circuit trial for each of the
238 circuits and each finite-shot method, using a fixed shot seed. For
\textsc{QMon}, the trial applies the complete deployment, including all
batches. For each assertion method, it applies the
method's complete checkpoint configuration.

The false-alarm comparison completes for all 238 circuits. MQT's required
correct-circuit pass is used to determine whether each assertion is
applicable to mutant evaluation. It is therefore not treated as a separate
MQT false-alarm experiment, and no MQT false-alarm rate is reported.

\paragraph{Matched-checkpoint execution}

The supplementary batched assertion configurations inherit the exact
deployed checkpoint locations, eligible final reads, and batch membership.
Each batch is executed for $8{,}192$ shots. Within each shot, all
checkpoints are measured sequentially during the same circuit execution. Batched statistical assertions measure the target in the
computational basis and continue without reset or replay. Projection
assertions insert the compiled basis rotation, measurement, and inverse
rotation on the data qubit. Dynamic assertions reuse one ancilla, measured
and reset after each check. Projection and dynamic assertions retain the
noise-free any-failure rule. The statistical family contains each deployed
location once. Runs and total shots are therefore equal by construction;
the inherited batches are not claimed to minimize assertion run counts.

All methods use the same 2,860 mutant identities and 238 correct circuits.
The QMon row comes from its complete two-read execution records; the assertion
rows come from the matched-batch runs. Equal locations and budgets do not
imply identical disturbance or gate exposure. In particular, computational
measurement without restoration can disturb a correct superposition, so
the batched statistical detection rate must be read together with its
false-alarm rate.

\subsection{MQT Debugger Applicability and Flag Composition}
\label{app:rq3details:mqt}

Using the applicability and flagging rules in
Appendix~\ref{app:rq3details:baselines}, MQT produces 1,688 combined flags:
534 evaluated state mismatches and 1,154 unevaluable outcomes.
Using the same unweighted five-seed reporting
convention as Table~\ref{tab:rq3-results}, the corresponding rates are
$18.7\pm6.8\%$, $40.4\pm6.6\%$, and $59.2\pm6.2\%$ for evaluated state
mismatches, unevaluable outcomes, and their combination,
respectively.

MQT returns a decision for all 2,860 non-output-equivalent mutants in the
single-fault evaluation. No timeout, tool-error, or budget-exhausted
outcome occurs in this population.

Because the decision-diagram backend directly accesses simulator-visible
quantum states, MQT's assertion checks are simulation-only and are not
treated as physical quantum-circuit executions. MQT is therefore omitted
from the planned circuit-run comparison in
Table~\ref{tab:rq3-fullscan-counts}.

\subsection{Detailed Miss Analysis}
\label{app:rq3-misses}

On the 2,860 non-output-equivalent mutants, \textsc{QMon} detects 2,031 and
misses 829. The pooled miss rate is 29.0\%, with a 95\% Wilson confidence
interval of $27.4\%$--$30.7\%$. This interval treats mutant outcomes as independent Bernoulli trials and does not account for clustering within circuits. The five-seed mean miss rate is
$28.9\pm4.5\%$.

Table~\ref{tab:rq3-misses} lists the circuit families and gate replacements
that contribute the largest numbers of misses. It also gives complete
partitions by the number of gate operands and by candidate or deployment
coverage. The circuit-family and replacement-operation rows are descriptive;
they are not statistical rankings.
The denominators count the evaluated non-output-equivalent mutants
in each stratum.
``Seed rate'' is the mean $\pm$ sample standard
deviation of the defined per-seed miss rates. $\dagger$ marks a row with
fewer than 20 evaluated non-output-equivalent mutants pooled across seeds,
or representation in fewer than five seeds.

\begin{table*}[!t]
\centering
\caption{Misses by circuit family, gate replacement, arity, and monitoring coverage.}
\label{tab:rq3-misses}
\scriptsize
\setlength{\tabcolsep}{4.0pt}
\renewcommand{\arraystretch}{1.02}
\begin{tabular}{@{}llrrrr@{}}
\toprule
Dimension & Stratum & \shortstack{Misses /\\evaluated mutants} & Pooled rate & Seed rate
& Share of misses \\
\midrule

Circuit family
 & QNN
 & 100/120 & 83.3\% & $83.3\pm7.8\%$ & 12.1\% \\

 & Portfolio QAOA
 & 82/120 & 68.3\% & $68.3\pm16.8\%$ & 9.9\% \\

 & Portfolio VQE
 & 79/120 & 65.8\% & $65.8\pm19.2\%$ & 9.5\% \\

 & Real-amplitude random
 & 79/120 & 65.8\% & $65.8\pm20.1\%$ & 9.5\% \\

 & SU(2) random
 & 78/120 & 65.0\% & $65.0\pm21.4\%$ & 9.4\% \\

 & Two-local random
 & 78/120 & 65.0\% & $65.0\pm19.7\%$ & 9.4\% \\
\midrule

Mutation operation
 & $\mathrm{CX}\!\rightarrow\!\mathrm{R_{ZX}}$
 & 133/237 & 56.1\% & $59.8\pm12.8\%^{\dagger}$ & 16.0\% \\

 & $\mathrm{CX}\!\rightarrow\!\mathrm{CP}$
 & 119/144 & 82.6\% & $89.0\pm11.2\%$ & 14.4\% \\

 & $\mathrm{CX}\!\rightarrow\!\mathrm{CZ}$
 & 104/124 & 83.9\% & $82.8\pm14.4\%$ & 12.5\% \\

 & $\mathrm{CX}\!\rightarrow\!\mathrm{R_{ZZ}}$
 & 70/108 & 64.8\% & $60.6\pm12.8\%^{\dagger}$ & 8.4\% \\
\midrule

Original gate arity
 & Two-qubit
 & 693/1,250 & 55.4\% & $56.9\pm8.1\%$ & 83.6\% \\

 & One-qubit
 & 136/1,610 & 8.4\% & $8.6\pm3.2\%$ & 16.4\% \\
\midrule

Monitoring coverage
 & Not candidate-covered
 & 638/643 & 99.2\% & $99.2\pm0.2\%$ & 77.0\% \\

 & Deployment-covered
 & 180/2,206 & 8.2\% & $8.2\pm0.8\%$ & 21.7\% \\

 & Deployment gap
 & 11/11 & 100.0\% & $100.0\pm0.0\%^{\dagger}$ & 1.3\% \\
\bottomrule
\end{tabular}
\end{table*}

The monitoring-coverage rows account for all 829 misses. Among the 643
mutants that are not candidate-covered, \textsc{QMon} misses 638. The other
five produce finite-shot false rejections at checkpoints that the mutation
cannot affect and therefore receive zero localization precision.
A further 180 misses are deployment-covered, and 11 fall into deployment
gaps. Among the 2,031
detections, 1,764 trigger both kinds of read somewhere in the circuit, 113
trigger only a second read, and 154 only a computational read.

A diagnostic of the 180 deployment-covered misses finds an affected
mid-circuit first read with an unadjusted binomial $p<0.05$ in 38
records, including 13 with $p<0.01$; none has an affected second read with
$p<0.05$. Three records have no affected mid-circuit read in this diagnostic,
which excludes final reads. These observed test outcomes illustrate the
effect of the corrected threshold but do not establish an exhaustive
mechanistic classification or prove that every miss has nonzero detection
power.

For mutations of two-qubit gates, \textsc{QMon} misses 693 of the
1,250 non-output-equivalent mutants (55.4\%). For mutations of one-qubit gates, \textsc{QMon} misses 136 of the
1,610 non-output-equivalent mutants (8.4\%).

Grouping the same mutants by circuit family reveals why misses
are concentrated in the sampled QNN and Portfolio QAOA circuits.
\textsc{QMon} selects a checkpoint only when the monitored qubit
satisfies the numerical separability condition with respect to the
rest of the original circuit. In these circuit instances, every
selected checkpoint occurs immediately after a qubit's first single-qubit preparation gate; no later acted-on location passes
this condition. All selected checkpoints are deployed, so the
limited coverage is not caused by the qubit budget. These checks
observe the initial preparations but cannot observe faults
introduced in the subsequent interaction operations. In the circuit's program order, some qubits undergo interaction
gates before other qubits receive their first single-qubit preparation gates. Consequently, a mutation may occur before the last
selected checkpoint while still lying outside its backward causal
cone: checkpoints on the affected qubits have already been read,
while a later checkpoint on another qubit cannot yet be influenced
by the mutation. This structure explains why 100 of the 100 QNN
misses and 79 of the 82 Portfolio QAOA misses are outside candidate
coverage.

\section{Extended Discussion}
\label{app:discussion}

\subsection{Scalability Protocol and Complexity}
\label{sec:scalability}

Scalability raises two separate questions. First, can \textsc{QMon} analyze
the source circuit and select numerical candidates? Second, can the resulting
monitor groups be deployed within a finite qubit budget at a reasonable
structural cost? Here, the source circuit is the uninstrumented original
circuit, and structural cost refers to qubit, batch, and operation counts
rather than hardware runtime or physical error. Completing candidate
selection does not imply that every selected group fits within the qubit
budget, and fitting within the budget does not imply that the resulting plan
is practical.

\paragraph{Deployment-budget sensitivity on the main evaluation set}

We first vary the deployment budget over
\[
Q_{\max}\in\{16,20,24,28,32,36,40\}
\]
for the 310-circuit evaluation set, giving 2,170 circuit--budget
evaluations. This sensitivity analysis is separate from the
$Q_{\max}=24$ results in Section~\ref{sec:rq2}.
Coverage uses the observable-point definitions in Section~\ref{sec:rq2}.

For this analysis, a circuit is \emph{fully deployable} when every
numerically selected monitor group can be included in a deployment plan under
the stated qubit budget. Full deployability does not mean that every
gate--qubit location in the source circuit passes the numerical selection
test.

Table~\ref{tab:qmax-sweep} reports the complete sweep. Entries are percentages, and the $Q_{\max}=24$ setting used in the main
evaluation is bold. Increasing the budget
from 16 to 24 raises the mean deployed gate coverage from 38.01\% to
44.54\% and the fraction of fully deployable circuits from 80.97\% to
94.52\%. At $Q_{\max}=28$, every numerically selected group is deployable in
every circuit. None of the reported coverage metrics changes further as
$Q_{\max}$ increases from 28 to 40.

\begin{table*}[!t]
\centering
\caption{Deployment-budget sensitivity on the 310-circuit evaluation set.}
\label{tab:qmax-sweep}
\scriptsize
\setlength{\tabcolsep}{5pt}
\renewcommand{\arraystretch}{1.02}
\begin{tabular}{@{}ccccc@{}}
\toprule
$Q_{\max}$
& Mean deployed gate coverage
& Mean qubit coverage
& Mean normalized depth reach
& Fully deployable circuits \\
\midrule
16 & 38.01 & 89.75 & 47.50 & 80.97 \\
20 & 41.74 & 89.75 & 48.06 & 88.06 \\
\textbf{24}
& \textbf{44.54}
& \textbf{89.75}
& \textbf{48.57}
& \textbf{94.52} \\
28 & 45.89 & 89.75 & 48.81 & 100.00 \\
32 & 45.89 & 89.75 & 48.81 & 100.00 \\
36 & 45.89 & 89.75 & 48.81 & 100.00 \\
40 & 45.89 & 89.75 & 48.81 & 100.00 \\
\bottomrule
\end{tabular}
\end{table*}

\paragraph{MPS-based candidate-selection cost}

Algorithm~\ref{alg:screen} analyzes only the original $n$-qubit source
circuit. Replay ancillas are not present during this selection pass. For a
circuit with $m$ one- and two-qubit gates, the dense statevector
implementation requires
$
O(m2^n)
$
time and
$
O(2^n+m)
$
memory.

To extend candidate selection beyond dense statevector simulation, we use a
matrix product state (MPS) representation. An MPS stores the $n$-qubit state
as a chain of local tensors connected by bond indices. The bond dimension at
each cut bounds the Schmidt rank across that cut. The required maximum bond
dimension $\chi$ therefore depends on the circuit's entanglement structure
and on the simulator's ordering of the qubits. In the worst case, $\chi$ can
grow exponentially with $n$.

With fixed local dimension and maximum bond dimension $\chi$, applying a
two-qubit gate whose operands may be far apart has a conservative
$O(n\chi^3)$ cost. The corresponding state-evolution bound over $m$ gates is
therefore
$
O(mn\chi^3).
$
Local two-qubit gates may be cheaper. The implementation also stores a
full-width MPS snapshot after every analyzed gate, so the stored data can
scale as
$
O(mn\chi^2).
$

These are worst-case bounds rather than predictions of runtime.
The experiment places no explicit upper bound on the bond dimension and uses
a numerical truncation threshold of $10^{-16}$. MPS-based selection therefore
does not have a general polynomial-time guarantee.

\paragraph{MPS scalability set and results}

The MPS scalability set contains 144 generated circuits from 27 circuit
families, with widths from 15 to 50 qubits. It contains 18 circuits at
15 qubits, 23 at 18 qubits, 22 at 20 qubits, 22 at 24 qubits, 19 at
30 qubits, 22 at 40 qubits, and 18 at 50 qubits.

At each source-gate index, all numerical candidates selected after that gate
form one monitor group. Across the 144 circuits, the selector identifies
568,697 numerical candidates, grouped into 446,678 monitor groups.
These counts exclude selected terminal locations already observed by
the original final measurements.

All 144 circuits complete MPS-based candidate selection. The largest observed bond dimension $\chi$ has a
median of 2, a 75th percentile of 14, and a maximum of 2,048.

\paragraph{Grouped reconstruction accounting}

For a selected monitor group $g_i=(i,M_i)$, let
\[
Q_{g_i}
:=
\bigcup_{q\in M_i}Q(i,q),
\qquad
\mathcal{C}_{g_i}
:=
\bigcup_{q\in M_i}\Cone(i,q),
\]
where $Q(i,q)$ is the set of source-circuit qubits touched by the backward
causal cone of gate--qubit location $(i,q)$. Each group uses two
reads and two replays of the same union cone, with local basis rotations before the
second read and reset of the group's ancillas before the second replay.

The ancilla demand and single-pass replay-gate count of the group are
\[
d_{g_i}
=
\left|Q_{g_i}\setminus M_i\right|,
\qquad
s_{g_i}
=
\left|\mathcal{C}_{g_i}\right|.
\]
A source gate that belongs to the cones of several monitored qubits at the
same gate index is counted only once in the group's union cone.

For one deployed batch $b$, different monitor groups use disjoint sets of
fresh ancillas. The structural totals for that batch are therefore
\begin{equation}
Q_{\mathrm{extra}}(b)
=
\sum_{g\in b}d_g,
\qquad
G_{\mathrm{replay}}(b)
=
2\sum_{g\in b}s_g.
\label{eq:grouped-overhead}
\end{equation}

For a plan containing $B\geq1$ separately executed batches, the largest number of
additional qubits required by any one batch is
\[
Q_{\mathrm{peak}}
=
\max_b Q_{\mathrm{extra}}(b).
\]
We also record
\[
Q_{\mathrm{alloc}}
=
\sum_b Q_{\mathrm{extra}}(b),
\]
the sum of the per-batch ancilla counts over all runs. Ancillas can be reused
between runs, so $Q_{\mathrm{alloc}}$ is not the number of additional qubits
required at one time.

Let
\[
G_{\mathrm{replay}}
=
\sum_b G_{\mathrm{replay}}(b)
=2G
\]
be the total number of replayed causal-cone gates across all batches.
Here, $G=\sum_b\sum_{g\in b}s_g$ counts one pass over each deployed union
cone, and $N_{\mathrm{reset}}=\sum_g d_g=Q_{\rm alloc}$ counts replay-ancilla resets before
the second replay. For a
plan with $B$ batches, let $m>0$ be the number of unitary operations in the
source circuit, excluding terminal measurements and barriers. Let
$E_{\mathrm{deploy}}$ be the number of deployed gate--qubit locations that
receive two inserted measurements and resets and one basis rotation.

To compare the plan's counted operations with the size of the original
circuit, we define the \emph{operation-count multiplier}
\begin{equation}
\mu_{\mathrm{op}}
=
\frac{
B\,m
+
2G
+
5E_{\mathrm{deploy}}+N_{\mathrm{reset}}
}{m}.
\label{eq:planned-operation-multiplier}
\end{equation}
The term $Bm$ counts one execution of the original unitary circuit in every
batch. The term $2G$ counts the replayed cone gates, and
$5E_{\mathrm{deploy}}$ counts one basis rotation, two measurements, and two resets
for each deployed gate--qubit location. Each basis rotation counts as one abstract
gate before native-gate decomposition. The term $N_{\mathrm{reset}}$ counts the ancilla resets
before the second replay. Existing terminal measurements, barriers, ancilla
allocation, and classical bookkeeping or control are not counted. Therefore,
$\mu_{\mathrm{op}}$ is a structural operation count, not an estimate of
hardware runtime or physical error.

A final-read-only plan instead executes the original circuit once,
with $Q_{\mathrm{peak}}=Q_{\mathrm{alloc}}=G_{\mathrm{replay}}=0$ and
$\mu_{\mathrm{op}}=1$ for $m>0$. This execution is not included in the
batch count returned by Algorithm~\ref{alg:batch}.

\paragraph{Large-circuit instrumentation planning}

Completing MPS-based candidate selection does not imply that the resulting
monitoring plan fits within a practical execution width. We therefore apply
the structural planner to each of the 144 circuits for
\[
Q_{\max}\in\{60,70,80,90,100\}.
\]

At each gate index, all selected qubits form one indivisible monitor
group. A group is excluded if its ancilla demand exceeds
$Q_{\max}-Q_{\mathrm{base}}$. We use the conditional FFD--ILP procedure
in Appendix~\ref{app:implementation-details}, with a 120-second limit
for each triggered ILP check.

For each circuit and budget, we record the numbers of selected and deployed
monitor groups, the numbers of selected and deployed gate--qubit locations,
whether the circuit is fully deployable, the batch count, the additional
ancilla requirements, and the operation-count multiplier.

Let $G_{\mathrm{candidate}}(C)$ and $G_{\mathrm{deploy}}(C)$ be the numbers
of selected and deployed monitor groups in circuit $C$. The pooled selected-group coverage is
\[
\frac{
\sum_C G_{\mathrm{deploy}}(C)
}{
\sum_C G_{\mathrm{candidate}}(C)
}.
\]

Similarly, let $E_{\mathrm{candidate}}(C)$ and
$E_{\mathrm{deploy}}(C)$ be the numbers of selected and deployed
gate--qubit locations. The pooled selected-location coverage is
\[
\frac{
\sum_C E_{\mathrm{deploy}}(C)
}{
\sum_C E_{\mathrm{candidate}}(C)
}.
\]

The final two columns of Table~\ref{tab:mps-planning-sweep} report
the mean, median, and 75th percentile over the same 144 circuits at
each budget. The 75th percentile uses the nearest-rank definition and is therefore the 108th
value after sorting the 144 circuit-level values. The operation counts include both replays and the basis rotations and are
reported to one decimal from structural counts of the fixed selections.
The denominators contain 446,678 selected monitor groups and 568,697 selected
mid-circuit gate--qubit locations. These metrics therefore measure how many numerical
candidates can be deployed. They do not measure coverage of all gates or all
gate--qubit locations in the source circuits.

\begin{table*}[!t]
\centering
\caption{Structural instrumentation planning on the 144-circuit MPS scalability set.}
\label{tab:mps-planning-sweep}

\begingroup
\scriptsize
\setlength{\tabcolsep}{1.8pt}
\renewcommand{\arraystretch}{1.00}

\begin{tabular*}{\textwidth}{@{\extracolsep{\fill}}rccccc@{}}
\toprule
$Q_{\max}$
& Fully deployable circuits
& Selected-group coverage
& Selected-location coverage
& Planned batches
& $\mu_{\mathrm{op}}$ ($\times$) \\
\multicolumn{6}{l}{Coverage is pooled; triples give mean / median / 75th percentile.} \\
\midrule

60
& 118/144 (81.94\%)
& 37.89\%
& 37.74\%
& $797.08\,/\,6\,/\,85$
& $1{,}595.3\,/\,12.7\,/\,242.6$ \\

70
& 118/144 (81.94\%)
& 41.88\%
& 42.41\%
& $761.98\,/\,6\,/\,85$
& $1{,}623.0\,/\,14.7\,/\,234.7$ \\

80
& 133/144 (92.36\%)
& 68.26\%
& 69.09\%
& $1{,}529.53\,/\,6\,/\,121$
& $3{,}088.2\,/\,24.8\,/\,354.0$ \\

90
& 133/144 (92.36\%)
& 68.51\%
& 69.43\%
& $1{,}299.02\,/\,5.5\,/\,97$
& $2{,}862.7\,/\,28.2\,/\,343.0$ \\

100
& 144/144 (100.00\%)
& 100.00\%
& 100.00\%
& $2{,}163.00\,/\,5\,/\,156$
& $4{,}424.4\,/\,28.9\,/\,436.4$ \\

\bottomrule
\end{tabular*}
\endgroup
\end{table*}

As shown in Table~\ref{tab:mps-planning-sweep}, increasing $Q_{\max}$ progressively removes deployment gaps caused by the
qubit limit. At $Q_{\max}=100$, every numerically selected monitor group and
every selected gate--qubit location is included in a deployment plan for all
144 circuits. This 100\% coverage is relative to the numerical candidates.
It does not mean that every gate--qubit location in the source circuits
passes the selection test.

The planning-cost distributions contain a small number of extremely large
cases, so their means are much larger than their medians. For example,
at $Q_{\max}=100$, the mean, median, and 75th percentile batch counts are 2,163, 5, and
156, respectively. The corresponding values of
$\mu_{\mathrm{op}}$ are $4{,}424.4\times$, $28.9\times$, and $436.4\times$.

The planning cost does not necessarily decrease as $Q_{\max}$ increases.
A larger qubit budget can make previously infeasible monitor groups
deployable, which increases the amount of monitoring included in the plan.
The largest plan at $Q_{\max}=100$ requires 134,893 batches and has an operation-count multiplier of $229{,}306.30\times$. Thus, making every
selected monitor group deployable does not guarantee a practical
instrumentation cost.

This 144-circuit study evaluates only MPS-based candidate selection and
structural deployment planning. It does not execute the resulting
instrumented circuits on either a simulator or a physical quantum backend.

\subsection{Multiple Simultaneous Faults}
\label{sec:multi-error}

The primary RQ3 experiment injects one fault at a time. A circuit may,
however, contain more than one fault. We therefore extend the RQ3 protocol
using the same 238 circuits and five mutation seeds.

For each circuit--seed pair, three mutations were generated in advance. For
fault multiplicity
\[
f\in\{1,2,3\},
\]
the jointly mutated circuit contains the first $f$ of these three mutations.
Thus, the one-fault, two-fault, and three-fault conditions use nested mutation
sets within each circuit--seed pair.

Each jointly mutated circuit is evaluated using the fixed monitoring plan and
expected one-qubit probabilities derived from the original circuit. The
checkpoint locations, expected states and basis rotations, backward causal cones, and batch
assignments are not recomputed after the mutations are introduced.
Both replays use
the circuit-under-test operations at the fixed cone instruction indices.

\textsc{QMon} and the three finite-shot assertion methods use
$8{,}192$ shots per explicitly executed run. \textsc{QMon} and the
statistical assertion method retain the same per-circuit Bonferroni
corrections as in the primary RQ3 experiment. Projection and dynamic ancilla
assertions retain their published any-failure rules, and MQT retains its
deterministic combined-flag rule.

The evaluation-only no-effect rules described in
Appendix~\ref{app:rq3details} also remain unchanged. With multiple faults, a
checkpoint is treated as unaffected only when none of the injected gates lies
in its backward causal cone.

For this extension, checkpoint scans continue after the first flag. This is
necessary because the localization analysis uses the union of all
\textsc{QMon} flagged cones. For every circuit, seed, and fault multiplicity,
we record whether each method completes, is inapplicable, times out, or
returns an error. The detection comparison uses the non-output-equivalent
cases completed by all methods.

We apply the output-equivalence criterion in
Appendix~\ref{app:rq3details:protocol} and exclude those cases from the
common-method detection population.

Within the common-method detection population, we evaluate localization
for cases detected by \textsc{QMon}. For such a case $r$, let
$\mathcal{F}_r$ be the set of injected gate indices. Let $\mathcal{S}_r$ be the union of
the original-circuit backward causal cones of all \textsc{QMon}
checkpoints in the fixed monitoring plan. A fault is \emph{structurally
covered} when its gate index belongs to $\mathcal{S}_r$.

\begin{itemize}
    \item \emph{Any-fault localization} is evaluated only when
    $\mathcal{F}_r\cap\mathcal{S}_r\neq\varnothing$. It succeeds when
    the backward causal cone of the earliest flagged checkpoint contains
    at least one gate in $\mathcal{F}_r$.

    \item \emph{All-fault localization} is evaluated only when
    $\mathcal{F}_r\subseteq\mathcal{S}_r$. It succeeds when the union of
    all flagged checkpoint cones contains every gate in $\mathcal{F}_r$.
\end{itemize}

Both measures use backward causal cones from the original circuit.
They evaluate the regions reported by \textsc{QMon}; they are
not additional fault-detection rules. A finite-shot flag can count as a
detection even when no fault is structurally covered. Such a case is
ineligible for both localization measures, so their denominators can be
smaller than the \textsc{QMon} detection numerators.

\paragraph{Results}

Detection entries in Table~\ref{tab:multi-error} are the unweighted mean
$\pm$ sample standard deviation of the five per-seed rates. The localization
rows use the same reporting format over their eligible detected cases.
Parenthesized fractions are pooled counts. The
MQT row reports a combined flag consisting of either an evaluated state
mismatch or an unevaluable outcome. Timeout, error, and
budget-exhausted outcomes are treated as incomplete.

\begin{table*}[!t]
\centering
\caption{Detection and conditional localization with one, two, or three simultaneous faults.}
\label{tab:multi-error}
\scriptsize
\setlength{\tabcolsep}{5pt}
\renewcommand{\arraystretch}{1.03}
\begin{tabular}{@{}lccc@{}}
\toprule
Method/measure & One fault & Two faults & Three faults \\
\midrule

\textsc{QMon} detection
 & $74.0\pm7.4\%$ (675/913)
 & $81.8\pm5.1\%$ (831/1,015)
 & $86.5\pm7.4\%$ (893/1,033) \\

Statistical assertion
& $70.5\pm7.5\%$ (643/913)
& $80.1\pm5.3\%$ (813/1,015)
& $85.4\pm6.1\%$ (881/1,033) \\

Projection assertion
& $73.1\pm6.6\%$ (667/913)
& $81.9\pm5.8\%$ (832/1,015)
& $86.5\pm7.4\%$ (893/1,033) \\

Dynamic ancilla
& $73.5\pm7.0\%$ (671/913)
& $81.9\pm5.8\%$ (832/1,015)
& $86.5\pm7.4\%$ (893/1,033) \\

MQT combined flag
& $59.4\pm10.7\%$ (540/913)
& $69.7\pm4.5\%$ (707/1,015)
& $73.9\pm6.5\%$ (762/1,033) \\

\midrule

\textsc{QMon} Any-fault localization
 & $99.7\pm0.4\%$ (671/673)
 & $100.0\pm0.0\%$ (828/828)
 & $100.0\pm0.0\%$ (893/893) \\

\textsc{QMon} All-fault localization
 & $99.9\pm0.3\%$ (672/673)
 & $94.4\pm2.2\%$ (632/669)
 & $94.8\pm1.4\%$ (618/652) \\
\bottomrule
\end{tabular}
\end{table*}

Each fault multiplicity contains 1,190 attempted circuit--seed cases.
\textsc{QMon} and the three finite-shot assertion methods return a decision
for every attempted case. MQT is the only method with incomplete outcomes:
it returns no decision for 78, 66, and 72 cases for
$f=1$, $f=2$, and $f=3$, respectively.

Among the 964, 1,070, and 1,092 non-output-equivalent cases, the
corresponding numbers of incomplete MQT outcomes are 51, 55, and 59.
The common-method detection populations therefore contain
$964-51=913$, $1{,}070-55=1{,}015$, and
$1{,}092-59=1{,}033$ cases, as reported in
Table~\ref{tab:multi-error}.

An MQT unevaluable outcome remains a combined flag and is not treated
as incomplete. Only a timeout, tool error, or budget-exhausted outcome
produces no decision and removes the case from the corresponding
common-method detection population.

The $f=1$ condition uses only the first of the three mutations generated for
each circuit--seed pair. It is therefore a different population from the
primary single-fault RQ3 experiment, which evaluates all three mutations as
separate mutants.

For $f=1$, two of the 675 detected cases have no structurally covered fault.
Both localization measures therefore use the remaining 673 cases.
Their success
counts can differ even with one fault: the earliest flagged cone may miss
the fault while a later flagged cone includes it.

For $f=2$, three of the 831 detected cases have no structurally covered fault,
leaving 828 cases for Any-fault localization. In 162 detected cases, at least one injected
fault is not structurally covered. All-fault localization therefore uses the remaining
669 cases.

For $f=3$, all 893 detected cases have at least one structurally covered
fault, so all are eligible for Any-fault localization. In 241 cases, at least one injected
fault is not structurally covered, leaving 652 cases for All-fault localization.

Across the three separate populations, the observed detection rate increases
with fault multiplicity for every method. These are descriptive
cross-population comparisons. Because the populations differ across
$f=1,2,3$, the results do not show what would happen if additional faults
were added to the same set of mutants.

Among detected cases with at least one structurally covered fault, the
earliest flagged \textsc{QMon} cone contains an injected fault in more than
99\% of cases at every multiplicity. Among detected cases in which all
injected faults are structurally covered, the union of flagged cones contains
every fault in $99.9\pm0.3\%$ of the one-fault cases,
$94.4\pm2.2\%$ of the two-fault cases, and
$94.8\pm1.4\%$ of the three-fault cases.

Thus, the union of flagged cones contains
every injected fault in approximately 94--95\% of the eligible two- and
three-fault detected cases.

\subsection{Use in the Quantum Software Workflow}
\label{sec:se-workflow}

\textsc{QMon} is intended for debugging and pre-deployment validation rather
than continuous production monitoring. A developer may use it after ordinary
circuit simulation, when a test output differs from the expected result,
during a dedicated validation run on a compatible backend, or after modifying
a circuit to check for regressions.

The purpose of
instrumentation is to collect checkpoint observations from a tested
execution environment using a previously constructed monitoring plan.

The present study evaluates the complete method in noise-free
simulation and with calibration-informed simulated noise. Running an instrumented circuit on physical hardware would
require compatible support for mid-circuit measurement and reset, as well as
enough qubits and gate support for causal-cone replay.

The developer does not need to provide a separate expected value for every
checkpoint. During analysis of the original circuit, \textsc{QMon} computes
the expected one-qubit states, computational-basis probabilities,
and basis rotations for the checkpoints included in the monitoring
plan. In the noise-free experiments, recorded outcomes are compared with
these expected probabilities. In RQ4, they are compared with calibration
outcomes collected from the original circuit under the same simulated
noise conditions. Calibration shots are independent of the test shots.

In the simulated setting evaluated here, the intermediate states
can also be obtained by directly simulating the original and tested
circuits. The monitoring operation could also use valid separable
checkpoints and expected states supplied by a programmer. This
alternative to automatic selection has not been evaluated here.

\textsc{QMon} operates on the gate-level circuit after decomposition into
one- and two-qubit gates. It inserts ordinary gates, measurements, and resets
while preserving the original gate order outside the inserted monitoring
operations. Our implementation targets Qiskit. Porting the method to another
SDK or backend requires equivalent gate behavior, qubit and classical-bit
mappings, and mid-circuit measurement and reset support. We do not evaluate
portability across software stacks in this work. The recorded outcomes are
used only for later analysis and do not control subsequent gates. Across repeated runs, the checkpoint observations provide a structured
record of intermediate circuit behavior. These observations could support
mutation testing~\cite{9678792,9678563} by providing internal evidence in
addition to final-output comparison.

\section{Noise Evaluation and Small-Angle Sensitivity}
\label{app:rq4details}

\subsection{Model and Decision Details}

The configurations and execution records for the noise evaluation and
small-angle sweep are indexed in the replication package linked in
Section~\ref{sec:introduction}.

RQ4 uses homogeneous medians from the FakeMontrealV2 calibration
snapshot dated 15 March 2021 (\texttt{qiskit-ibm-runtime} version 0.47.0), applied to all-to-all
logical circuits. The base $T_1$ and $T_2$ are 99.60 and 77.93\,$\mu$s.
The native $\mathrm{id}$, $\mathrm{sx}$, and $\mathrm{x}$ durations are
35.56\,ns and their depolarizing parameter is $1.5123\times10^{-4}$;
for $\mathrm{cx}$ these values are 405.33\,ns and $4.3365\times10^{-3}$.
Frozen $\mathrm{rz}$ gates have zero duration and no error channel. Thermal
relaxation accompanies executed gates and scheduled idle intervals.

Measurement records an effective assignment error at the start of the
read, followed by relaxation of the postmeasurement state during the
5.202\,$\mu$s measurement interval; the base assignment probabilities are
$p(1\mid0)=0.0116$ and $p(0\mid1)=0.0232$. Reset duration 1\,$\mu$s and
post-reset $X$-error probability 0.001 are explicit assumptions, not snapshot
measurements. At positive scale $s$, the measurement-assignment error probabilities,
post-reset $X$-error probability, and depolarizing parameters are
multiplied by $s$, while $T_1$ and $T_2$ are divided by $s$.
Gate, measurement, and reset durations remain fixed at every scale.
Scale zero disables all noise channels while retaining these durations. Native compilation
uses optimization level zero and transpiler seed 1729, after removing
barriers from every configuration. Dynamic-ancilla assertion templates
first undergo Qiskit level-3 resynthesis using the Cartan (KAK) decomposition
to avoid excess CNOTs from the controlled-$U$ library expansion, consistent
with Liu et al.~\cite{10.1145/3373376.3378488}. The KAK decomposition factors
a two-qubit unitary, up to global phase, into local single-qubit operations
before and after a canonical two-qubit interaction~\cite{zhang2003geometric}.
The unitary body is checked
for equivalence up to global phase and a non-increasing CNOT count;
measurements and resets are unchanged. Only dynamic-ancilla templates
receive this preprocessing, before the common level-zero compilation.
The noise comparisons concern these compiled configurations; ideal-unitary
equivalence does not imply identical noisy behavior. Instructions start when all operand
qubits are available; waiting operands undergo thermal relaxation.
Reset applies an ideal reset, relaxation during the reset interval, and
then the specified $X$ error. No routing, device
heterogeneity, crosstalk, leakage, or drift is modeled.

For each method and noise scale, we collect calibration outcomes from
its execution configuration of the original circuit, assumed to be correct.
Calibration shots are independent of the test shots. A batch receives 32,768 calibration shots and 8,192 shots
per tested mutant or correct-circuit repetition. Statistical assertions
receive these budgets at every separately executed checkpoint prefix. Twenty correct tests
per circuit and scale give 2,000 tests on the 100-circuit noise set.
Calibration is shared across that circuit's tests, so these repeated
decisions are not independent observations.

QMon and statistical assertions apply two-sided Fisher tests to test
versus calibration 0/1 counts. Projection and dynamic assertions apply
one-sided Fisher tests to first failures among shots still at risk (no
previous recorded assertion failure). A later read after a failure does not
contribute a fresh at-risk observation. The strict rejection threshold is
$0.01/K_{\mathrm{QMon}}$ for QMon, with
$K_{\mathrm{QMon}}=2K_{\mathrm{mid}}+K_{\mathrm{final}}$ as defined in
Appendix~\ref{app:rq3details:protocol}; eligible final readouts use the
same calibrated test once. Each assertion method divides $0.01$ by its
complete declared checkpoint count. These calibrated
decisions are explicit adaptations of every method, including QMon; the
noise-free rules are retained only as diagnostics. All planned shots execute; there is
no automatic top-up after postselection. A checkpoint with no usable test
or calibration samples supplies no fault flag; the circuit is flagged only
if at least one valid checkpoint comparison flags it.

Across correct-circuit tests, every calibration checkpoint retained
usable samples. Zero usable test samples occurred only for dynamic-ancilla
assertions at scale~2, in 102 of 16,080 checkpoint comparisons. At that
scale, the minimum/median/maximum usable test counts were
3,277/7,432/8,192 for projection and 0/6,428.5/8,192 for dynamic-ancilla
assertions (checkpoint-weighted, including zeros). QMon and statistical
assertions used all 8,192 test shots per checkpoint at every scale.
These checkpoint counts do not change the 2,000-test false-alarm denominator.

\subsection{Full Detection Results and Paired Comparisons}

Table~\ref{tab:rq4-full} gives all-mutant results on the same 1,373
mutants at each scale. Rates are the unweighted five-seed mean $\pm$ sample
standard deviation, with pooled detection counts in parentheses. The
candidate-covered and correct-circuit results are in
Table~\ref{tab:rq4-noise}. All evaluated cases remain in the denominators;
only explicit fault flags count as detections or false alarms.

\begin{table*}[!t]
\centering
\caption{All-mutant detection under calibration-informed simulated noise.}
\label{tab:rq4-full}
\scriptsize
\setlength{\tabcolsep}{3pt}
\begin{tabular*}{\textwidth}{@{\extracolsep{\fill}}lcccc@{}}
\toprule
Scale & \textsc{QMon} & Projection & Dynamic ancilla & Statistical \\
\midrule
0 & $54.9\pm10.1\%$ (754/1,373) & $54.4\pm10.2\%$ (747/1,373) & $54.5\pm10.1\%$ (748/1,373) & $52.4\pm8.6\%$ (720/1,373) \\
0.25 & $54.3\pm10.1\%$ (745/1,373) & $54.1\pm10.2\%$ (743/1,373) & $53.6\pm11.0\%$ (735/1,373) & $52.2\pm8.9\%$ (716/1,373) \\
0.5 & $54.2\pm9.7\%$ (744/1,373) & $54.1\pm10.3\%$ (743/1,373) & $53.1\pm10.9\%$ (728/1,373) & $51.9\pm9.1\%$ (712/1,373) \\
1 & $54.1\pm10.5\%$ (743/1,373) & $53.7\pm10.6\%$ (737/1,373) & $52.7\pm10.4\%$ (723/1,373) & $51.9\pm9.1\%$ (712/1,373) \\
2 & $54.0\pm10.4\%$ (741/1,373) & $53.3\pm11.0\%$ (731/1,373) & $50.9\pm11.2\%$ (699/1,373) & $52.0\pm9.5\%$ (714/1,373) \\
\bottomrule
\end{tabular*}
\end{table*}

For exploratory paired comparisons between \textsc{QMon} and
each of the three assertion methods, Table~\ref{tab:rq4-mcnemar}
reports discordant counts and two-sided exact McNemar $p$-values on the
same 762 candidate-covered mutants at every scale.
For discordant counts $b,c$, the calculation is
$\min\{1,2\Pr[\operatorname{Bin}(b+c,1/2)\leq\min(b,c)]\}$.
These nominal $p$-values are not adjusted across methods or scales;
the checkpoint correction above controls a different testing family.
The tests also do not account for dependence among mutants of the
same circuit or tests sharing calibration data.
At the nominal 0.01 level, no comparison with projection assertions
shows a statistically significant difference; this does not establish
equivalence.
\textsc{QMon} achieves significantly higher candidate-covered detection
than statistical assertions at all five scales, with nominal $p$-values
from $1.46\times10^{-11}$ to $7.92\times10^{-9}$, and
dynamic-ancilla assertions at all four positive scales, with nominal
$p$-values from $1.82\times10^{-12}$ to 0.0034.

\begin{table*}[!t]
\centering
\caption{Paired candidate-covered detections: QMon-only/other-only counts and exact McNemar $p$.}
\label{tab:rq4-mcnemar}
\scriptsize
\setlength{\tabcolsep}{5pt}
\begin{tabular*}{\textwidth}{@{\extracolsep{\fill}}lcccccc@{}}
\toprule
Scale & \multicolumn{2}{c}{Projection} & \multicolumn{2}{c}{Dynamic ancilla} & \multicolumn{2}{c}{Statistical} \\
\cmidrule(lr){2-3}\cmidrule(lr){4-5}\cmidrule(l){6-7}
& QMon-only / other-only & $p$ & QMon-only / other-only & $p$ & QMon-only / other-only & $p$ \\
\midrule
0 & 0/0 & $1$ & 0/1 & $1$ & 37/0 & $1.46\times10^{-11}$ \\
0.25 & 4/3 & $1$ & 12/1 & $0.0034$ & 32/1 & $7.92\times10^{-9}$ \\
0.5 & 7/3 & $0.344$ & 17/1 & $1.45\times10^{-4}$ & 33/1 & $4.07\times10^{-9}$ \\
1 & 8/2 & $0.109$ & 22/2 & $3.59\times10^{-5}$ & 32/1 & $7.92\times10^{-9}$ \\
2 & 12/2 & $0.0129$ & 40/0 & $1.82\times10^{-12}$ & 28/0 & $7.45\times10^{-9}$ \\
\bottomrule
\end{tabular*}
\end{table*}

\subsection{Small-Angle and Phase-Fault Sweep}
\label{app:small-angle}

An additional sweep uses 14 circuits. For each circuit, we list RX/RY/RZ
instructions in source order. If at least two are themselves planned
checkpoints, including eligible final readouts, we select the first and
last of these, then the last remaining rotation outside all planned
checkpoint backward causal cones, or the last remaining rotation if none
is outside. Otherwise, we select the first, middle (the later middle
instruction when the number is even), and last rotation.
A gate is in a checkpoint's backward cone when it can structurally affect
the target qubit immediately after that checkpoint's source instruction,
including the checkpoint instruction itself. Unique source indices
determine the order; selection does not inspect detection outcomes.
We perturb each selected angle by
$\pm\delta$ for $\delta\in\{0.02,0.05,0.1,0.2,0.5\}$ radians. Phase
faults replace an RX/RY instruction by the equivalent single-instruction
composition $R_z(\delta)R_{x/y}(\theta)$ for $\delta\in\{0.2,0.5\}$.
All 42 selected sites are RX/RY gates. Phase faults
use only positive $\delta$ at these same sites. The full sweep therefore
contains $14\times3\times(2\times5+2)=504$ variants: 420 parameter
faults and 84 phase faults. Table~\ref{tab:delta-sweep} reports
only candidate-covered variants: 28 parameter faults per magnitude and 14
phase faults per magnitude. These arise from 14 covered RY sites in
seven circuits, giving 168 covered and 336 uncovered variants overall.
The noise model, calibration, and shot budgets
are those above. Each scale also includes 280 correct tests.

At scale 1 and $\delta=0.05$, QMon detects 19/28 parameter faults,
compared with 0/28 for projection and dynamic assertions and 17/28 for
statistical assertions. In the ideal pure-state model, these parameter changes can alter
the computational-basis probability to first order, while the
expected-state overlap loss begins at second order. This sensitivity is shared with statistical
assertions, is not universal for all rotation types, and does not imply
that an $R_z$ perturbation changes a computational-basis probability.
The phase-fault rows instead illustrate the second read;
they do not establish a uniform advantage over projection.

\begin{table*}[!t]
\centering
\caption{Small-angle sweep: detected/candidate-covered variants; FA rows give correct-test flags/280.}
\label{tab:delta-sweep}
\scriptsize
\setlength{\tabcolsep}{3pt}
\renewcommand{\arraystretch}{1.02}
\begin{tabular*}{\textwidth}{@{\extracolsep{\fill}}llcccc@{}}
\toprule
Scale & Fault / $\delta$ (rad) & \textsc{QMon} & Projection & Dynamic ancilla & Statistical \\
\midrule
0 & Parameter / 0.02 & 0/28 & 0/28 & 0/28 & 1/28 \\
 & Parameter / 0.05 & 24/28 & 21/28 & 18/28 & 23/28 \\
 & Parameter / 0.1 & 28/28 & 28/28 & 28/28 & 28/28 \\
 & Parameter / 0.2 & 28/28 & 28/28 & 28/28 & 28/28 \\
 & Parameter / 0.5 & 28/28 & 28/28 & 28/28 & 28/28 \\
 & Phase / 0.2 & 12/14 & 12/14 & 12/14 & 0/14 \\
 & Phase / 0.5 & 13/14 & 13/14 & 13/14 & 0/14 \\
 & FA & 0/280 & 0/280 & 0/280 & 5/280 \\
\midrule
0.25 & Parameter / 0.02 & 2/28 & 0/28 & 1/28 & 1/28 \\
 & Parameter / 0.05 & 16/28 & 0/28 & 0/28 & 19/28 \\
 & Parameter / 0.1 & 27/28 & 20/28 & 7/28 & 26/28 \\
 & Parameter / 0.2 & 28/28 & 28/28 & 24/28 & 28/28 \\
 & Parameter / 0.5 & 28/28 & 28/28 & 28/28 & 28/28 \\
 & Phase / 0.2 & 11/14 & 12/14 & 7/14 & 0/14 \\
 & Phase / 0.5 & 12/14 & 13/14 & 12/14 & 0/14 \\
 & FA & 2/280 & 1/280 & 4/280 & 3/280 \\
\midrule
0.5 & Parameter / 0.02 & 3/28 & 0/28 & 0/28 & 3/28 \\
 & Parameter / 0.05 & 20/28 & 0/28 & 0/28 & 23/28 \\
 & Parameter / 0.1 & 26/28 & 7/28 & 2/28 & 26/28 \\
 & Parameter / 0.2 & 28/28 & 28/28 & 23/28 & 28/28 \\
 & Parameter / 0.5 & 28/28 & 28/28 & 28/28 & 28/28 \\
 & Phase / 0.2 & 9/14 & 9/14 & 6/14 & 0/14 \\
 & Phase / 0.5 & 12/14 & 12/14 & 12/14 & 0/14 \\
 & FA & 1/280 & 1/280 & 0/280 & 1/280 \\
\midrule
1 & Parameter / 0.02 & 0/28 & 0/28 & 0/28 & 4/28 \\
 & Parameter / 0.05 & 19/28 & 0/28 & 0/28 & 17/28 \\
 & Parameter / 0.1 & 25/28 & 3/28 & 1/28 & 26/28 \\
 & Parameter / 0.2 & 28/28 & 28/28 & 18/28 & 28/28 \\
 & Parameter / 0.5 & 28/28 & 28/28 & 28/28 & 28/28 \\
 & Phase / 0.2 & 9/14 & 10/14 & 5/14 & 0/14 \\
 & Phase / 0.5 & 12/14 & 12/14 & 11/14 & 0/14 \\
 & FA & 1/280 & 3/280 & 2/280 & 3/280 \\
\midrule
2 & Parameter / 0.02 & 0/28 & 1/28 & 0/28 & 2/28 \\
 & Parameter / 0.05 & 14/28 & 0/28 & 0/28 & 14/28 \\
 & Parameter / 0.1 & 25/28 & 0/28 & 2/28 & 24/28 \\
 & Parameter / 0.2 & 27/28 & 26/28 & 11/28 & 26/28 \\
 & Parameter / 0.5 & 28/28 & 28/28 & 25/28 & 28/28 \\
 & Phase / 0.2 & 6/14 & 7/14 & 2/14 & 0/14 \\
 & Phase / 0.5 & 12/14 & 12/14 & 11/14 & 0/14 \\
 & FA & 6/280 & 1/280 & 8/280 & 2/280 \\
\bottomrule
\end{tabular*}
\end{table*}

\end{document}